\documentclass[aps,onecolumn]{revtex4}
\usepackage{amssymb,epsf}
\usepackage{latexsym}

\begin{document}

\title{Black hole thermodynamics in Lovelock gravity's
rainbow with (A)dS asymptote }
\author{Seyed Hossein Hendi$^{1,2}$\footnote{email address: hendi@shirazu.ac.ir},
Ali Dehghani$^1$\footnote{email address: ali.dehghani.phys@gmail.com}, and
Mir Faizal$^{3, 4}$%
\footnote{ email address: f2mir@uwaterloo.ca}} \affiliation{$^1$
Physics Department and Biruni Observatory, College of Sciences,
Shiraz University, Shiraz 71454, Iran \\
$^2$ Research Institute for Astrophysics and Astronomy of Maragha
(RIAAM), P.O. Box 55134-441, Maragha, Iran\\
$^3$ Irving K. Barber School of Arts and Sciences, \\
University of British Columbia - Okanagan,
Kelowna,   BC V1V 1V7, Canada 
\\$^4$ Department of Physics and Astronomy, University of
Lethbridge, Lethbridge, AB T1K 3M4, Canada}

\begin{abstract}
In this paper, we combine Lovelock gravity with gravity's rainbow
to construct Lovelock gravity's rainbow. Considering the Lovelock
gravity's rainbow coupled to linear and also nonlinear
electromagnetic gauge fields, we present two new classes of
topological black hole solutions. We compute conserved and
thermodynamic quantities of these black holes (such as
temperature, entropy, electric potential, charge and mass) and
show that these quantities satisfy the first law of
thermodynamics.  In order to study the thermal stability in
canonical ensemble, we calculate the heat capacity and determinant
of the Hessian matrix and show in what regions there are thermally
stable phases for black holes. Also, we discuss the dependence of
thermodynamic behavior and thermal stability of black holes on
rainbow functions. Finally, we investigate the critical behavior
of black holes in the extended phase space and study their
interesting properties.

\end{abstract}

 \maketitle

\section{Introduction}

It is important to understand the UV behavior of general relativity, and
various attempts have been made to obtain the UV completion of general
relativity such that it reduces to the general relativity in the IR limit.
This has been the main motivation behind the development of the
Horava-Lifshitz gravity \cite{HoravaPRD,HoravaPRL}. In the Horava-Lifshitz
gravity, space and time are made to have different Lifshitz scaling. Thus,
the Horava-Lifshitz gravity reduces to general relativity in the IR limit.
However, its behavior in the UV limit is different from that of general
relativity. Motivated by the development of the Horava-Lifshitz gravity, the
UV completion of various geometric structures in the string theory has been
studied by taking a different Lifshitz scaling for space and time. In fact,
such a UV completion has been studied for geometries that occur in the type
IIA string theory \cite{A} and type IIB string theory \cite{B}. The AdS/CFT
correspondence has also been used for analyzing the geometries where the
space and time have different Lifshitz scaling \cite{ho,h1,h2,oh}. This
formalism has also been applied for analyzing the UV completion of dilaton
black branes \cite{d,d1} and dilaton black holes \cite{dh,hd}. The
Horava-Lifshitz gravity is based on a deformation of the usual
energy-momentum dispersion relation in the UV limit, such that it reduces to
the usual energy-momentum dispersion relation in the IR limit. Such a
deformation of usual energy-momentum dispersion in the UV limit has been
observed to occur in ghost condensation \cite{FaizalJPA} and non-commutative
geometry \cite{Carroll,FaizalMPLA}.

There is another approach for obtaining the UV completion of general
relativity based on the deformation of the usual energy-momentum dispersion
relation in the UV limit, and this approach is called gravity's rainbow \cite%
{Magueijo:2002xx}. In gravity's rainbow, the geometry of spacetime is made
energy dependent and this energy dependence of the spacetime metric is
incorporated through the introduction of rainbow functions. It has been
suggested that the deflection of light and gravitational red-shift can be
used to test various choices of rainbow functions \cite{rf}.

It has been demonstrated that for a certain choice of rainbow functions, the
gravity's rainbow is related to the Horava-Lifshitz gravity \cite{re}.
String theory can also be used as a motivation for gravity's rainbow. This
is because the background fluxes in string theory produce a noncommutative
deformation of the geometry \cite{st,st1}, and noncommutativity has also
been used to motivate one of the most important rainbow functions in
gravity's rainbow \cite{a,j}. In string field theory, a tachyon field can
have the wrong sign for its mass squared, and the perturbative string vacuum
become unstable in the presence of such a tachyon field \cite{Samuel1}. This
existence of such an unstable perturbative string vacuum spontaneously
breaks the Lorentz symmetry. It may be noted that a gravitational Higgs
mechanism in supergravity theories also spontaneously breaks the Lorentz
symmetry \cite{Samuel2}. The spontaneous breaking of the Lorentz symmetry
deforms the usual energy-momentum relations, and this in turn can be used as
a motivation for introducing gravity's rainbow.

In gravity's rainbow a one-parameter family of energy dependent
orthonormal frame fields introduced. This gives rise to a
one-parameter family of energy dependent metrics as, $g^{\mu \nu
}(\varepsilon )=e_{a}^{\mu }(\varepsilon )e^{a\nu }(\varepsilon )$
{in which $\varepsilon =E/E_{p}$ is dimensionless energy
ratio, $E$ is the energy of the test particle and $E_{P}$ is the
Planck energy \cite{Magueijo:2002xx}.} Here, the rainbow functions
are used to relate these new tetrad fields to
the usual frame fields $\tilde{e}_{a}^{\mu }$ of general relativity, $%
f(\varepsilon )e_{0}^{\mu }(\varepsilon )=\tilde{e}_{0}^{\mu }$ and $%
g(\varepsilon )e_{i}^{\mu }(\varepsilon )=\tilde{e}_{i}^{\mu }$, where $i$
is the spatial index. In the IR limit, $\varepsilon \rightarrow 0$, we have $%
\lim_{\varepsilon \rightarrow 0}f(\varepsilon )=\lim_{\varepsilon
\rightarrow 0}g(\varepsilon )=1$, and so in the IR limit one recovers the
general relativity. It may be noted that if the energy $E$ was just a
non-dynamical parameter in the theory then we could gauge it away by
rescaling. However, it dynamically depends on the coordinates, and it
originally breaks the diffeomorphisms symmetry of the full metric. In fact,
even the local symmetry in gravity's rainbow is not Lorentz symmetry, as it
is based on deformation of the usual energy-momentum dispersion relation,
which breaks the Lorentz symmetry in the UV limit (similar to
Horava-Lifshitz gravity) \cite{Magueijo:2002xx}. As the energy $\varepsilon $
is an implicit function of the coordinates, and the the rainbow functions
are explicit functions of the energy ratio $\varepsilon $, they are also
dynamical functions of the coordinates, and so they cannot be gauged away.
It may be noted that for certain systems, an explicit dependence of the
energy on the coordinates has been constructed \cite{re}. Even though it is
difficult to find such an explicit dependence of rainbow functions on energy
for different systems, it is important to know that these rainbow functions
are implicitly dynamical functions of the coordinates, and so they cannot be
gauged away. Thus, these functions are expected to produce physically
different results from general relativity \cite{a1}. In this paper, we will
explicitly demonstrate that the thermodynamics of the rainbow deformed
Lovelock black hole is different from a black hole in the usual Lovelock
theory.

It may be noted that the rainbow functions physically change the
thermodynamics of black holes \cite{a1}. In fact, the rainbow deformation of
a black hole produces non-trivial effects like the existence of a black hole
remnant at the last stage of the evaporation of a black hole, and this has
been observed to have phenomenological consequences for the detection of
mini black holes at the LHC \cite{f1}. Thus, in gravity's rainbow, the last
stages of the evaporation of a black hole is very different from general
relativity. It has been explicitly demonstrated that such a remnant also
occurs for black rings \cite{f2}. In fact, it has been proposed that such a
black remnant will form for all black objects in the gravity's rainbow \cite%
{f4}. This has also been explicitly demonstrated for Kerr, Kerr-Newman-dS,
charged-AdS, higher dimensional Kerr-AdS black holes and a black Saturn \cite%
{f4}. The black holes in the Gauss-Bonnet gravity have also been studied
using gravity's rainbow \cite{f5}. It was demonstrate that even though the
thermodynamics of the black holes get modified in the Gauss-Bonnet gravity's
rainbow, the first law of thermodynamics still holds for this modified
thermodynamics. As the Gauss-Bonnet gravity generalizes to Lovelock gravity,
it is interesting to analyze the black hole solutions in Lovelock gravity
using gravity's rainbow. Thus, in this paper, we will analyze certain
aspects of black holes in Lovelock gravity using gravity's rainbow.
Furthermore, string theory is one of the motivations for studding gravity's
rainbow, and the low energy effective action for the string theory produces
Lovelock gravity \cite{L1,L2,L4,L5}. It means, in the context of string
theory, higher curvature terms of Lovelock Lagrangian are string corrections
on gravity side at the classical level. Thus, theoretically, it seems
natural to study the effect of higher curvature terms in the context of
gravity's rainbow. We also extend the investigation of Lovelock gravity's
rainbow to the case of coupling with linear (and nonlinear) electromagnetic
field. The Lovelock gravity coupled to different classes of the
electromagnetic fields has also been studied \cite%
{CaiLovelock,e1,DehghaniPourhasan,Zou,DehghaniAsnafi,1110.0064,e2,e4,e5,PVlovelock1,PVlovelock2,LovelockGTD,HendiAliDehghani}%
. It has been observed that this can have interesting phenomenological
consequences \cite{g2}. Motivated by these subjects, we will analyze the
black holes in Lovelock gravity's rainbow coupled to the Maxwell field, and
then extend our investigations to the case of nonlinear electrodynamics.

We have organized this paper as follows: First, in Sec. \ref%
{Lovelock-Maxwell}, we give a brief review of the proper field equations of
Lovelock gravity in the presence of linear electromagnetic field, and then
present a new class of charged black hole solutions in gravity's rainbow
formalism. After that, in subsection \ref{Thermodynamics-M}, we study the
effect of rainbow functions on thermodynamic quantities and show that the
first law of thermodynamics still holds in the context of this formalism.
Also, in subsection \ref{Stabilty-M}, we perform a thermal stability
analysis for these black holes in the canonical ensemble. Next, in Sec. \ref%
{Lovelock-BI}, we extend our consideration to the case of Born-Infeld
nonlinear electrodynamics (BI) and present a new class of charged black
holes in Lovelock-BI gravity's rainbow, and then investigate their
thermodynamic properties. In Sec. \ref{pv}, in order to complete our
discussion of thermodynamic properties, we study the critical behavior of
black holes in the extended phase space by regarding the cosmological
constant as a thermodynamical pressure. We finish our paper with some
concluding remarks.

\section{Lovelock-Maxwell gravity's rainbow \label{Lovelock-Maxwell}}

Lovelock theory of gravity is one of the standard extensions of general
relativity in higher dimensional spacetimes. One can consider the two first
terms (zero and first term) of Lovelock Lagrangian to obtain a constant term
related to the cosmological constant and also Ricci scalar, respectively.
Higher curvature terms in the gravitational field equations appear as a
result of adding the second and third terms (and also higher order terms) of
Lovelock Lagrangian in the gravitational action. The theory obtained by
considering first three terms of the Lovelock Lagrangian (and ignoring
higher order curvature terms) is called Gauss-Bonnet gravity. Here, we want
to expand our study to the third term (first four terms of Lovelock
Lagrangian) in the presence of Maxwell electrodynamics; the so-called third
order Lovelock-Maxwell gravity. The Lagrangian of the third order
Lovelock-Maxwell gravity can be written as
\begin{equation}
{\mathcal{L}={\alpha _{0}\mathcal{L}_{0}+{\alpha _{1}}\mathcal{L}_{1}}+{%
\alpha _{2}}{\mathcal{L}_{2}}+{\alpha _{3}}{\mathcal{L}_{3}}-\mathcal{F}},
\label{Lagrangian LM}
\end{equation}%
where ${{\mathcal{L}_{0}=-2\Lambda }}$ and ${{\mathcal{L}_{1}=}}R$ are,
respectively, the cosmological constant and the Ricci scalar and ${{\alpha
_{0}=\alpha _{1}=1}}$. The second and third order Lovelock coefficients $%
\alpha _{2}$ and $\alpha _{3}$ indicate the strength of the second and third
order curvature terms. Also, ${{\mathcal{L}_{2}}}$ and ${{\mathcal{L}_{3}}}$
are, respectively, the Gauss-Bonnet Lagrangian and the third order Lovelock
term given as%
\begin{equation}
\mathcal{L}_{2}=R_{\mu \nu \gamma \delta }R^{\mu \nu \gamma \delta }-4R_{\mu
\nu }R^{\mu \nu }+R^{2},  \label{L2}
\end{equation}%
\begin{eqnarray}
\mathcal{L}_{3} &=&2R^{\mu \nu \sigma \kappa }R_{\sigma \kappa \rho \tau }R_{%
\phantom{\rho \tau }{\mu \nu }}^{\rho \tau }+8R_{\phantom{\mu \nu}{\sigma
\rho}}^{\mu \nu }R_{\phantom {\sigma \kappa} {\nu \tau}}^{\sigma \kappa }R_{%
\phantom{\rho \tau}{ \mu \kappa}}^{\rho \tau }+24R^{\mu \nu \sigma \kappa
}R_{\sigma \kappa \nu \rho }R_{\phantom{\rho}{\mu}}^{\rho }  \nonumber \\
&&+3RR^{\mu \nu \sigma \kappa }R_{\sigma \kappa \mu \nu }-12RR_{\mu \nu
}R^{\mu \nu }+24R^{\mu \nu \sigma \kappa }R_{\sigma \mu }R_{\kappa \nu
}+16R^{\mu \nu }R_{\nu \sigma }R_{\phantom{\sigma}{\mu}}^{\sigma }+R^{3}.
\label{L3}
\end{eqnarray}

In addition, the last term of Eq. (\ref{Lagrangian LM}) is the Maxwell
invariant $\mathcal{F}={F^{\mu \nu }}{F_{\mu \nu }}$, where ${F_{\mu \nu }}={%
\partial _{\mu }}{A_{\nu }}-{\partial _{\nu }}{A_{\mu }}$ is the
electromagnetic tensor related to the gauge potential $A_{\mu }$. Using the
variational principle, the electromagnetic and gravitational field equations
of third order Lovelock-Maxwell gravity can be obtained as
\begin{equation}  \label{gravitational Eq}
G_{\mu \nu }^{(0)}+G_{\mu \nu }^{(1)}+G_{\mu \nu }^{(2)}+G_{\mu \nu }^{(3)}=-%
\frac{1}{2}{g_{\mu \nu }}\mathcal{F}+{F_{\mu \lambda }}{F_{\nu }}^{\lambda },
\label{eqEM}
\end{equation}%
\begin{equation}  \label{Maxwell Eq}
{\partial _{\mu }}\left( {\sqrt{-g}{F^{\mu \nu }}}\right) =0,  \label{eqM}
\end{equation}%
where the cosmological constant term and Einstein tensor are, respectively,
denoted by $G_{\mu \nu }^{(0)}=-\frac{1}{2}g_{\mu \nu }\mathcal{L}_{0}$ and $%
G_{\mu \nu }^{(1)}=R_{\mu \nu }-\frac{1}{2}g_{\mu \nu }\mathcal{L}_{1}$. In
addition, $G_{\mu \nu }^{(2)}$ and $G_{\mu \nu }^{(3)}$ are, respectively,
the second and third orders Lovelock tensor given as
\begin{equation}  \label{GB}
G_{\mu \nu }^{(2)}=2(R_{\mu \sigma \kappa \tau }R_{\nu }^{\phantom{\nu}%
\sigma \kappa \tau }-2R_{\mu \rho \nu \sigma }R^{\rho \sigma }-2R_{\mu
\sigma }R_{\phantom{\sigma}\nu }^{\sigma }+RR_{\mu \nu })-\frac{1}{2}g_{\mu
\nu }\mathcal{L}_{2},  \label{Love2}
\end{equation}%
\begin{eqnarray}
G_{\mu \nu }^{(3)} &=&-3[4R^{\tau \rho \sigma \kappa }R_{\sigma \kappa
\lambda \rho }R_{\phantom{\lambda }{\nu \tau \mu}}^{\lambda }-8R_{%
\phantom{\tau \rho}{\lambda \sigma}}^{\tau \rho }R_{\phantom{\sigma
\kappa}{\tau \mu}}^{\sigma \kappa }R_{\phantom{\lambda }{\nu \rho \kappa}%
}^{\lambda }+2R_{\nu }^{\phantom{\nu}{\tau \sigma \kappa}}R_{\sigma \kappa
\lambda \rho }R_{\phantom{\lambda \rho}{\tau \mu}}^{\lambda \rho }  \nonumber
\label{TOL} \\
&&-R^{\tau \rho \sigma \kappa }R_{\sigma \kappa \tau \rho }R_{\nu \mu }+8R_{%
\phantom{\tau}{\nu \sigma \rho}}^{\tau }R_{\phantom{\sigma \kappa}{\tau \mu}%
}^{\sigma \kappa }R_{\phantom{\rho}\kappa }^{\rho }+8R_{\phantom
{\sigma}{\nu \tau \kappa}}^{\sigma }R_{\phantom {\tau \rho}{\sigma \mu}%
}^{\tau \rho }R_{\phantom{\kappa}{\rho}}^{\kappa }  \nonumber \\
&&+4R_{\nu }^{\phantom{\nu}{\tau \sigma \kappa}}R_{\sigma \kappa \mu \rho
}R_{\phantom{\rho}{\tau}}^{\rho }-4R_{\nu }^{\phantom{\nu}{\tau \sigma
\kappa }}R_{\sigma \kappa \tau \rho }R_{\phantom{\rho}{\mu}}^{\rho
}+4R^{\tau \rho \sigma \kappa }R_{\sigma \kappa \tau \mu }R_{\nu \rho
}+2RR_{\nu }^{\phantom{\nu}{\kappa \tau \rho}}R_{\tau \rho \kappa \mu }
\nonumber \\
&&+8R_{\phantom{\tau}{\nu \mu \rho }}^{\tau }R_{\phantom{\rho}{\sigma}%
}^{\rho }R_{\phantom{\sigma}{\tau}}^{\sigma }-8R_{\phantom{\sigma}{\nu \tau
\rho }}^{\sigma }R_{\phantom{\tau}{\sigma}}^{\tau }R_{\mu }^{\rho }-8R_{%
\phantom{\tau }{\sigma \mu}}^{\tau \rho }R_{\phantom{\sigma}{\tau }}^{\sigma
}R_{\nu \rho }-4RR_{\phantom{\tau}{\nu \mu \rho }}^{\tau }R_{\phantom{\rho}%
\tau }^{\rho }  \nonumber \\
&&+4R^{\tau \rho }R_{\rho \tau }R_{\nu \mu }-8R_{\phantom{\tau}{\nu}}^{\tau
}R_{\tau \rho }R_{\phantom{\rho}{\mu}}^{\rho }+4RR_{\nu \rho }R_{%
\phantom{\rho}{\mu }}^{\rho }-R^{2}R_{\nu \mu }]-\frac{1}{2}g_{\mu \nu }%
\mathcal{L}_{3}.  \label{Love3}
\end{eqnarray}

Now, we consider the following line element to obtain the $d$-dimensional
static topological black hole solutions in the context of gravity's rainbow
\begin{equation}
d{\tau ^{2}}=-d{s^{2}}=\frac{{\Psi (r)}}{{f{{(\varepsilon )}^{2}}}}d{t^{2}}-%
\frac{1}{{g{{(\varepsilon )}^{2}}}}\left( {\frac{{d{r^{2}}}}{{\Psi {{(r)}^{2}%
}}}+{r^{2}}d{\Omega _{k}^{2}}}\right) ,  \label{Metric}
\end{equation}%
where ${d{\Omega _{k}^{2}}}$ is the metric of a $(d-2)$-dimensional
hypersurface with constant curvature $(d-2)(d-3)k$ and volume $V_{d-2}$ with
the following explicit forms
\begin{equation}
d{\Omega _{k}^{2}}=\left\{
\begin{array}{cc}
d\theta _{1}^{2}+\sum\limits_{i=2}^{d-2}\prod\limits_{j=1}^{i-1}\sin
^{2}\theta _{j}d\theta _{i}^{2} & k=1 \\
d\theta _{1}^{2}+\sinh ^{2}\theta _{1}d\theta _{2}^{2}+\sinh ^{2}\theta
_{1}\sum\limits_{i=3}^{d-2}\prod\limits_{j=2}^{i-1}\sin ^{2}\theta
_{j}d\theta _{i}^{2} & k=-1 \\
\sum\limits_{i=1}^{d-2}d\phi _{i}^{2} & k=0%
\end{array}%
\right. .  \label{met2}
\end{equation}

Using Eq. (\ref{Maxwell Eq}) the gauge potential is obtained as follows
\begin{equation}
{A_{\mu }}=\frac{{-q}}{{(d-3){r^{(d-3)}}}}\delta _{\mu }^{0},
\label{Amu-Maxwell}
\end{equation}
where $q$ is an integration constant which is related to the electric
charge. It is easy to show that the nonzero components of the
electromagnetic tensor are
\begin{equation}
{F_{tr}}=-{F_{rt}}=\frac{q}{{{r^{d-2}}}}.  \label{Ftr}
\end{equation}

Here, to have a practical solutions for gravitational field equations, we
consider a special class, in which ${\alpha _{2}}$ and ${\alpha _{3}}$ are
related to each other such as ${\alpha _{3}}=\frac{{{\alpha ^{2}}}}{{%
3(d-3)(d-4)(d-5)(d-6)}}$ and ${\alpha _{2}}=\frac{\alpha }{{(d-3)(d-4)}}$.
The gravitational filed equations with this condition have one real and two
complex (conjugate) solutions for the metric function $\Psi (r)$ . The real
metric function $\Psi (r)$ has the following form
\begin{equation}
\Psi (r)=k+\frac{{{r^{2}}}}{{\alpha g{{(\varepsilon )}^{2}}}}\left( {1-{{%
\left[ {1+\frac{{3\alpha m}}{{{r^{d-1}}}}+\frac{{6\alpha \Lambda }}{{%
(d-1)(d-2)}}-\frac{{6\alpha {q^{2}}f{{(\varepsilon )}^{2}}g{{(\varepsilon )}%
^{2}}}}{{(d-2)(d-3){r^{2(d-2)}}}}}\right] }^{\frac{1}{3}}}}\right) ,
\label{Psi-Maxwell}
\end{equation}%
where the parameter $m$ is an integration constant which is related to the
finite mass. It is obvious that the obtained solutions reduce to the
Lovelock-Maxwell solutions with condition $f(\varepsilon )=g(\varepsilon )=1$
(or correspondingly $\varepsilon \longrightarrow 0$). In order to
investigate the possible curvature singularity, we can calculate the
Kretschmann scalar. The Kretschmann scalar has the following form
\begin{equation}
{R^{\alpha \beta \gamma \delta }}{R_{\alpha \beta \gamma \delta }}=g{%
(\varepsilon )^{4}}\left[ {{{\left( {\frac{{{d^{2}}\Psi (r)}}{d{{r^{2}}}}}%
\right) }^{2}}+2(d-2){{\left( {\frac{1}{r}\frac{d{\Psi (r)}}{{dr}}}\right) }%
^{2}}+2(d-2)(d-3){{\left( {\frac{{\Psi (r)-k}}{{{r^{2}}}}}\right) }^{2}}}%
\right] .  \label{RR}
\end{equation}

Inserting Eq. (\ref{Psi-Maxwell}) into Eq. (\ref{RR}), one finds an
essential singularity at the origin and for $r\neq 0$ all curvature
invariants are nonsingular. Equation (\ref{RR}) confirms that the strengths
of singularity and other finite values of curvature can be drastically
affected by rainbow functions. Numerical calculations show that, depending
on the values of parameters, the metric function has two real positive
roots, one extreme root or it may be positive definite. Hence, obtained
solutions can be covered with an event horizon and the solutions may be
interpreted as the black holes with two horizons (Cauchy and event), extreme
black holes or naked singularity. The asymptotical behavior of the solution
(i.e. $r\rightarrow \infty $) is obtained as follows
\begin{equation}
{\left. \Psi {(r)}\right\vert _{asymp.}}=k+\frac{{{r^{2}}}}{{\alpha g{{%
(\varepsilon )}^{2}}}}\left( {1-{{\left[ {1+\frac{{6\alpha \Lambda }}{{%
(d-1)(d-2)}}}\right] }^{\frac{1}{3}}}}\right) ,  \label{PsiAsymp}
\end{equation}%
and therefore these solutions are asymptotically AdS (${\Lambda _{eff}}<0$),
dS (${\Lambda _{eff}}>0$) or flat (${\Lambda _{eff}}=0$) with an effective
energy dependent cosmological constant ${\Lambda _{eff}}=\frac{{\Lambda
\lbrack (d-1)(d-2)-2\alpha \Lambda ]}}{{(d-1)(d-2)g{{(\varepsilon )}^{2}}}}$.

{In order to study the effects of energy dependency, we
have to use a specific functional form of the rainbow functions.
There are different phenomenologically motivations for considering
different forms  of the rainbow functions. A specific form of the
rainbow functions has been motivated  from results obtained in the
loop quantum gravity and noncommutative geometry \cite{a,j}. The
hard spectra from gamma-ray burster has also been used to motivate
the construction of a different form of rainbow functions
\cite{Ellis}. Another interesting form of rainbow functions is
based on the modified dispersion relations, such that the
constancy of velocity of the light is not violated \cite
{MagueijoSPRL}, and it can be  explicitly written as
\begin{equation}
f(\varepsilon )=g(\varepsilon )=\frac{1}{1+\lambda \varepsilon },
\label{RainbowFunctions}
\end{equation}
where $\lambda$ is an arbitrary parameter. Hereafter, we focus on
Eq. (\ref{RainbowFunctions}) as a typical form of rainbow
functions.}
\begin{table}
\begin{center}
\begin{tabular}{|c|c|}
\hline
\begin{tabular}{cc}
\hline\hline
${\varepsilon }$ & $f({\varepsilon })\&g({\varepsilon })$ \\ \hline\hline
$0.00$ \  & $1.00$ \  \\ \hline
$0.10$ & $0.95$ \\ \hline
$0.50$ & $0.80$ \\ \hline
$0.90$ & $0.69$ \\ \hline
\end{tabular}
&
\begin{tabular}{cc}
\hline\hline
${\varepsilon }$ & $f({\varepsilon })\&g({\varepsilon })$ \\ \hline\hline
$0.00$ \  & $1.00$ \  \\ \hline
$0.10$ & $1.05$ \\ \hline
$0.50$ & $1.33$ \\ \hline
$0.90$ & $1.82$ \\ \hline
\end{tabular}%
\  \\ \hline
\end{tabular}
\end{center}
\caption{Typical values for the rainbow functions in Eq. (%
\textbf{\ref{RainbowFunctions}}): left panel: $\lambda =0.5$ and
right panel: $\lambda =-0.5$.} \label{tabI}
\end{table}

{It may be noted  that depending on the energy ratio
$\varepsilon$, one can obtain different values for the rainbow
functions (see table \ref{tabI}). So, in this paper, we relax
dynamic determination of these functions and analyze the behavior
of the system from this set of phenomenologically motivate rainbow
functions (motivated by the modified dispersion relation with
constant velocity of light).}


\subsection{Thermodynamics of black holes in Lovelock-Maxwell gravity's
rainbow \label{Thermodynamics-M}}

This section is devoted to calculation of conserved and thermodynamic
quantities, and examination of the first law of thermodynamics. At first, we
use the definition of surface gravity to calculate the temperature
\begin{equation}
T=\frac{1}{{2\pi }}\sqrt{-\frac{1}{2}({\nabla _{\mu }}{\chi _{\nu }})({%
\nabla ^{\mu }}{\chi ^{\nu }})}=\left. \frac{{g(\varepsilon )}}{{4\pi
f(\varepsilon )}}\frac{{d\Psi (r)}}{{dr}}\right\vert _{r=r_{+}},
\label{Hawking Temperature}
\end{equation}%
where after simplification, we obtain
\begin{equation}
T=\frac{{(d-2)kg{{(\varepsilon )}^{2}}r_{+}^{2(d-5)}\left( {(d-7){\alpha ^{2}%
}g{{(\varepsilon )}^{4}}+3(d-5)k\alpha g{{(\varepsilon )}^{2}}%
r_{+}^{2}+3(d-3)r_{+}^{4}}\right) -6\Lambda r_{+}^{2(d-2)}-6{q^{2}}g{{%
(\varepsilon )}^{2}}f{{(\varepsilon )}^{2}}}}{{12\pi (d-2)g(\varepsilon
)f(\varepsilon ){{\left( {k\alpha g{{(\varepsilon )}^{2}}+r_{+}^{2}}\right) }%
^{2}}r_{+}^{2d-9}}}.  \label{Temperature}
\end{equation}%
Regarding Eq. (\ref{Temperature}), we find that black hole temperature is
modified due to the presence of rainbow functions $g(\varepsilon)$ and $%
f(\varepsilon)$.

Entropy is an extensive quantity corresponds to the temperature as an
intensive quantity. In higher derivative gravity the area law of entropy is
not valid, generally. Therefore, we calculate it by Wald method which is
valid in higher derivative gravity such as Lovelock gravity. It is shown in
third order Lovelock gravity the entropy can be written as
\begin{equation}
S=\frac{1}{4}\int {{d^{d-2}}x\sqrt{\tilde{g}}\left( {1+2{\alpha _{2}}\tilde{R%
}+3{\alpha _{3}}\left( {{{\tilde{R}}^{\mu \nu \gamma \delta }}{{\tilde{R}}%
_{\mu \nu \gamma \delta }}-4{{\tilde{R}}^{\mu \nu }}{{\tilde{R}}_{\mu \nu }}+%
{{\tilde{R}}^{2}}}\right) }\right) },
\end{equation}%
where ${{{\tilde{R}}_{\mu \nu \gamma \delta }}}$, ${{{\tilde{R}}_{\mu \nu }}}
$ and ${\tilde{R}}$ are, respectively, the Riemann tensor, the Ricci tensor
and the Ricci scalar of the induced metric ${\tilde{g}_{\mu \nu }}$ on $%
(d-2) $-dimensional horizon. Calculation shows that the modified entropy of
third order Lovelock gravity is obtained as follows
\begin{equation}
S=\frac{{{V_{d-2}}r_{+}^{d-2}}}{{4g{{(\varepsilon )}^{d-2}}}}\left( {1+\frac{%
{2\left( {d-2}\right) k\alpha g{{(\varepsilon )}^{2}}}}{{\left( {d-4}\right)
r_{+}^{2}}}+\frac{{\left( {d-2}\right) {k^{2}}{\alpha ^{2}}g{{(\varepsilon )}%
^{4}}}}{{\left( {d-6}\right) r_{+}^{4}}}}\right) .  \label{Entropy}
\end{equation}%
in which shows that the area law is valid only for the Ricci flat solutions (%
$k=0$).

In order to establish the first law of thermodynamics for charged black
holes, we have to calculate quantities related to the electrodynamics.
First, in order to obtain its extensive quantity (i.e. the electric charge),
we calculate the flux of the electromagnetic field at infinity. The electric
charge is obtained as
\begin{equation}
Q=\frac{{{V_{d-2}}}}{{4\pi }}\frac{{qf(\varepsilon )}}{{g{{(\varepsilon )}%
^{d-3}}}}.  \label{Charge}
\end{equation}

Now, we should calculate the electric potential of black hole solutions.
Electric potential is an intensive quantity corresponds the electric charge,
in which measured at the horizon with respect to infinity as a reference
\begin{equation}  \label{Potential}
\Phi = {\left. {{A_\mu }{\chi ^\mu }} \right|_{r \to \infty }} - {\left. {{%
A_\mu }{\chi ^\mu }} \right|_{r \to r_ {+ }}} = \frac{q}{{(d - 3)r_ {+ }^{(d
- 3)}}}.
\end{equation}

The finite mass of the black hole can be obtained by using different methods
that all of them have the same result. Using the behavior of the metric at
large $r$ (for asymptotically flat black hole) or counterterm method (for
asymptotically AdS black hole), it is easy to show that the finite mass of
the black hole is
\begin{equation}
M=\frac{{{V_{d-2}}}}{{16\pi }}\frac{{(d-2)m}}{{f(\varepsilon )g{{%
(\varepsilon )}^{(d-1)}}}}.  \label{finite mass}
\end{equation}

Now, we rewrite the finite mass $M$ as a function of the extended quantities
(the entropy and electric charge). Straightforward calculations show that
\begin{equation}
M(S,Q)=\frac{{(d-2)}}{{16\pi f(\varepsilon )g{{(\varepsilon )}^{(d-1)}}}}%
\left( {\frac{1}{3}{k^{3}}{\alpha ^{2}}g{{(\varepsilon )}^{6}}r_{+}^{d-7}+{%
k^{2}}\alpha g{{(\varepsilon )}^{4}}r_{+}^{d-5}+kg{{(\varepsilon )}^{2}}%
r_{+}^{d-3}+{\mathcal{O}_{+}}}\right) ,  \label{Mass}
\end{equation}%
where $r_{+}=r_{+}(S)$ and
\[
\mathcal{O}_{+}=\frac{{32{\pi ^{2}}{Q^{2}}g{{(\varepsilon )}^{2(d-2)}}}}{{%
(d-2)(d-3)r_{+}^{d-3}}}-\frac{{2\Lambda r_{+}^{d-1}}}{{(d-1)(d-2)}}.
\]

Now, we use the first law of thermodynamics to define temperature and
electric potential as the intensive parameters conjugate to the entropy and
electric charge
\begin{eqnarray}
&&T=\left( \frac{\partial M}{\partial S}\right) _{Q}=\frac{\left( \frac{%
\partial M}{\partial r_{+}}\right) _{Q}}{\left( \frac{\partial S}{\partial
r_{+}}\right) _{Q}},  \label{T} \\
&&\Phi =\left( \frac{\partial M}{\partial Q}\right) _{S}=\frac{\left( \frac{%
\partial M}{\partial q}\right) _{r_{+}}}{\left( \frac{\partial Q}{\partial q}%
\right) _{r_{+}}}.  \label{Phi}
\end{eqnarray}

Using a complete numerical analysis, it is easy to show that Eqs. (\ref{T})
and (\ref{Phi}) coincide with Eqs. (\ref{Temperature}) and (\ref{Potential}%
), respectively, and therefore we deduce that all intensive and
corresponding extensive parameters satisfy the first law of thermodynamics
with the following form
\begin{equation}
dM=TdS+\Phi dQ.  \label{FirstLaw}
\end{equation}


\subsection{Thermal stability of black holes in Lovelock-Maxwell gravity's
rainbow\label{Stabilty-M}}

Here, we investigate local stability of the black hole solutions in
Lovelock-Maxwell gravity's rainbow by using determinant of Hessian matrix.
The local stability requires that the behavior of energy (i.e. $M(S,Q)$) be
a convex function of its extensive quantities, $S$ and $Q$. In canonical
ensemble the electric charge is a fixed variable and so determinant of
Hessian matrix is a function of the heat capacity as follows
\begin{equation}
H_{S,S}^M = {\left( {\frac{{{\partial ^2}M}}{{\partial {S^2}}}} \right)_Q} =
\frac{T}{{C_Q}} \ge 0.
\end{equation}


\begin{figure}[tbp]
$%
\begin{array}{ccc}
\epsfxsize=5.5cm \epsffile{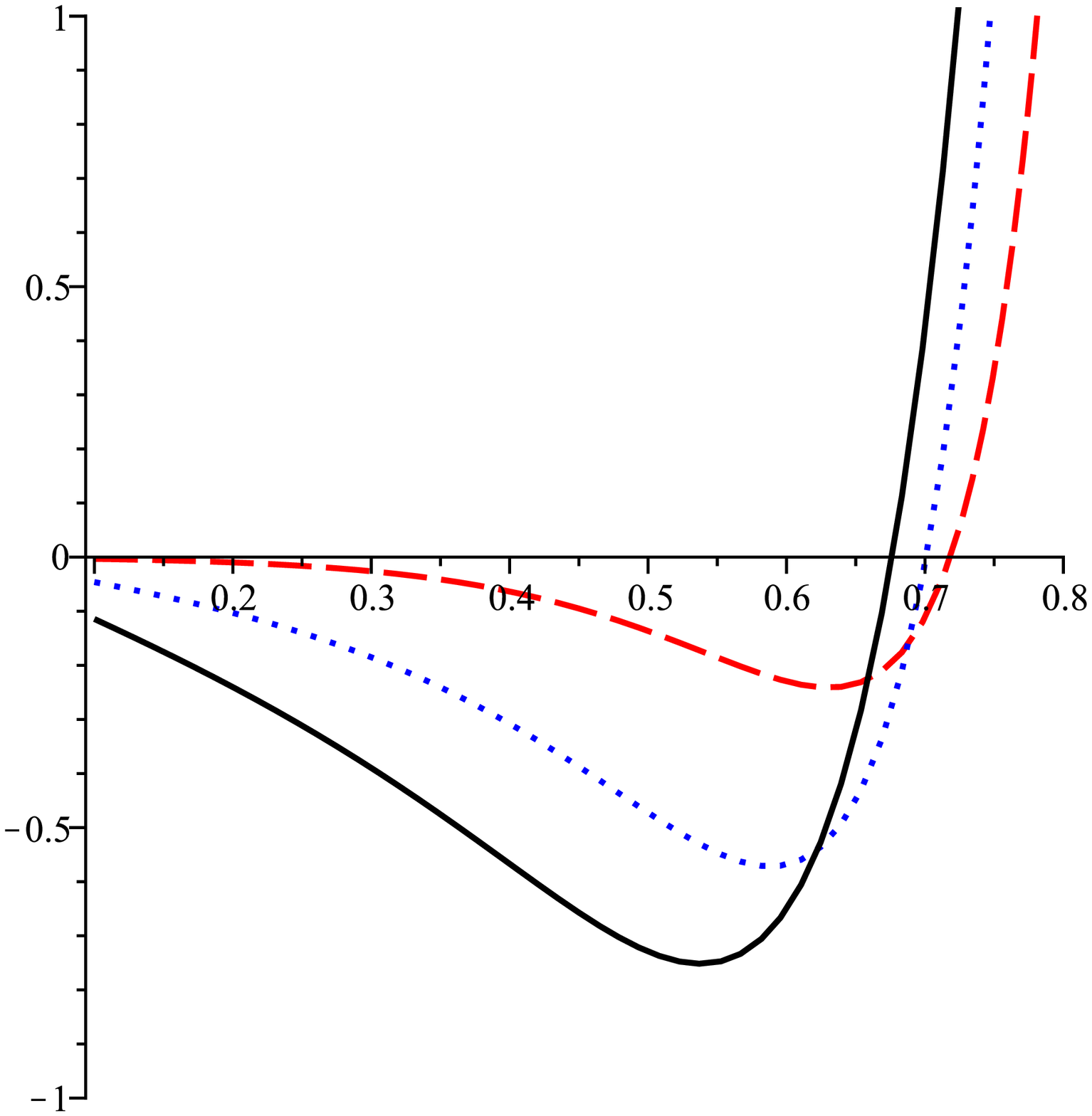} \epsfxsize=5.5cm %
\epsffile{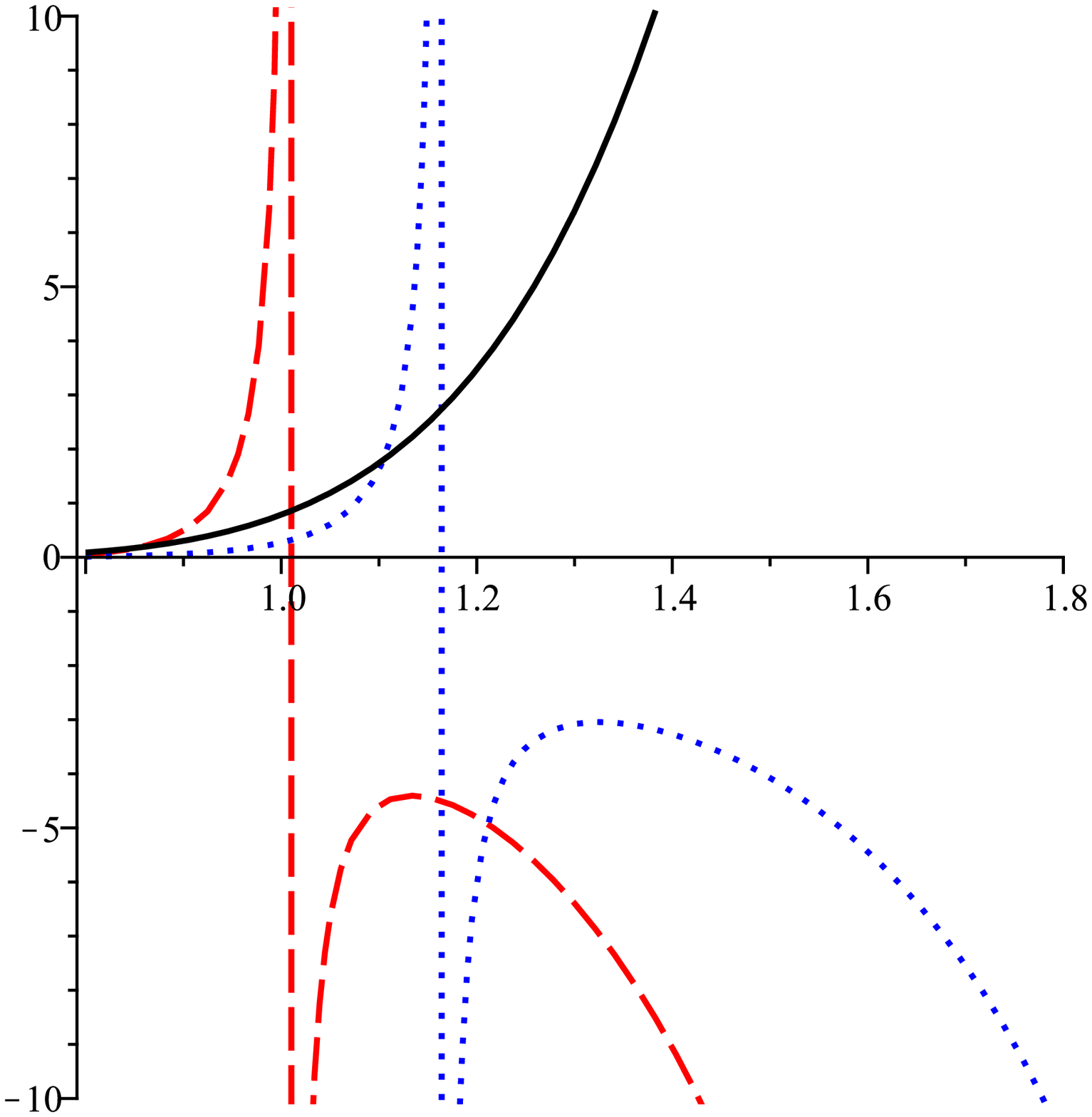}\epsfxsize=5.5cm %
\epsffile{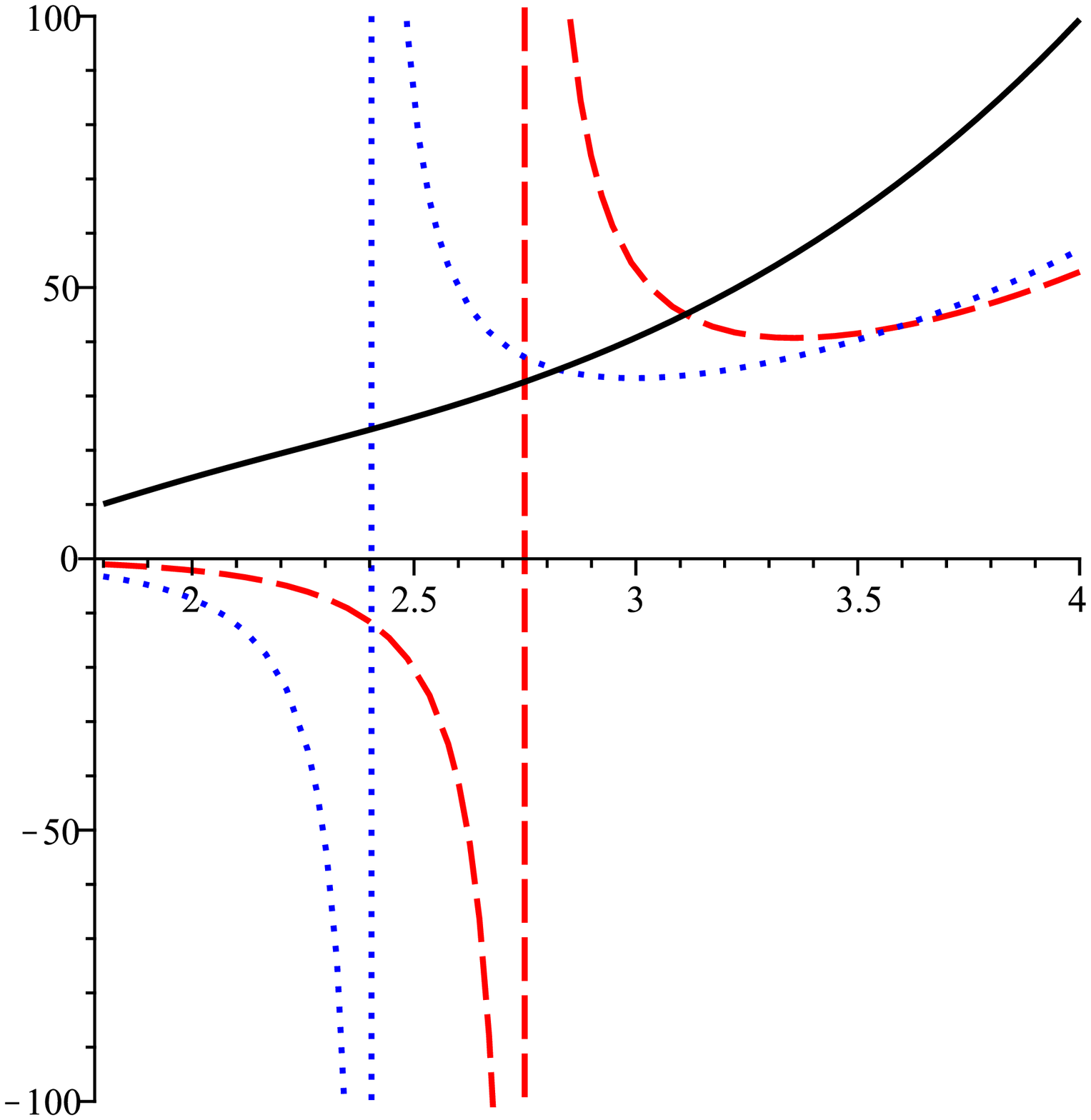} &  &
\end{array}
$%
\caption{ $C_{Q}$ versus ${r_ + }$ for $k=1$, $d=7$, $q=1$, $f(\protect%
\varepsilon) = g(\protect\varepsilon) =0 .9$, $\Lambda = - 1$ and $\protect%
\alpha = 0.1$ ({dashed line}), $\protect\alpha = 0.4$ ({dotted line}) and $%
\protect\alpha = 0.9$ ({continuous line}). \textbf{Different scales:} \emph{%
\ left panel ($0<r_{+}<0.8$), middle panel ($0.8<r_{+}<1.8$) and right panel
($1.8<r_{+}<4$).}}
\label{different-alpha}
\end{figure}


\begin{figure}[tbp]
$%
\begin{array}{ccc}
\epsfxsize=5.5cm \epsffile{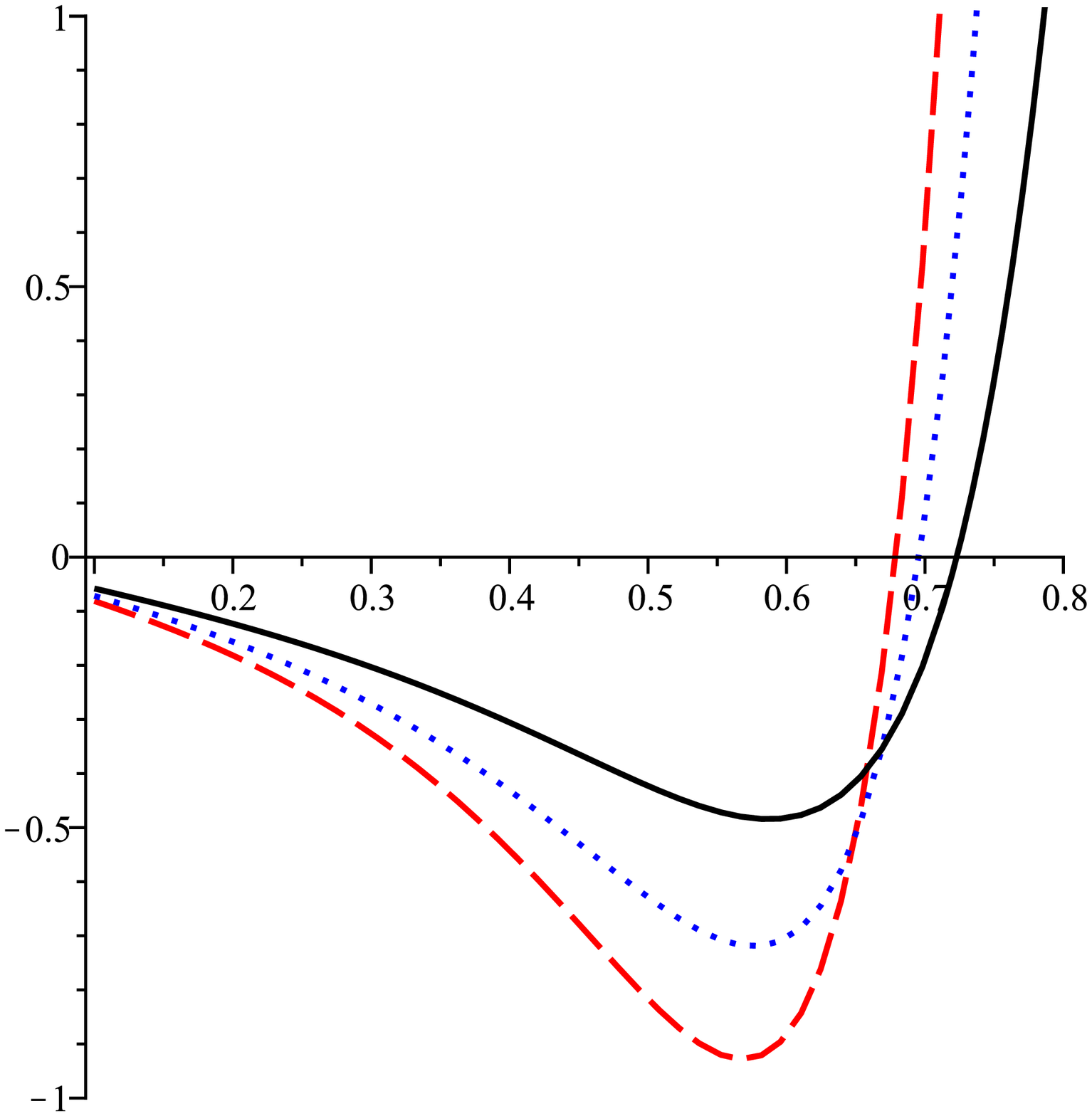} \epsfxsize=5.5cm %
\epsffile{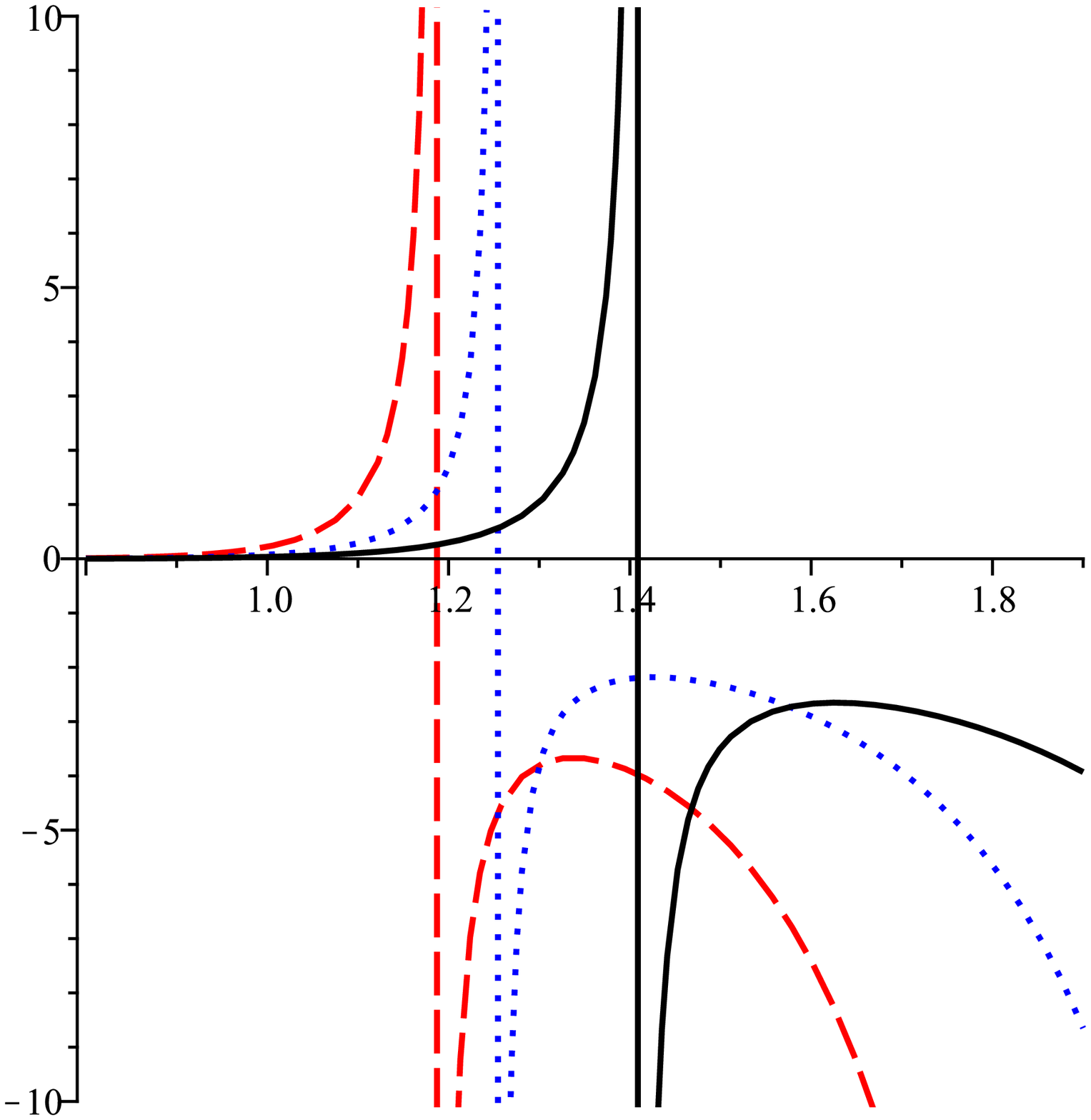}\epsfxsize=5.5cm %
\epsffile{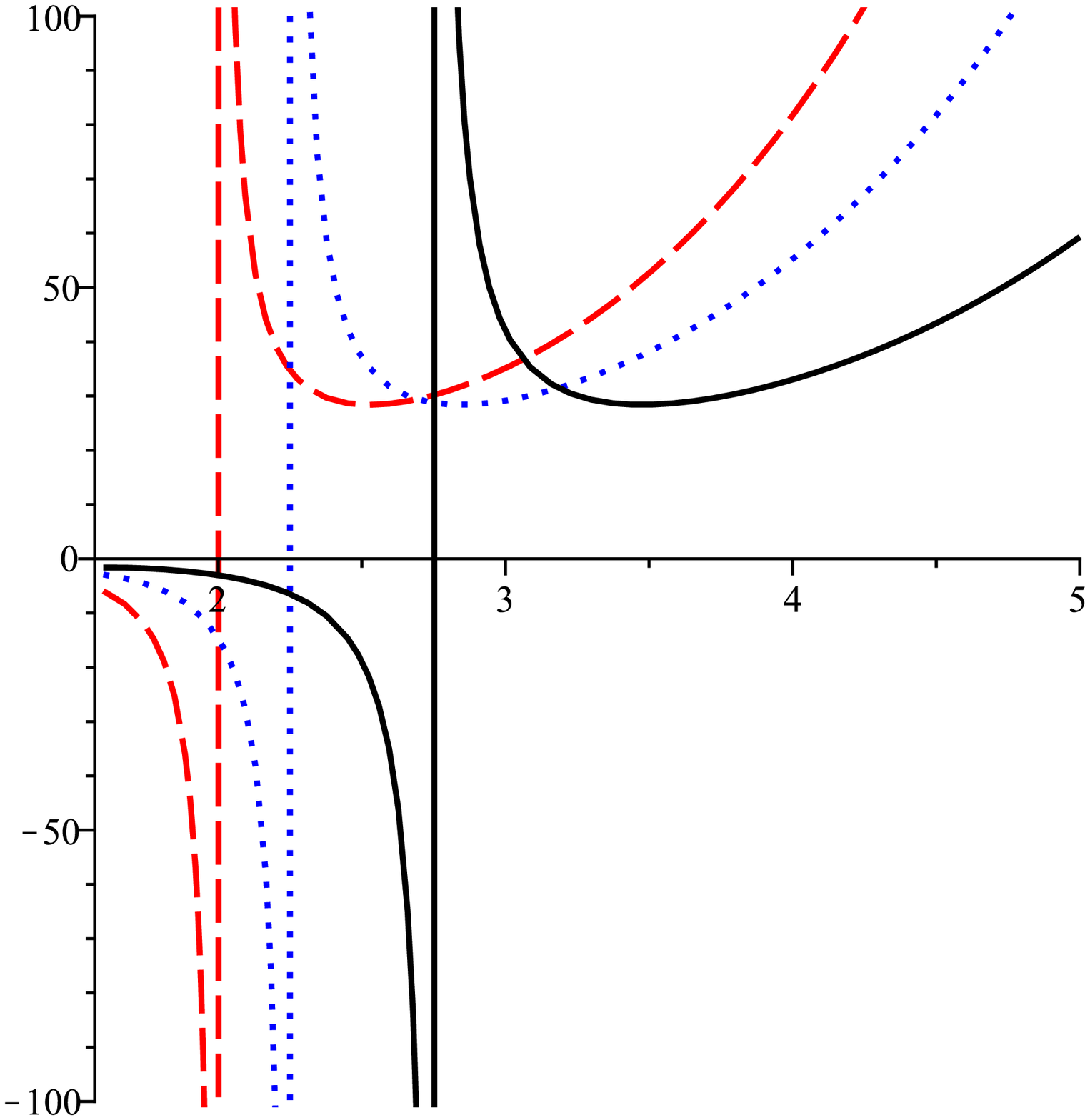} &  &
\end{array}
$%
\caption{ $C_{Q}$ versus ${r_ + }$ for $k=1$, $d=7$, $q=1$, $\protect\alpha %
= 0.5$, $\Lambda = - 1$ and $f(\protect\varepsilon) = g(\protect\varepsilon)
= 0.8$ ({dashed line}), $f(\protect\varepsilon) = g(\protect\varepsilon) =
0.9$ ({dotted line}) and $f(\protect\varepsilon) = g(\protect\varepsilon)
=1.1$ ({continuous line}). \textbf{Different scales:} \emph{\ left panel ($%
0<r_{+}<0.8$), middle panel ($0.8<r_{+}<1.9$) and right panel ($1.5<r_{+}<5$%
).}}
\label{different-f=g}
\end{figure}

\begin{figure}[tbp]
$%
\begin{array}{ccc}
\epsfxsize=5.5cm \epsffile{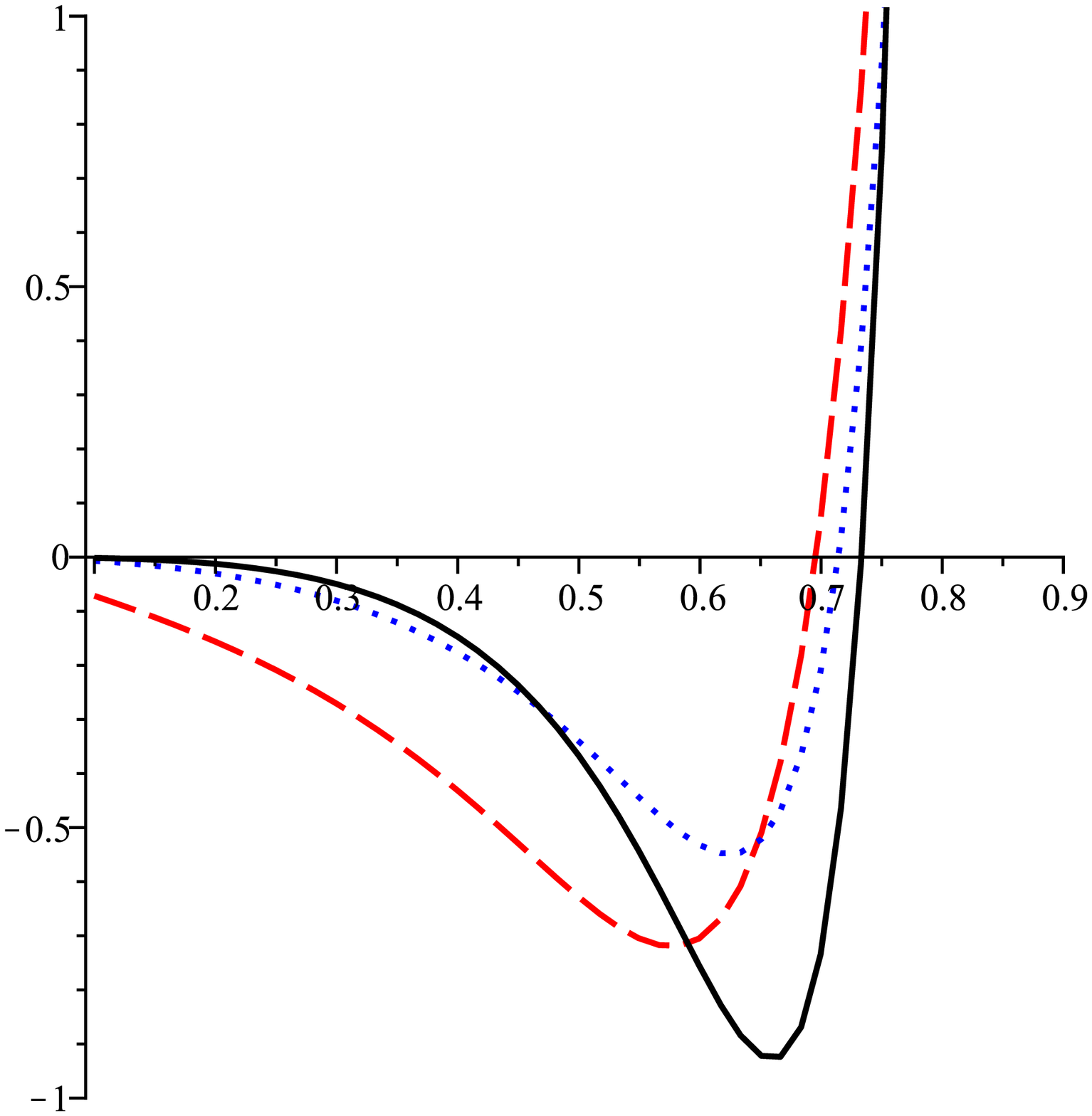} \epsfxsize=5.5cm %
\epsffile{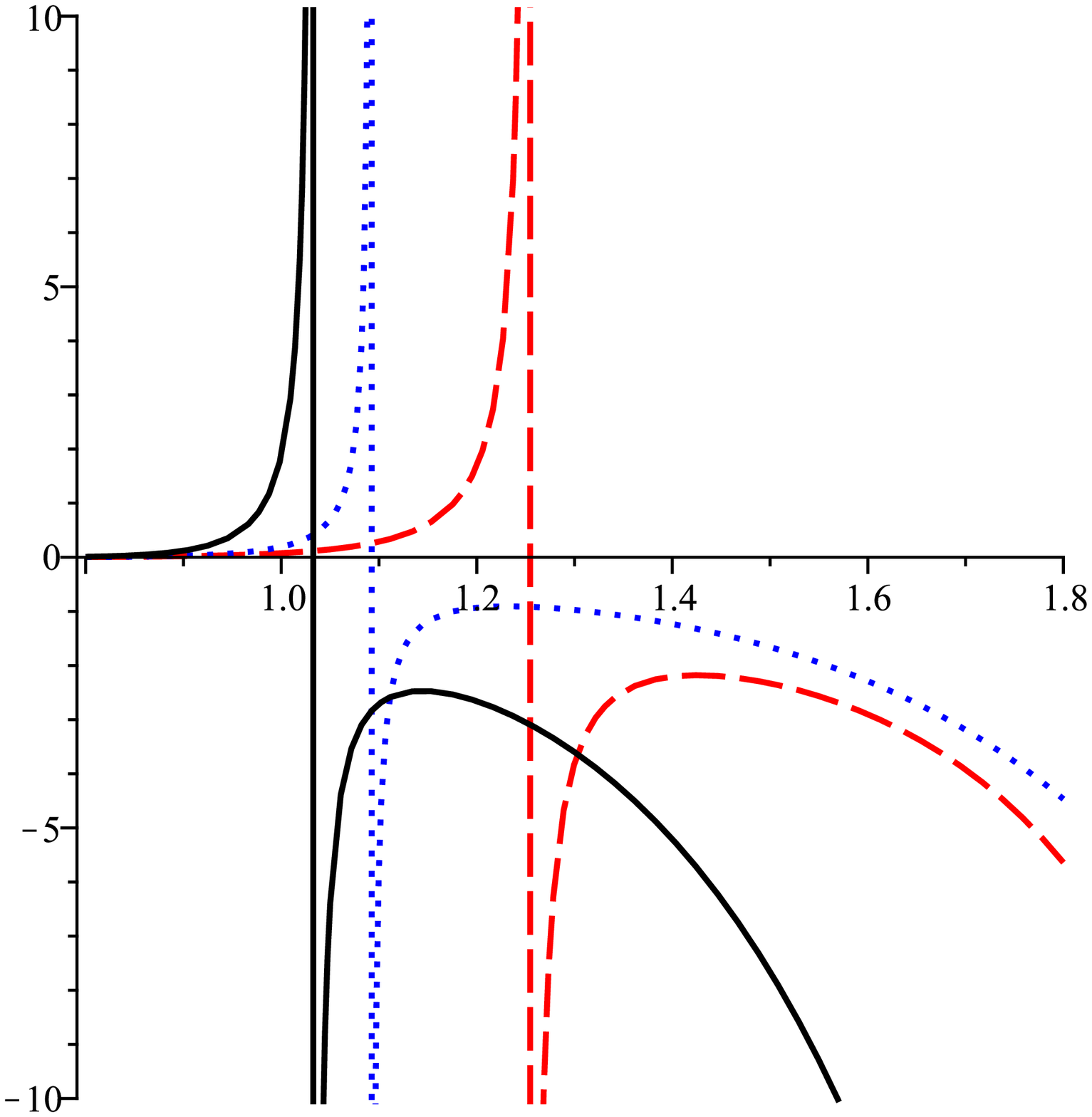}\epsfxsize=5.5cm \epsffile{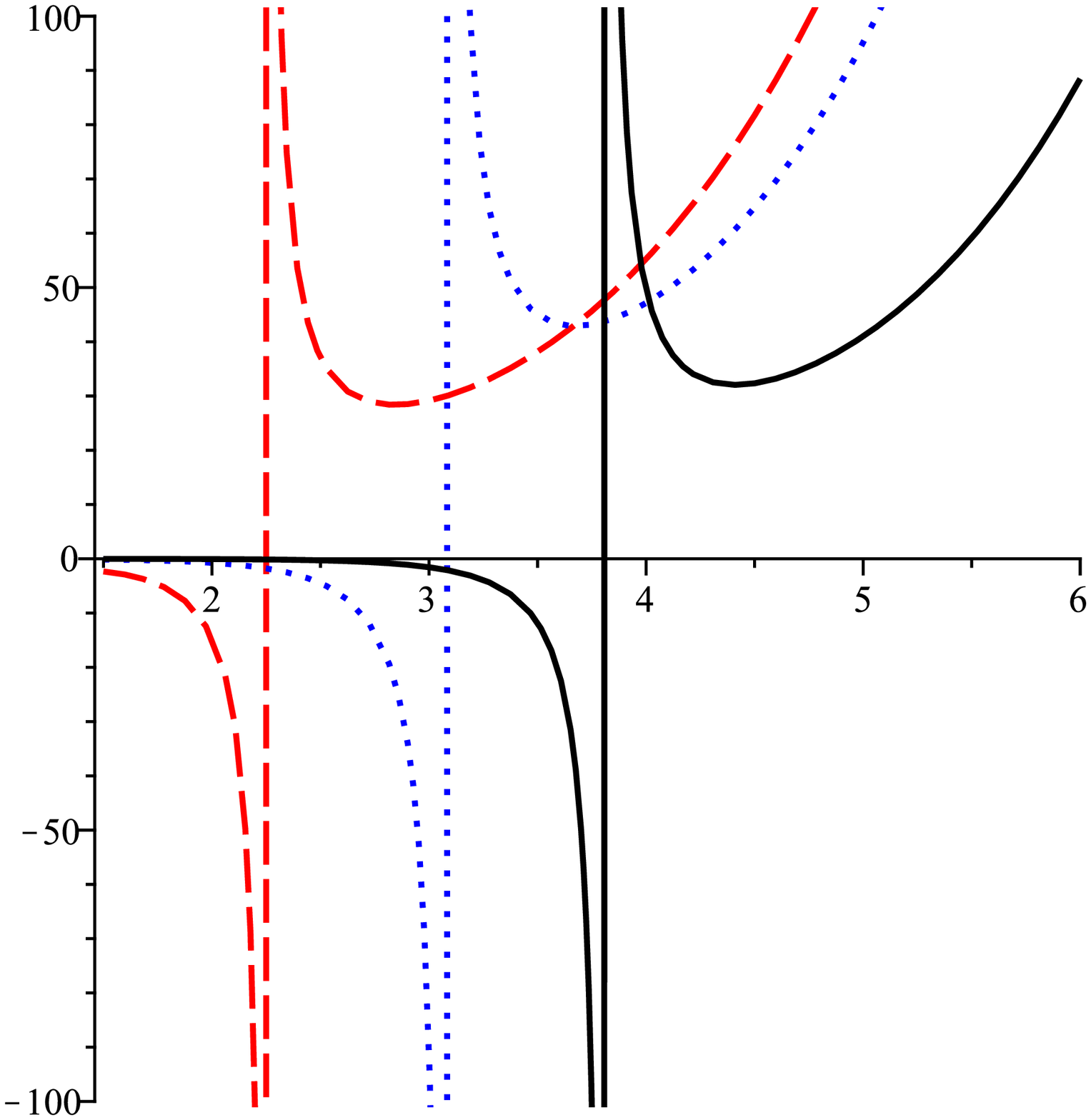}
&  &
\end{array}
$%
\caption{ $C_{Q}$ versus ${r_ + }$ for $k=1$, $q=1$, $\protect\alpha = 0.5$,
$f(\protect\varepsilon) = g(\protect\varepsilon) =0 .9$, $\Lambda = - 1$ and
$d=7$ ({dashed line}), $d=8$ ({dotted line}) and $d=9$ ({continuous line}).
\textbf{Different scales:} \emph{\ left panel ($0<r_{+}<0.9$), middle panel (%
$0.8<r_{+}<1.8$) and right panel ($1.5<r_{+}<6$).}}
\label{different-dimensions}
\end{figure}


For allowed physical black hole parameters (such as positive temperature $T$
and finite mass $M$), the positivity of the heat capacity ensures the local
stability. Analytical calculations of the heat capacity (or determinant of
the Hessian matrix) are too large and, therefore, we leave out the
analytical results for reasons of economy and brevity, and would rather
offer a more practical discussion with some figures. Numerical calculations
for thermal analysis show that for a black hole with definite mass $M$,
there always is a lower critical value for the radius of event horizon, $r_{
+ c}$. The black hole temperature is always positive for $r_{ + } > r_{ + c}$%
.

Now, we can discuss the heat capacity of black holes. Taking into account
the heat capacity, we find that there is a critical value $\alpha _{c}$ for
the Lovelock parameter. In the cases of allowable temperature, i.e. $r_ {+ }
> r_{ + c}$, black holes are thermally stable for large enough values of $%
\alpha $, i.e. $\alpha > \alpha _{c}$ (see Fig. \ref{different-alpha}). In
addition, we can find that for $\alpha < \alpha _{c}$, there is an extreme
radius for the event horizon ($r_{ + ext}$) where in regions $r_{+c} < r_{+}
< r_{+ext}$ the heat capacity is positive and therefore black holes are
stable. Also, there is an upper limit for the radius of event horizon ($r_{
+ u}$) where in regions $r_{ + ext} < r_{ + } < r_{ + u}$ black holes are
unstable. Finally, for regions $r_{ + } > r_{ + u}$, we find a stable
behavior for black holes (see Figs. \ref{different-alpha}-\ref%
{different-dimensions} for more details).

\section{Lovelock-BI gravity's rainbow \label{Lovelock-BI}}

In this section, we generalize our consideration to the case of nonlinear
electrodynamics (NED). We choose the Born-Infeld field (BI field) of NED to
obtain the Lovelock-BI gravity's rainbow black holes. To construct this
theory one can replace Maxwell Lagrangian $(-\mathcal{F})$ with the BI
Lagrangian $L(\mathcal{F})$, which is a function of Maxwell invariant, $%
\mathcal{F}$. The BI Lagrangian is
\begin{equation}
L(\mathcal{F})=4{\beta ^{2}}\left( {1-\sqrt{1+\frac{\mathcal{F}}{{2{\beta
^{2}}}}}}\right),  \label{BI Lagrangian}
\end{equation}%
where $\beta $ is BI parameter with the dimension of $(mass)^{2}$ in natural
units and as one expects, in the limit $\beta \rightarrow \infty $, $L(%
\mathcal{F}) $ reduces to the Maxwell one $(-\mathcal{F)}$. Varying the
following total Lagrangian with respect to the metric tensor ${g_{\mu \nu }}$
and gauge potential $A_{\mu }$, one can obtain the gravitational and
electromagnetic field equations.
\begin{equation}
{\mathcal{L}={\alpha _{0}\mathcal{L}_{0}+{\alpha _{1}}\mathcal{L}_{1}}+{%
\alpha _{2}}{\mathcal{L}_{2}}+{\alpha _{3}}{\mathcal{L}_{3}}+{\mathcal{L}}(%
\mathcal{F})},  \label{Lagrangian Lovelock-BI}
\end{equation}%
where $\alpha _{i}$'s and $\mathcal{L}_{i}$'s are defined before.
Calculations show that the proper field equations are
\begin{equation}  \label{gravitational-BI Eq}
G_{\mu \nu }^{(0)}+G_{\mu \nu }^{(1)}+G_{\mu \nu }^{(2)}+G_{\mu \nu }^{(3)}=%
\frac{1}{2}{g_{\mu \nu }}\mathcal{L}(\mathcal{F})-2{F_{\mu \lambda }}{F_{\nu
}}^{\lambda }{\mathcal{L}_{\mathcal{F}}},  \label{eqGrav}
\end{equation}

\begin{equation}  \label{BI equation}
{\partial _\mu }\left( {\sqrt { - g} {\mathcal{L}_{\mathcal{F}}}{F^{\mu \nu
} }} \right) = 0,
\end{equation}
where ${\mathcal{L}_{\mathcal{F}}} = \frac{{d\mathcal{L}(\mathcal{F})}}{{d%
\mathcal{F}}}$, and $G_{\mu \nu }^{(i)}$'s are defined in section \ref%
{Lovelock-Maxwell}. In order to study Lovelock-BI gravity's rainbow, we use
the defined rainbow metric (Eq. (\ref{Metric})), and find the gauge
potential and the metric function as
\begin{equation}  \label{Amu-BI}
{A_\mu } = \frac{{\ - q}}{{(d - 3)r^{d-3}}}\mathrm{H}({\eta })\delta _\mu ^0,
\end{equation}

\begin{equation}
{\Psi _{BI}}(r)=k+\frac{{r^{2}}}{{\alpha g{{(\varepsilon )}^{2}}}}\left( {1-{%
{\left[ {1+\frac{{3\alpha m}}{{{r^{d-1}}}}+\frac{{6\alpha \Lambda }}{{%
(d-1)(d-2)}}-\frac{{12\alpha {\beta ^{2}}}}{{(d-1)(d-2)}}\left( {1-\sqrt{%
1+\eta }+\frac{{(d-2)}}{{(d-3)}}\eta \mathrm{H}(\eta )}\right) }\right] }^{%
\frac{1}{3}}}}\right) ,  \label{Psi-BI}
\end{equation}%
where
\[
\mathrm{H}(\eta )={}_{2}{F_{1}}\left( {\left[ {\frac{1}{2},\frac{{d-3}}{{%
2(d-2)}}}\right] ,\left[ {\frac{{3d-7}}{{2(d-2)}}}\right] ,-\eta }\right) ,
\]%
\[
\eta =\frac{{{q^{2}}f{{(\varepsilon )}^{2}}g{{(\varepsilon )}^{2}}}}{{{\beta
^{2}}{r^{2(d-2)}}}},
\]%
in which $\mathrm{H}(\eta )$ is a hypergeometric function, and in the limit $%
\beta \rightarrow \infty $ the obtained gauge potential and metric function
reduce to those of the Maxwell gauge potential and Lovelock-Maxwell metric
function (Eqs. (\ref{Amu-Maxwell}) and (\ref{Psi-Maxwell})), respectively.
The Kretschmann scalar diverges only at $r=0$ for the new solutions, and
therefore, there is an essential singularity at the origin. It is notable
that the mentioned singularity can be covered with an event horizon and thus
one can interpret the singularity as a black hole. It was shown that for
nonlinearly charged black holes, depending on the values of $\alpha $ and $%
\beta $, the metric function may have two different behaviors;
Reissner--Nordstr\"{o}m like and Schwarzschild-like (for more details see
\cite{HendiJHEP,HendiAnn}). In addition, for large values of $r$, these
solutions reduce to previous Maxwell case, and therefore, the nonlinearity
parameter does not change the asymptotical behavior.

\begin{figure}[tbp]
$%
\begin{array}{cc}
\epsfxsize=8cm \epsffile{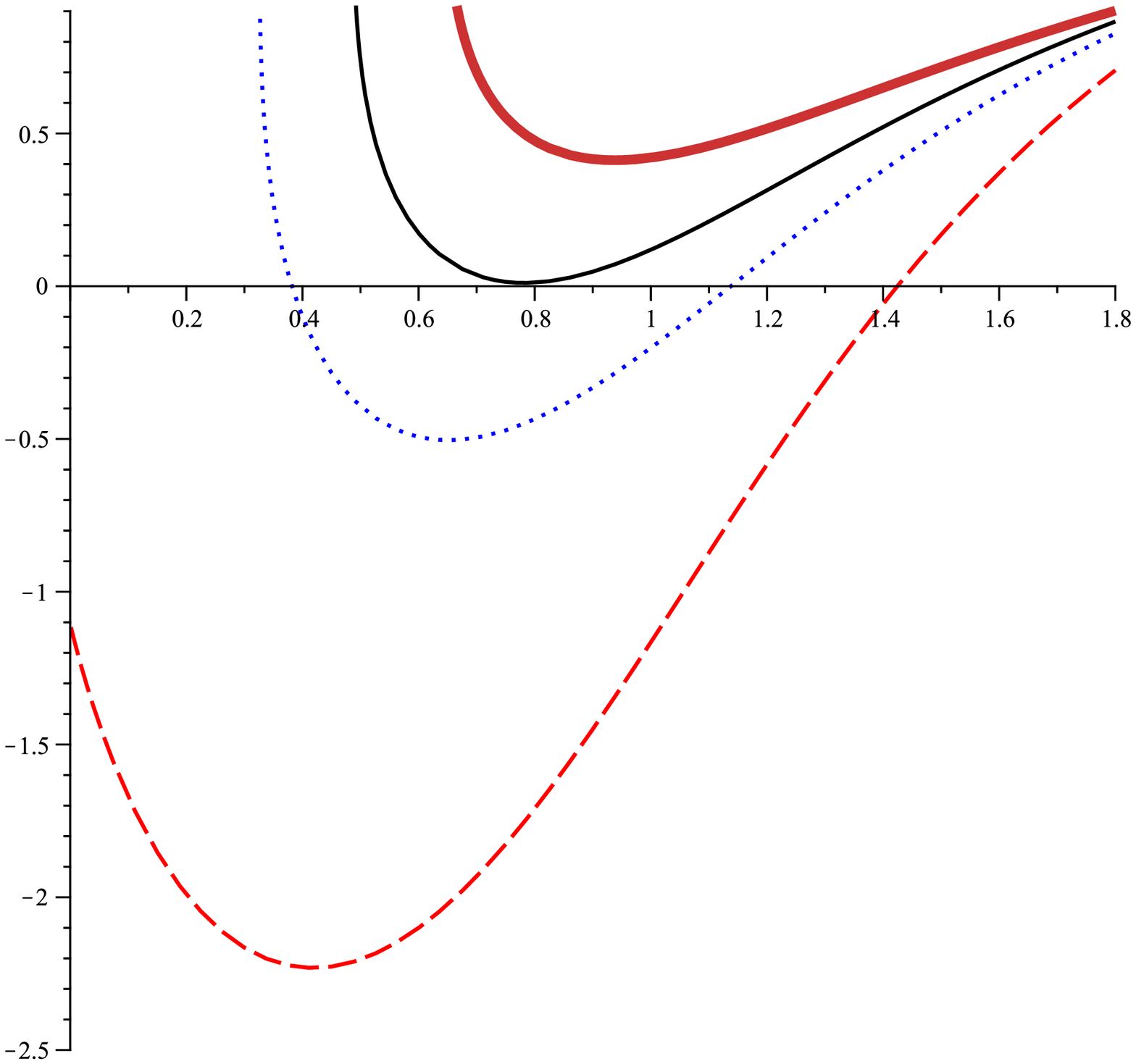} \epsfxsize=8cm \epsffile{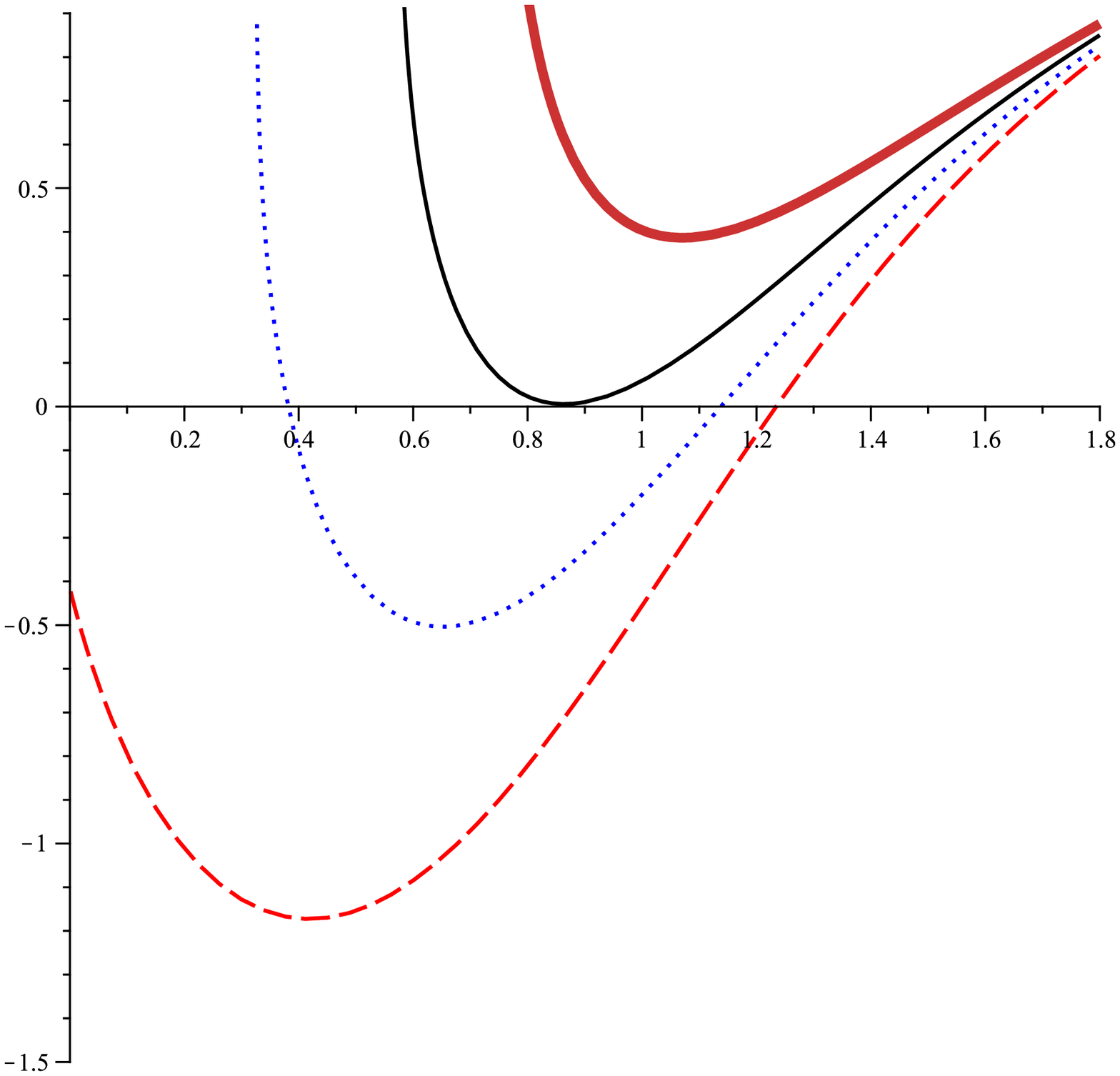} &
\end{array}
$%
\caption{\textbf{Eq. (\protect\ref{Psi-BI}):} ${\Psi _{BI}}(r) $ versus $r$
for $k=1$, $d=7$, $q=10$, $\protect\alpha = 0.5$, $\protect\beta=0.4$, $M=5$
and $\Lambda = - 1$. \newline
\textbf{Left panel:} $f(\protect\varepsilon) =1$ with $g(\protect\varepsilon%
)=0.82$ (dashed line), $g(\protect\varepsilon) = 1$ (dotted line), $g(%
\protect\varepsilon) =1.09$ (continuous line) and $g(\protect\varepsilon%
)=1.2 $ (bold line). \newline
\textbf{Right panel:} $g(\protect\varepsilon) =1$ with $f(\protect\varepsilon%
)=0.82$ (dashed line), $f(\protect\varepsilon) = 1$ (dotted line), $f(%
\protect\varepsilon) =1.15$ (continuous line) and $f(\protect\varepsilon%
)=1.3 $ (bold line).}
\label{Psi-gE-fE}
\end{figure}

In order to investigate the effect of rainbow functions on the metric, we
plot Fig. \ref{Psi-gE-fE}. According to this figure, one finds that both $%
f(\varepsilon)$ and $g(\varepsilon)$ not only affect the location of event
horizon, but also their values characterize the type of horizon. In other
words, decreasing rainbow functions leads to increasing the event horizon.
In addition, regarding different values of rainbow functions, one can obtain
timelike singularity with two horizons, an extreme horizon or naked
singularity. Also there is a minimum value of rainbow function, in which for
its lower values, singularity will be spacelike with a non-extreme event
horizon.


\subsection{Thermodynamics of black holes in Lovelock-BI gravity's rainbow
\label{Thermodynamics-BI}}

Here, we are going to check the first law of thermodynamics for the black
hole solutions in Lovelock-BI gravity's rainbow. At the first step, we
calculate the conserved and thermodynamic quantities. Surface gravity
interpretation leads to the following Hawking temperature
\begin{equation}
T=\frac{{(d-2)kg{{(\varepsilon )}^{2}}}\left[ {{(d-7){\alpha ^{2}}g{{%
(\varepsilon )}^{4}}+3(d-5)k\alpha g{{(\varepsilon )}^{2}}{r_{+}^{2}}+3(d-3){%
r_{+}^{4}}}}\right] {+6\left( {-\Lambda +2{\beta ^{2}}}\right) {r_{+}^{6}}-12%
{\beta ^{2}}\sqrt{1+{\eta _{+}}}{r_{+}^{6}}}}{{12\pi (d-2)g(\varepsilon
)f(\varepsilon ){{\left( {k\alpha g{{(\varepsilon )}^{2}}+{r_{+}^{2}}}%
\right) }^{2}}}}.  \label{Temperature-BI}
\end{equation}

\begin{figure}[tbp]
$%
\begin{array}{cc}
\epsfxsize=8cm \epsffile{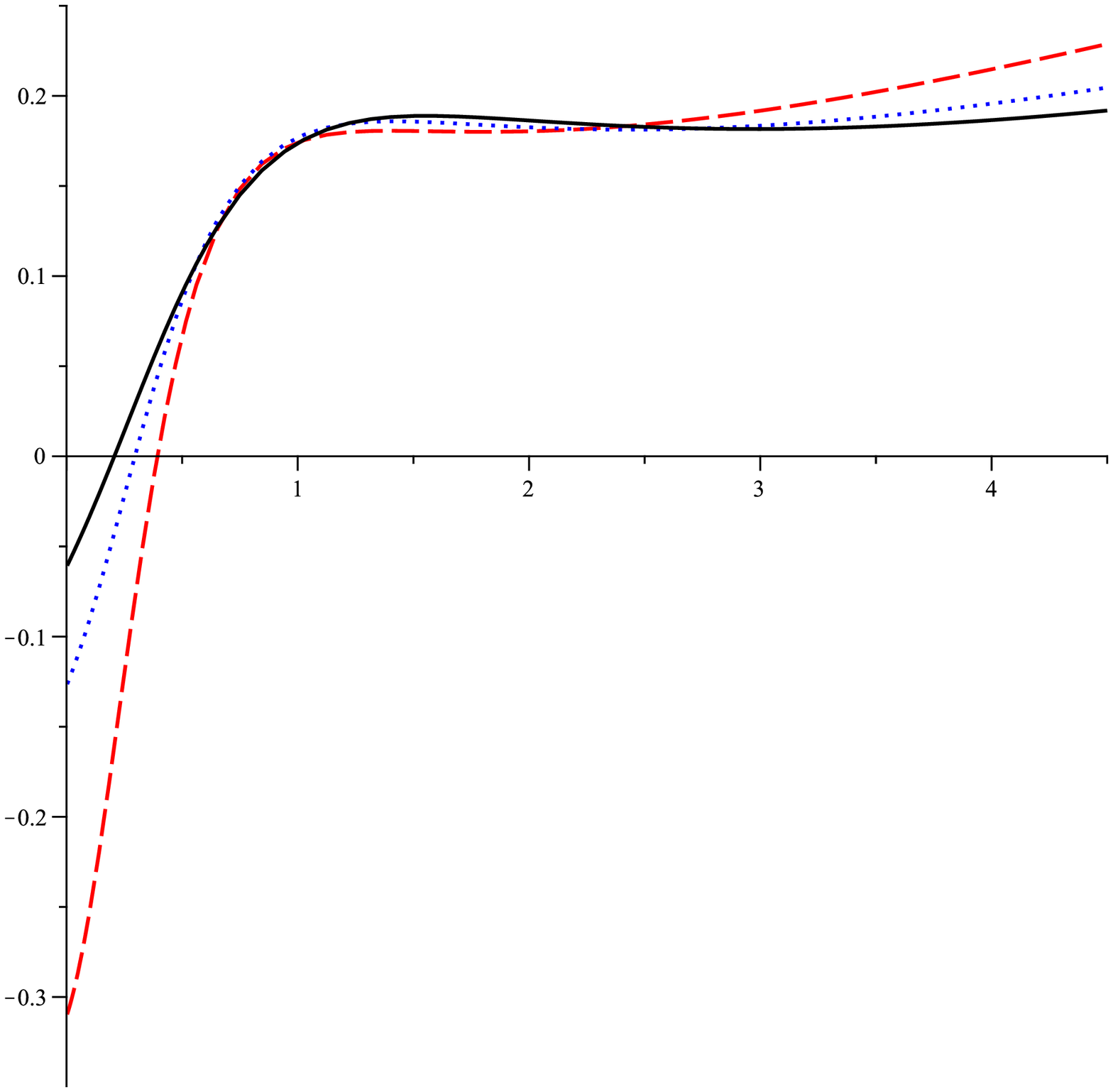} \epsfxsize=8cm \epsffile{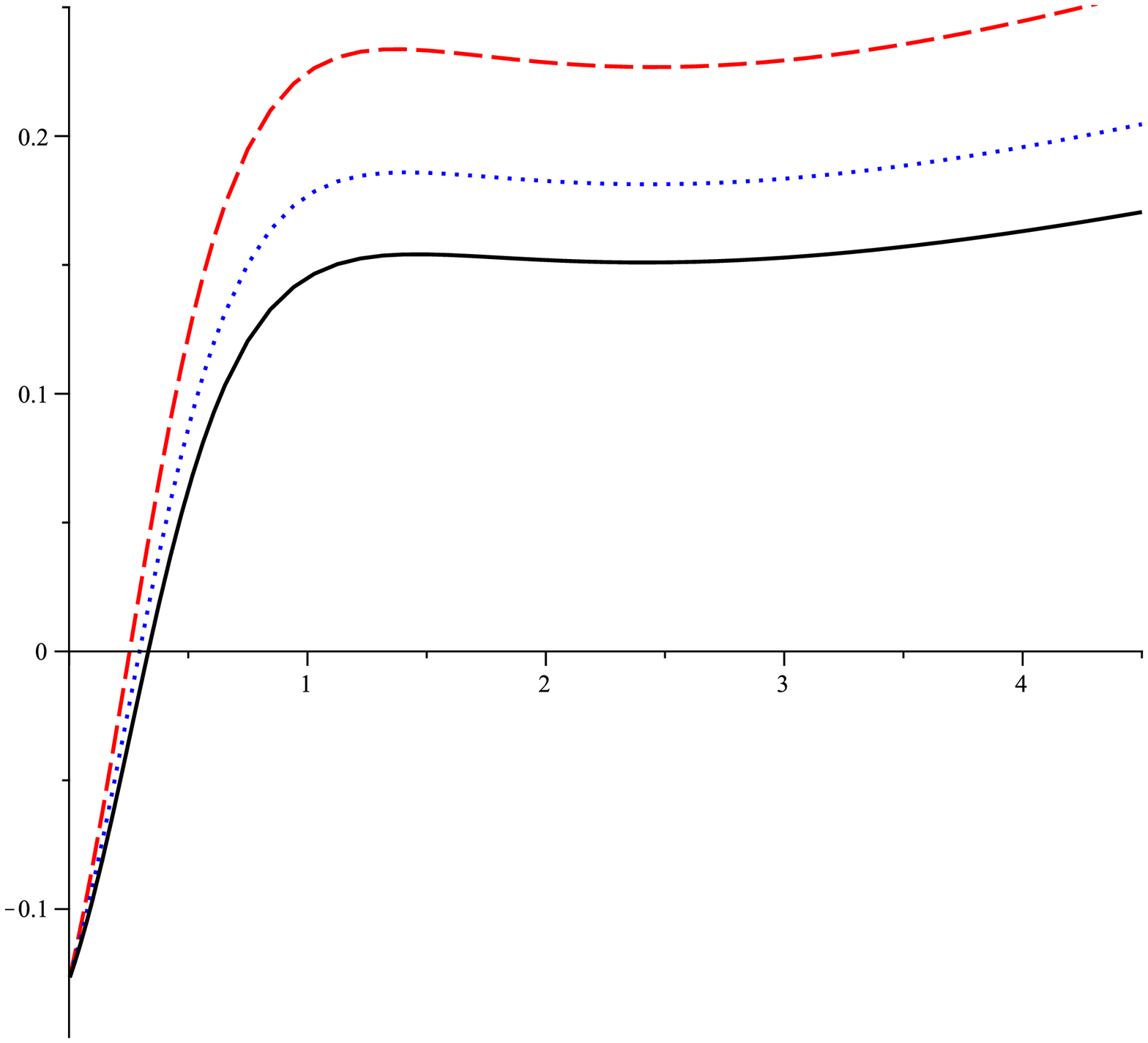} &
\end{array}
$%
\caption{\textbf{Eq. (\protect\ref{Temperature-BI}):} $T$ versus ${r_ + }$
for $k=1$, $d=7$, $q=10$, $\protect\alpha = 0.5$, $\protect\beta=0.05$ and $%
\Lambda = - 1$. \newline
\textbf{Left panel:} $f(\protect\varepsilon) =1$ with $g(\protect\varepsilon%
)=0.8$ (dashed line), $g(\protect\varepsilon) = 1$ (dotted line) and $g(%
\protect\varepsilon) =1.2$ (continuous line). \newline
\textbf{Right panel:} $g(\protect\varepsilon) =1$ with $f(\protect\varepsilon%
)=0.8$ (dashed line), $f(\protect\varepsilon) = 1$ (dotted line) and $f(%
\protect\varepsilon) =1.2$ (continuous line).}
\label{T-gE-fE}
\end{figure}

According to Eq. (\ref{Temperature-BI}), we find that rainbow functions
affect the Hawking temperature of the black holes. In addition, based on
Fig. \ref{T-gE-fE}, one finds that changing rainbow functions affect the
value of temperature and also increasing $g(\varepsilon)$ (or decreasing $%
f(\varepsilon)$) leads to increasing the root of $T$. It is notable that the
root of temperature is a bond point for the radius of event horizon, in
which there is no physical black hole with horizon radius smaller than such
bond point.

Entropy of the black hole can be calculated as the previous relationship
which leads to the same relation as Eq. (\ref{Entropy}). The electric
potential and charge of the black hole solutions are obtained as
\begin{equation}
\Phi ={\left. {{A_{\mu }}{\chi ^{\mu }}}\right\vert _{r\rightarrow \infty }}-%
{\left. {{A_{\mu }}{\chi ^{\mu }}}\right\vert _{r\rightarrow {r_{+}}}}=\frac{%
q}{{(d-3)r_{+}^{d-3}}}\mathrm{H}({\eta _{+}}),  \label{BI Potential}
\end{equation}

\begin{equation}
Q=\frac{{{V_{d-2}}}}{{4\pi }}\frac{{qf(\varepsilon )}}{{g{{(\varepsilon )}%
^{d-3}}}}.  \label{BI charge}
\end{equation}

Since the finite mass of the solutions is the same as that of Sec. \ref%
{Lovelock-Maxwell}, we can find the black hole mass as a function of the
entropy and electric charge as
\begin{equation}
M(S,Q)=\frac{{(d-2)}}{{16\pi f(\varepsilon )g{{(\varepsilon )}^{d-1}}}}%
\left( {\frac{1}{3}{k^{3}}{\alpha ^{2}}g{{(\varepsilon )}^{6}}r_{+}^{d-7}+{%
k^{2}}\alpha g{{(\varepsilon )}^{4}}r_{+}^{d-5}+kg{{(\varepsilon )}^{2}}%
r_{+}^{d-3}+{\Theta _{+}}}\right) ,  \label{Mass BI}
\end{equation}%
where
\[
{\Theta _{+}}=\frac{{4{\beta ^{2}}r_{+}^{d-1}}}{{(d-1)(d-2)}}\left( {1-\sqrt{%
1+{\varsigma _{+}}}+\frac{{(d-2)}}{{(d-3)}}{\varsigma _{+}}\mathrm{H}({%
\varsigma _{+}})}\right) -\frac{{2\Lambda r_{+}^{d-1}}}{{(d-1)(d-2)}},
\]%
and
\[
{\varsigma _{+}}=\frac{{16{\pi ^{2}}{Q^{2}}g{{(\varepsilon )}^{2(d-2)}}}}{{{%
\beta ^{2}}r_{+}^{2(d-2)}}}.
\]

Now, we are in a position to check the first law of thermodynamics. By
computing $\partial M/\partial {r_{+}}$, $\partial S/\partial {r_{+}}$ and $%
\partial Q/\partial {r_{+}}$ and using chain rule, one can show that the
following quantities
\begin{equation}
T={\left( {\frac{{\partial M}}{{\partial S}}}\right) _{Q}}%
,\,\,\,\,\,\,\,\,\,\,\,\Phi =\,{\left( {\frac{{\partial M}}{{\partial Q}}}%
\right) _{Q},}
\end{equation}%
are the same as the temperature and electric potential given in Eqs. (\ref%
{Temperature-BI}) and (\ref{BI Potential}), respectively. So, the obtained
thermodynamic and conserved quantities satisfy the first law of
thermodynamics, $dM=TdS+\Phi dQ$.


\subsection{Thermal stability of black holes in Lovelock-BI gravity's
rainbow \label{Stability-BI}}

Taking into account thermodynamic quantities, we are in a position to
investigate the local stability of black hole solutions in Lovelock-BI
gravity's rainbow. As we have mentioned before in Sec. \ref{Stabilty-M}, the
local stability requires that the behavior of energy (i.e. $M(S,Q)$) be a
convex function of its extensive quantities $S$ and $Q$, and the positivity
of the heat capacity guarantees the local stability in the canonical
ensemble. Here, we continue our discussion of thermal stability with
numerical analysis again. The only difference between this case and the case
of Lovelock-Maxwell gravity's rainbow is adding the nonlinearity. In this
case, the nonlinearity parameter affects the value of $r_{ + c}$.
Calculations show that regardless of the values of $\beta$, there is always
an unstable phase of black hole solutions for $\alpha < \alpha _{c}$. So as
before, there are two limits for the small enough values of $\alpha$, in
which black holes are unstable for $r_{ + ext} < r_ {+ } < r_{ + u}$. Also,
Numerical calculations for thermal analysis show that the black hole
temperature is always positive for $r_ {+ } > r_{ + c}$ depending on the
parameter $\beta$. Therefore, stability condition for $\alpha < \alpha _{c}$
shows that the heat capacity is positive for $r_{ + c} < r_ {+ } < r_{ +
ext} $ and thus black holes are thermally stable in this region.
Furthermore, for $r_ {+ } > r_{ + u}$, we find a stable phase for black
holes again. Finally, it is worthwhile to mention that for $r_ {+ } > r_{ +
c}$ and $\alpha > \alpha _{c} $, calculations show a thermally stable phase
for the black hole solutions of Lovelock-BI gravity's rainbow (see Fig. \ref%
{different-beta} for more details).

\begin{figure}[tbp]
$%
\begin{array}{cc}
\epsfxsize=5.5cm \epsffile{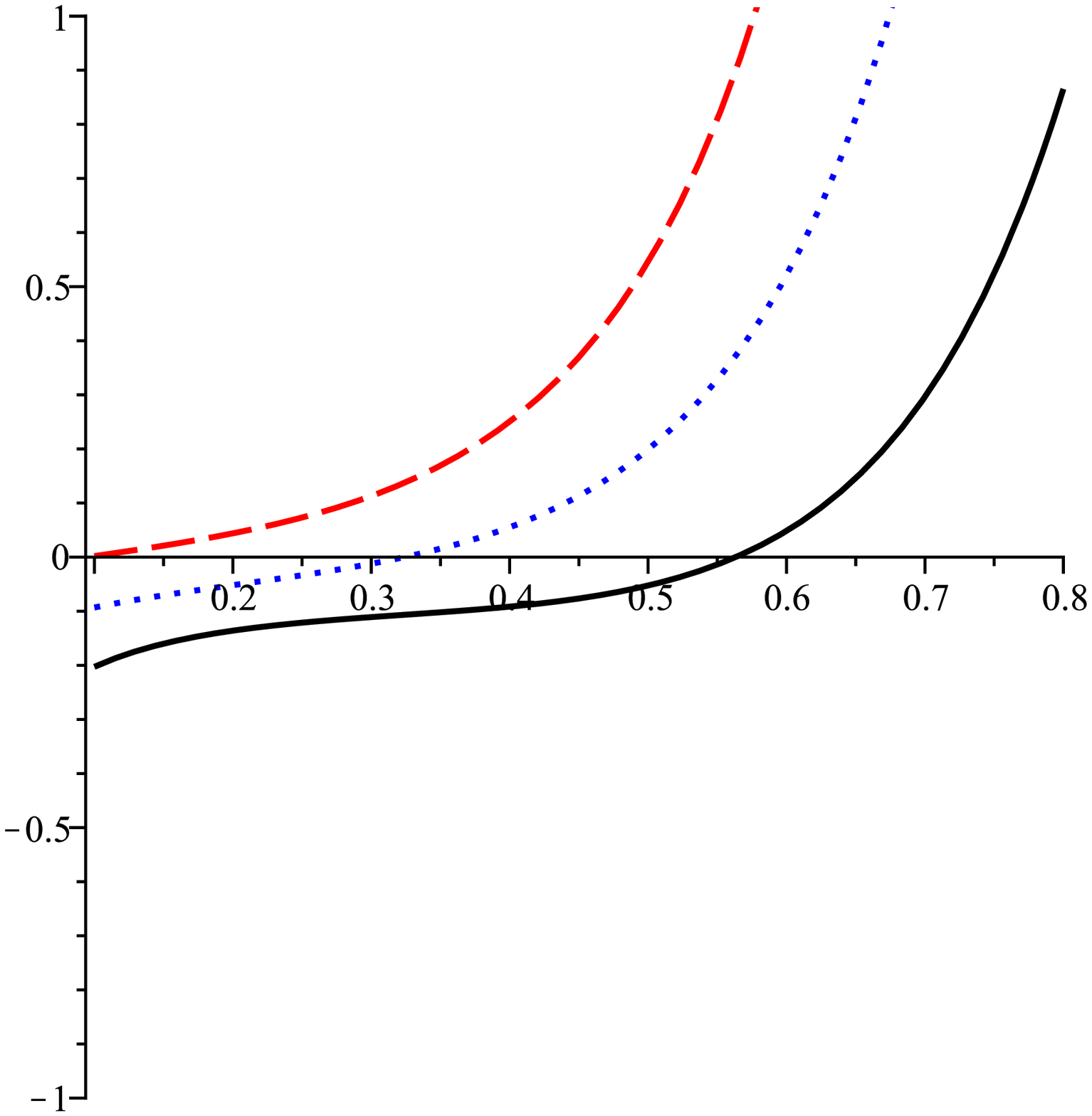} \epsfxsize=5.5cm %
\epsffile{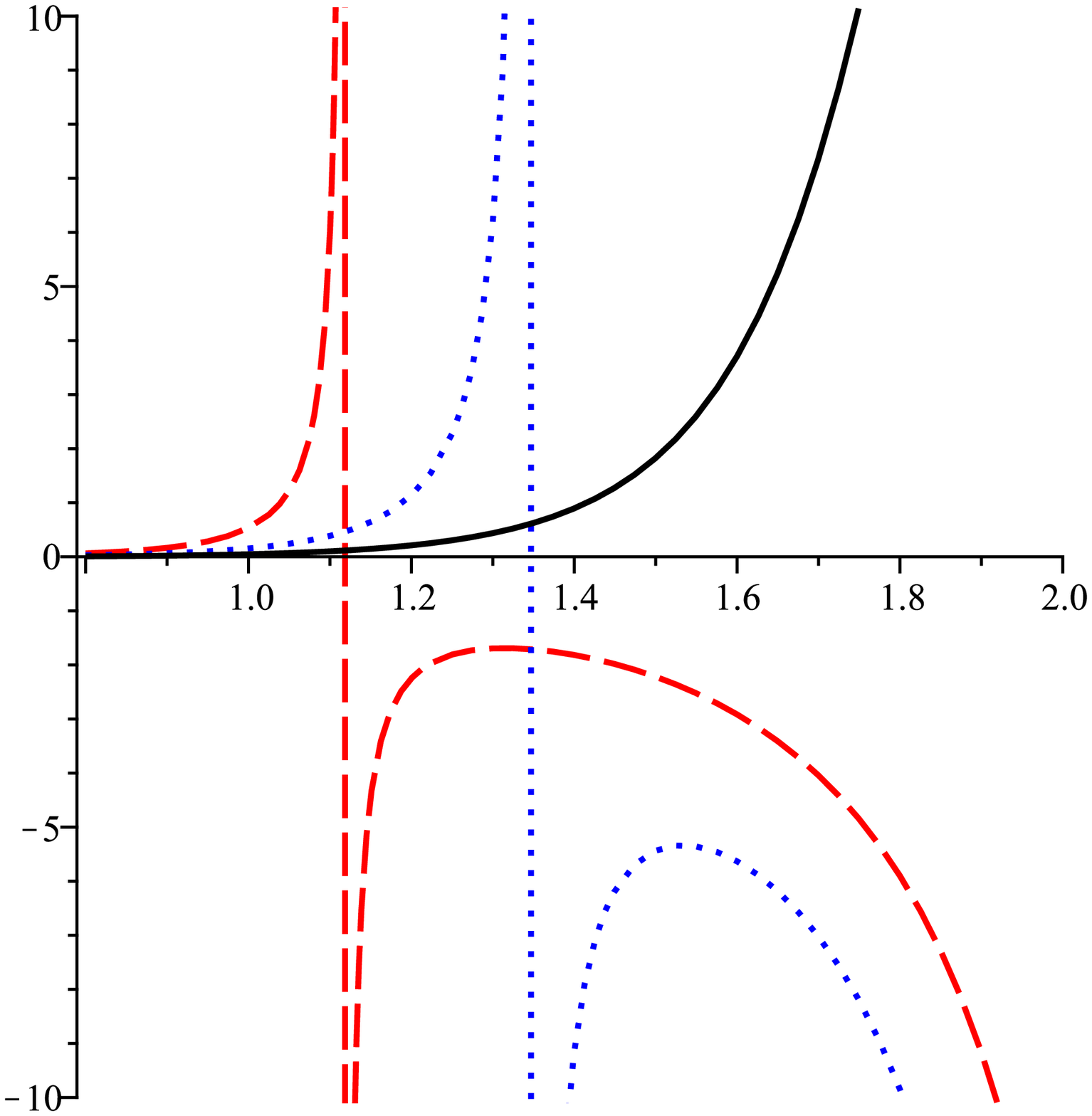}\epsfxsize=5.5cm %
\epsffile{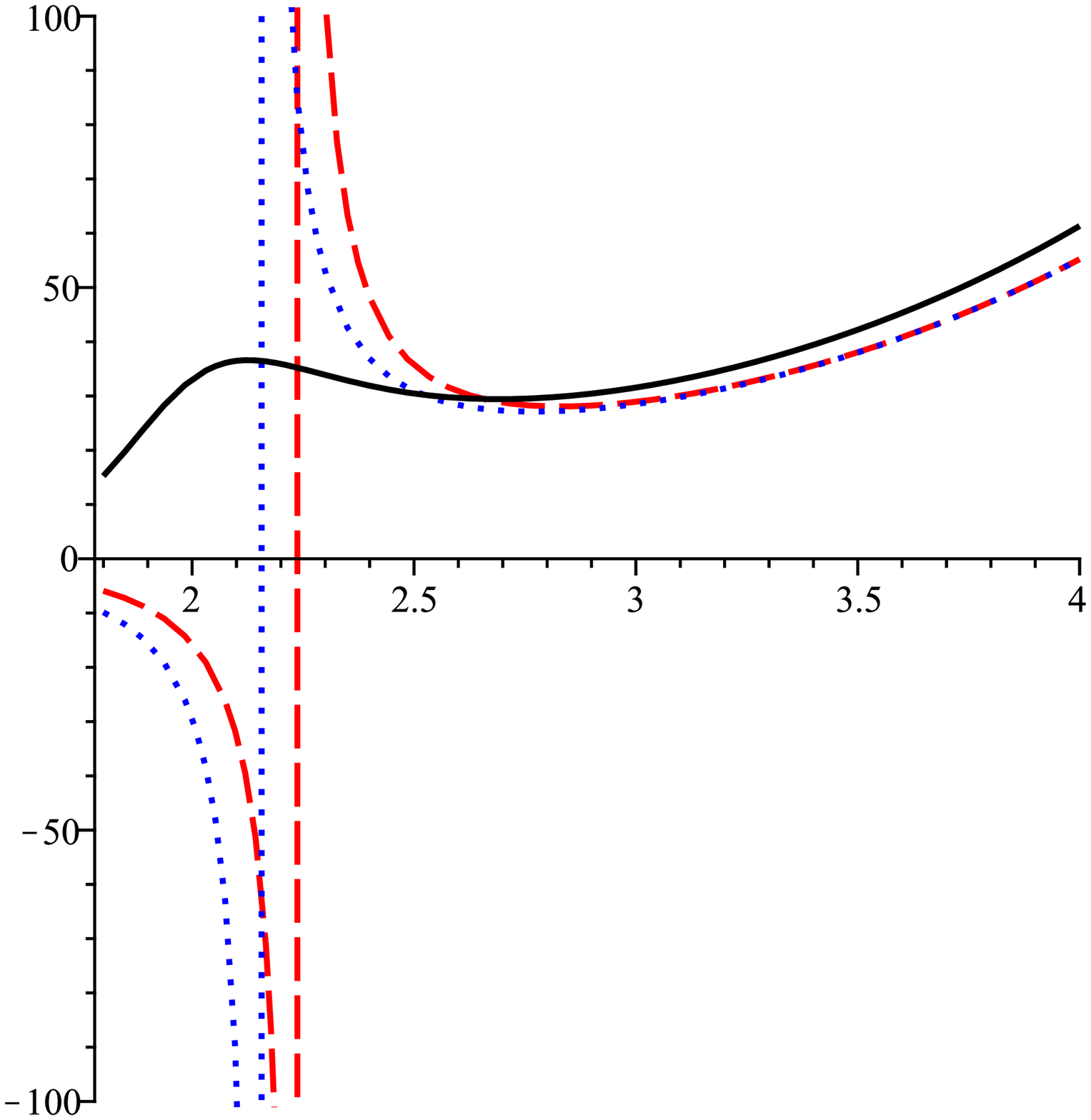} &
\end{array}
$%
\caption{$C_{Q}$ versus ${r_ + }$ for $k=1$, $d=7$, $q=10$, $f(\protect%
\varepsilon) = g(\protect\varepsilon) =0 .9$, $\protect\alpha = 0.5$, $%
\Lambda = - 1$ and $\protect\beta = 0.01$ ({dashed line}), $\protect\beta =
0.05$ ({dotted line}) and $\protect\beta = 0.15$ ({continuous line}).
\textbf{Different scales:} \emph{\ left panel ($0.1<r_{+}<0.8$), middle
panel ($0.8<r_{+}<2$) and right panel ($1.8<r_{+}<4$).}}
\label{different-beta}
\end{figure}



\begin{figure}[tbp]
$%
\begin{array}{cc}
\epsfxsize=8cm \epsffile{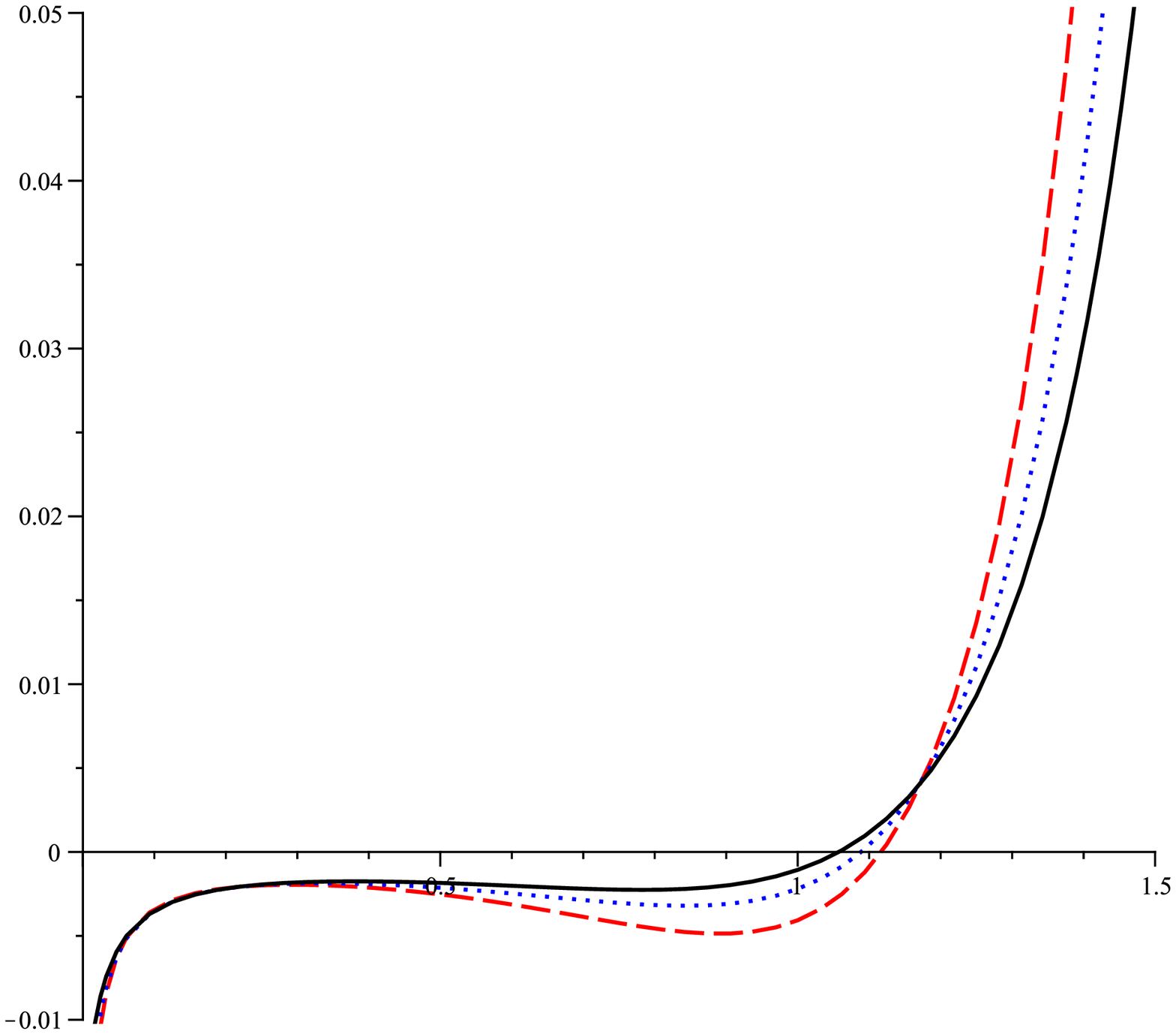} \epsfxsize=8cm \epsffile{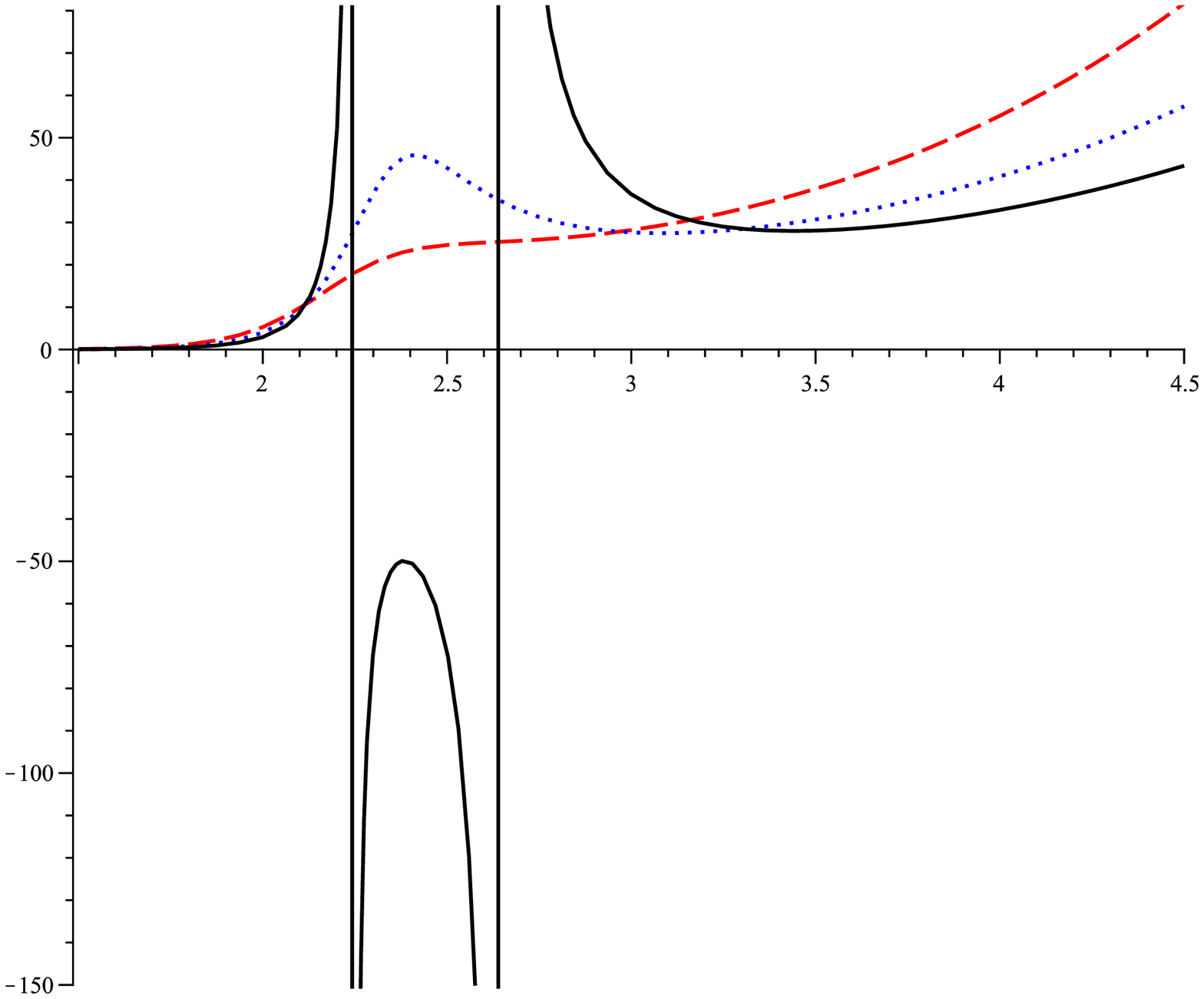} &
\end{array}
$%
\caption{$C_{Q}$ versus ${r_ + }$ for $k=1$, $d=7$, $q=10$, $\protect\alpha %
= 0.5$, $\protect\beta=1$, $\Lambda = - 1$, $f(\protect\varepsilon) =1$ and $%
g(\protect\varepsilon)=0.9$ (dashed line), $g(\protect\varepsilon) = 1$
(dotted line) and $g(\protect\varepsilon) =1.1$ (continuous line). \textbf{%
Different scales:} \emph{\ left panel ($0<r_{+}<1.5$) and right panel ($%
1.5<r_{+}<4.5$).} }
\label{HC-gE}
\end{figure}

\begin{figure}[tbp]
$%
\begin{array}{cc}
\epsfxsize=8cm \epsffile{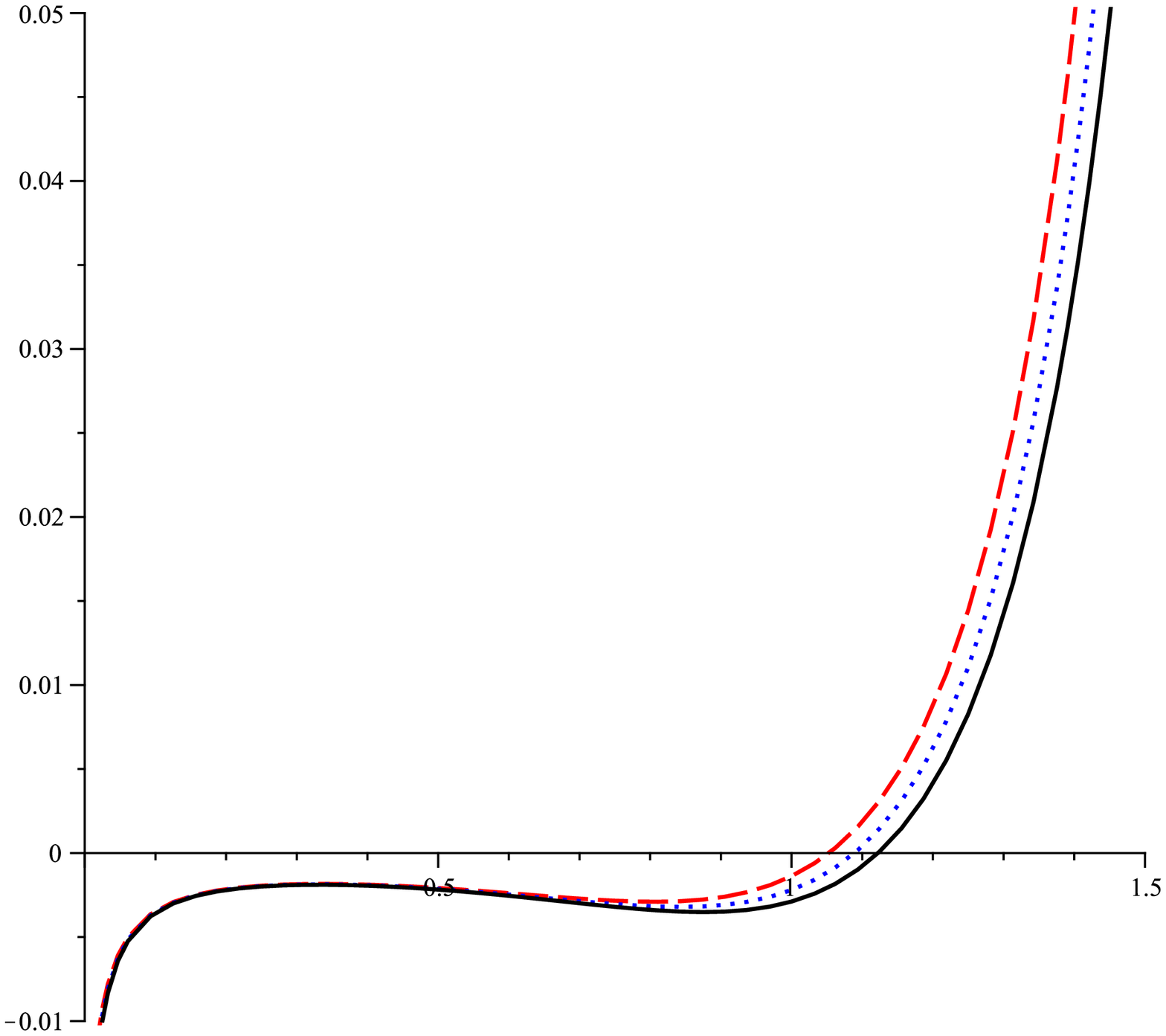} \epsfxsize=8cm \epsffile{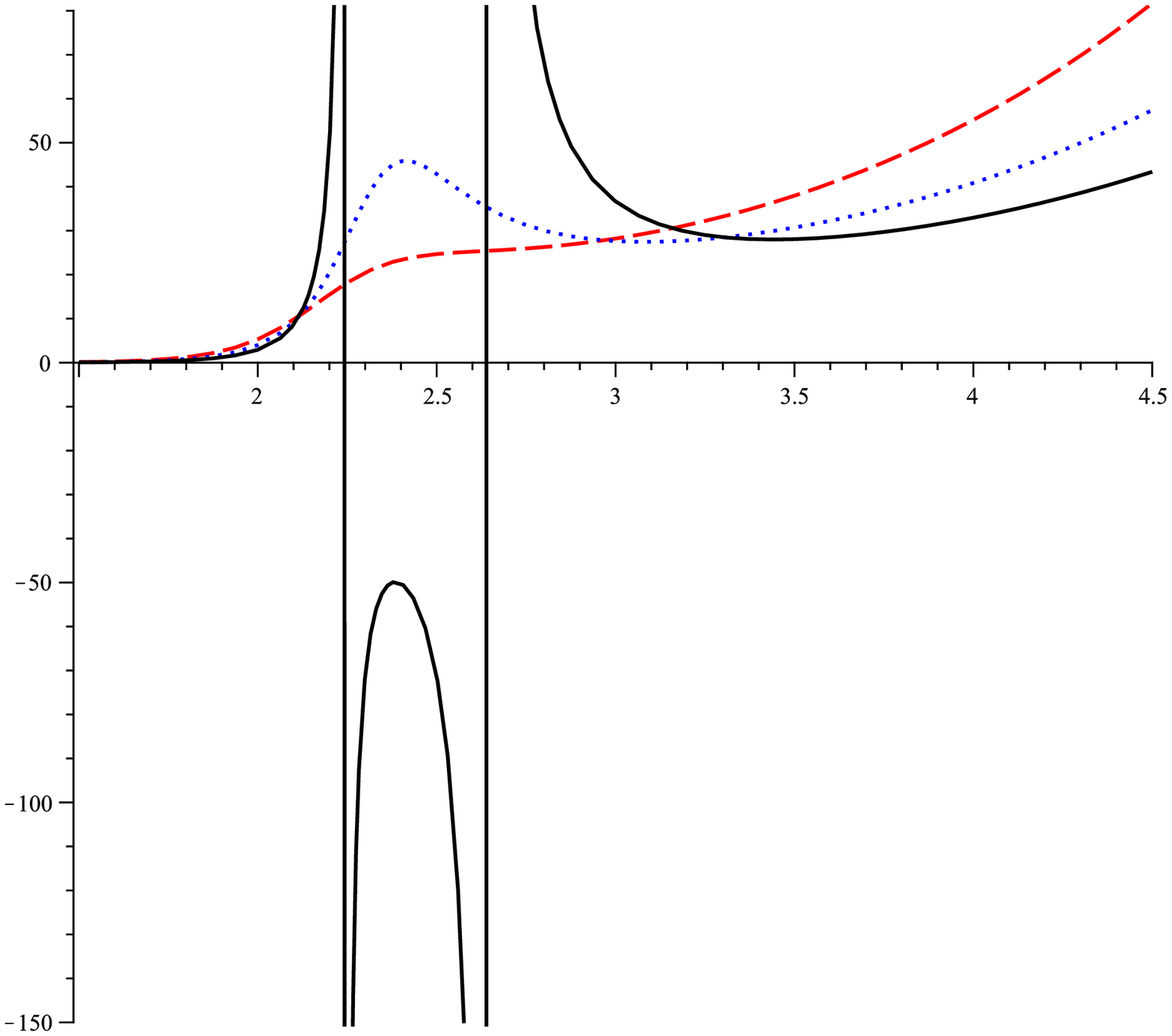} &
\end{array}
$%
\caption{$C_{Q}$ versus ${r_ + }$ for $k=1$, $d=7$, $q=10$, $\protect\alpha %
= 0.5$, $\protect\beta=1$, $\Lambda = - 1$, $g(\protect\varepsilon) =1$ and $%
f(\protect\varepsilon)=0.9$ (dashed line), $f(\protect\varepsilon) = 1$
(dotted line) and $f(\protect\varepsilon) =1.1$ (continuous line). \textbf{%
Different scales:} \emph{\ left panel ($0<r_{+}<1.5$) and right panel ($%
1.5<r_{+}<4.5$).} }
\label{HC-fE}
\end{figure}

For the sake of completeness, we plot Figs. \ref{HC-gE} and \ref{HC-fE} to
investigate the effects of rainbow functions on thermal stability of the
solutions. Left panels of these figures indicate the root of heat capacity
which is coincident with the place of vanishing temperature. In addition,
these figures show that there is a minimum value for the rainbow functions ($%
f_{min}(\varepsilon )$ and $g_{min}(\varepsilon )$) which separates two
different behaviors. In other words, black holes are thermally stable for $%
f(\varepsilon )<f_{min}(\varepsilon )$ ($g(\varepsilon )<g_{min}(\varepsilon
)$). Otherwise for $f(\varepsilon )>f_{min}(\varepsilon )$ ($g(\varepsilon
)>g_{min}(\varepsilon )$) there are two divergences for the heat capacity
which indicate a phase transition. It is notable that the values of $%
f_{min}(\varepsilon )$ and $g_{min}(\varepsilon )$ can be numerically
calculated as functions of other parameters.

\section{PV Critically of Lovelock gravity's rainbow \label{pv}}

In order to study the critical behavior of black holes in the extended phase
space, we treat the cosmological constant as a thermodynamics pressure. In
other words, we do not work in a fixed AdS or dS background. In fact, the
cosmological constant is not a constant anymore, but a variable. According
to the AdS/CFT correspondence, of interesting case is asymptotically AdS
black holes, and in this section, we consider AdS black holes with
spherically topological horizons (there is not any phase transition for $k=0$
and $k=-1$ in this gravity model). If we treat the negative cosmological
constant proportional to thermodynamic pressure, its conjugate quantity will
be thermodynamic volume. This new assumption has considered by many authors
in recent years (see \cite%
{PV1,PV2,PVwork1,PVwork2,PVwork3,PVwork4,PVwork5,PVwork6,PVwork7,PVwork8,PVwork9,PVwork10,PVwork12,PVwork13}
and references therein). Already, using scaling argument, it has shown that
the Smarr relation is consistent with first law of thermodynamics by
assuming the cosmological constant as a thermodynamic variable \cite%
{PVwork8,HendiAliDehghani,Kastor1,Kastor2,Kastor3}. Generalization of
extended phase space thermodynamics to higher derivative gravity theories
and investigation of triple points, reentrant phase transitions and equal
area law have been studied before \cite%
{PVlovelock1,PVlovelock2,LovelockPV1,LovelockPV2,LovelockPV3,LovelockPV4,LovelockPV5,LovelockPV6,LovelockPV7,LovelockPV8,LovelockPV9,LovelockPV10,LovelockPV11}%
. Here, we are going to study the phase transition and critical behavior of
black hole solutions in a more complicated gravitational background, i.e.
for our interest, Lovelock gravity's rainbow. In the geometric units ${G_{N}}%
=\hbar =c={k_{B}}=1$, one can identify the cosmological constant with the
pressure as
\begin{equation}
P=-\frac{\Lambda }{{8\pi }}=\frac{{(d-1)(d-2)}}{{16\pi {\ell ^{2}}}}.
\label{Pressure}
\end{equation}

It was seen that by considering the cosmological constant as thermodynamic
pressure, the black hole mass $M$ can be explained as enthalpy rather than
internal energy of the system, i.e. $M=H$ \cite{Kastor1,Kastor2,Kastor3},
and so the Gibbs free energy will be in the form of $G=M-TS$. Using the
chain rule, thermodynamic volume of black hole is obtained as
\begin{equation}
V={\left( {\frac{{\partial H}}{{\partial P}}}\right) _{{X_{i}}}}={\left( {%
\frac{{\partial M}}{{\partial P}}}\right) _{{X_{i}}}}=\frac{{{w_{d-2}}{%
r_{+}^{d-1}}}}{{(d-1)f(\varepsilon )g{{(\varepsilon )}^{d-1}}}},
\label{Volume}
\end{equation}%
where ${{X_{i}}}$ denotes all other extensive quantities and ${{w_{d-2}=}}%
\left. V_{d-2}\right\vert _{k=1}$ is the volume of $(d-2)-$dimensional unit
sphere (see Eq. \ref{met2} with $k=1$). The critical point in an isothermal $%
P-V$ diagram is an inflection point, and therefore, it can be obtained from
the following equations
\begin{equation}
{\left( {\frac{{\partial P}}{{\partial {V}}}}\right) _{T}}=0,  \label{dPdV}
\end{equation}%
\begin{equation}
{\left( {\frac{{\partial }^{2}{P}}{{\partial {V}}^{2}}}\right) _{T}}=0.
\label{d2PdV2}
\end{equation}

Investigation of the critical behavior is not possible, analytically, and so
we have to use numerical analysis. In order to study the mentioned critical
behavior, we present various tables and figures. It is worthwhile to mention
that the characteristic swallow-tail form of $G-T$ diagrams, inflection
point of isothermal $P-V$ plots and also subcritical isobar of $T-V$
diagrams guarantee the existence of the phase transition. The associated $%
P-V $, $T-V$ and $G-T$ diagrams for both types of the obtained black hole
solutions (Lovelock-Maxwell and Lovelock-BI) are displayed and the related
critical points are obtained in various tables. The effects of higher
derivative Lovelock gravity, nonlinearity parameter of BI theory and rainbow
functions are evident in presented tables and figures.

\begin{figure}[tbp]
$%
\begin{array}{ccc}
\epsfxsize=6cm \epsffile{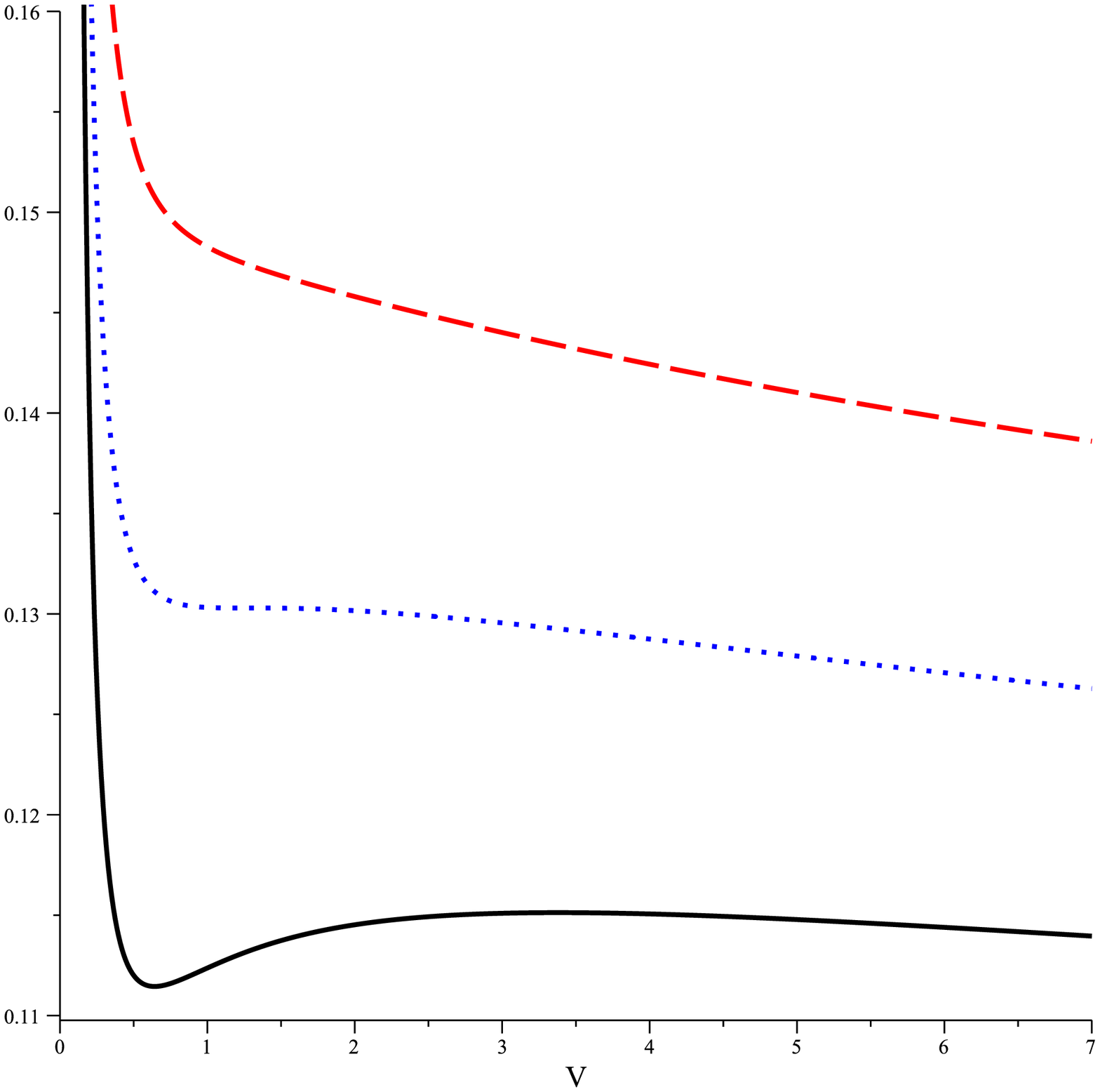} & \epsfxsize=6cm \epsffile{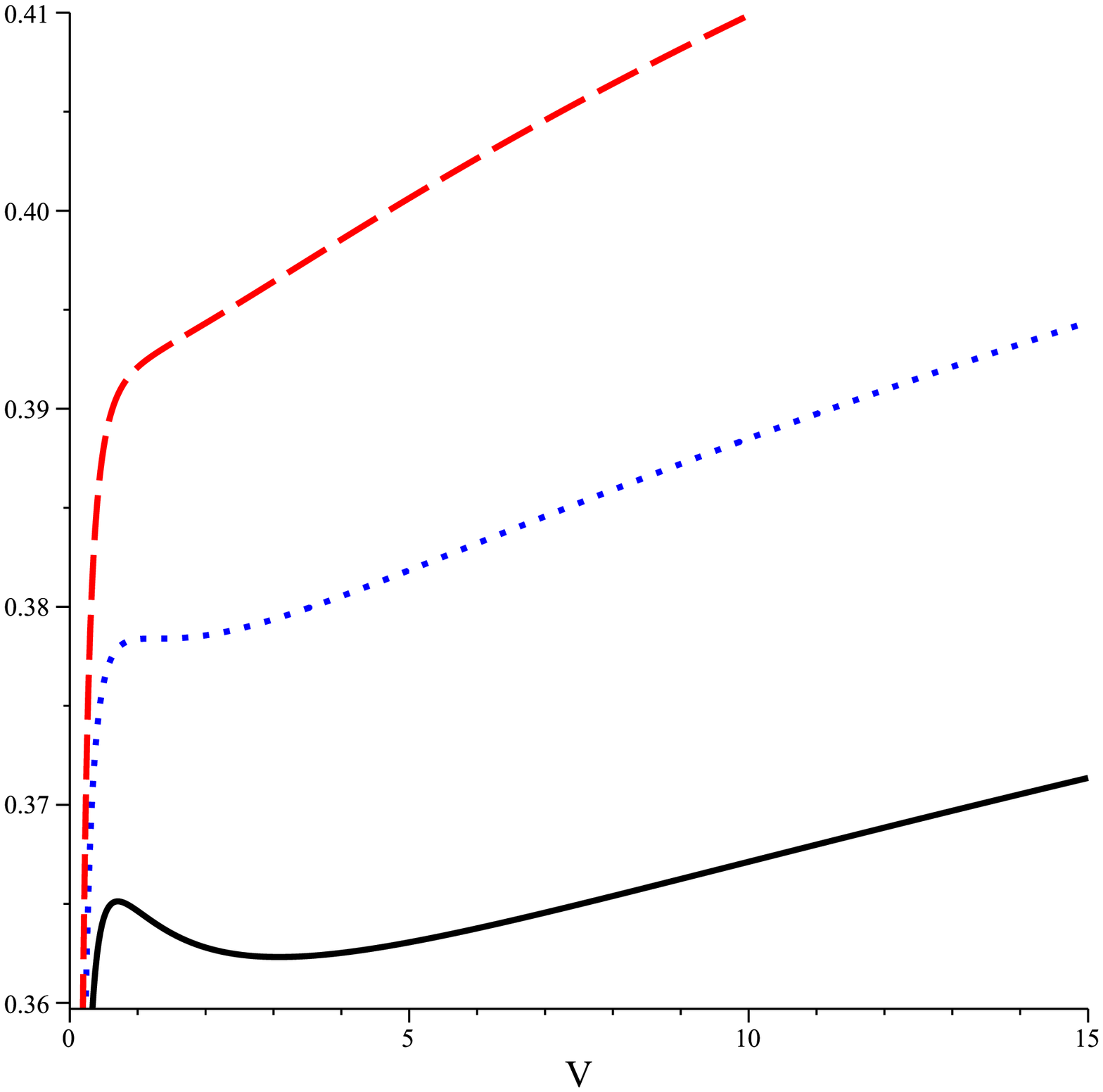} & %
\epsfxsize=6cm \epsffile{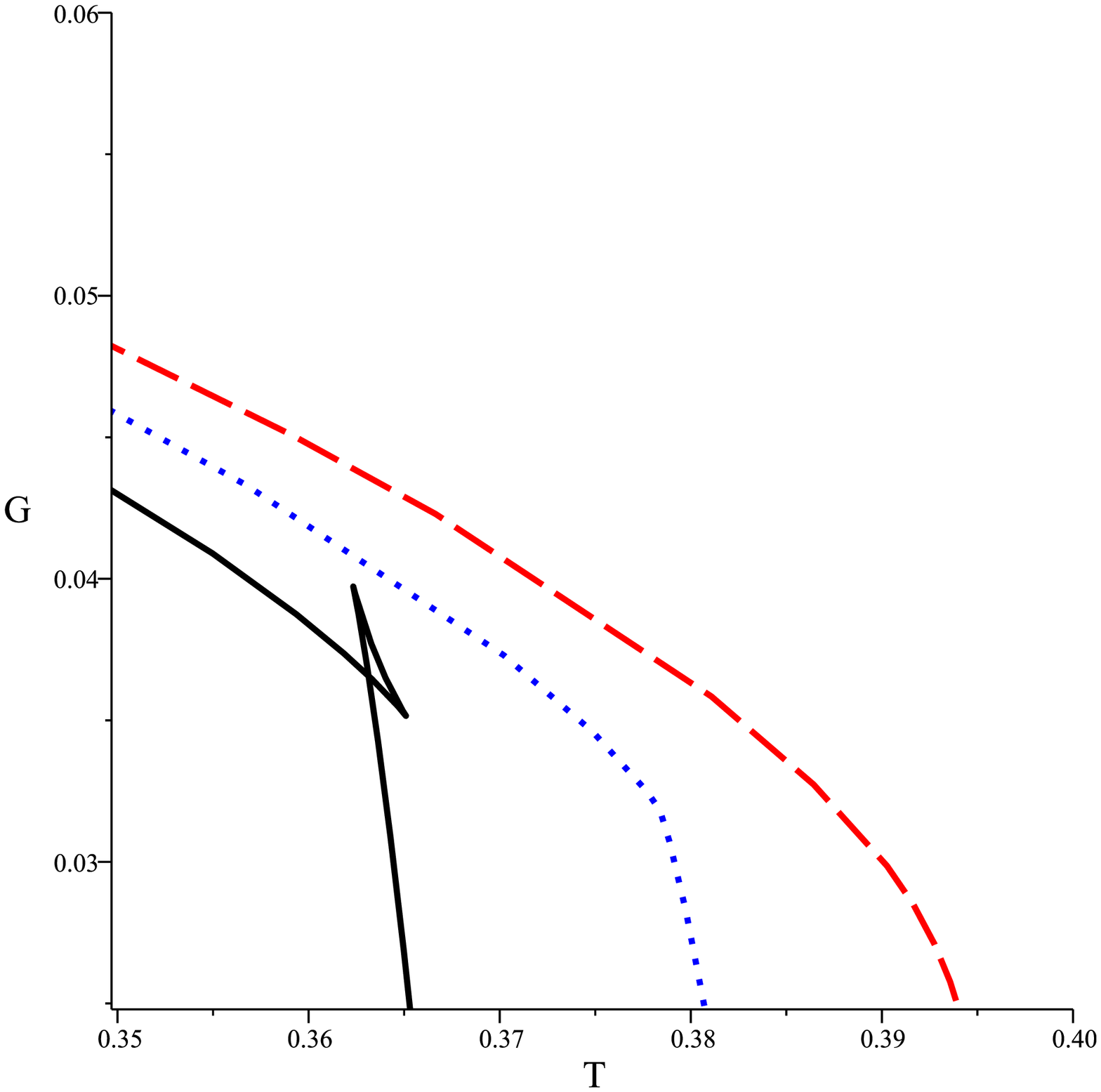}%
\end{array}
$%
\caption{\textbf{Lovelock-Maxwell gravity's rainbow:} $P-V$ (left), $T-V$
(middle) and $G-T$ (right) diagrams for $k=1$, $d=7$, $q=1$, $f(\protect%
\varepsilon)=g(\protect\varepsilon)=0.9$ and $\protect\alpha=0.1$. \newline
\textbf{Left panel:} $T<T_{c}$ (continuous line), $T=T_{c}$ (dotted line)
and $T>T_{c}$ (dashed line).\newline
\textbf{Middle and right panels:} $P<P_{c}$ (continuous line), $P=P_{c}$
(dotted line) and $P>P_{c}$ (dashed line).}
\label{PV-Max}
\end{figure}

\begin{figure}[tbp]
$%
\begin{array}{ccc}
\epsfxsize=6cm \epsffile{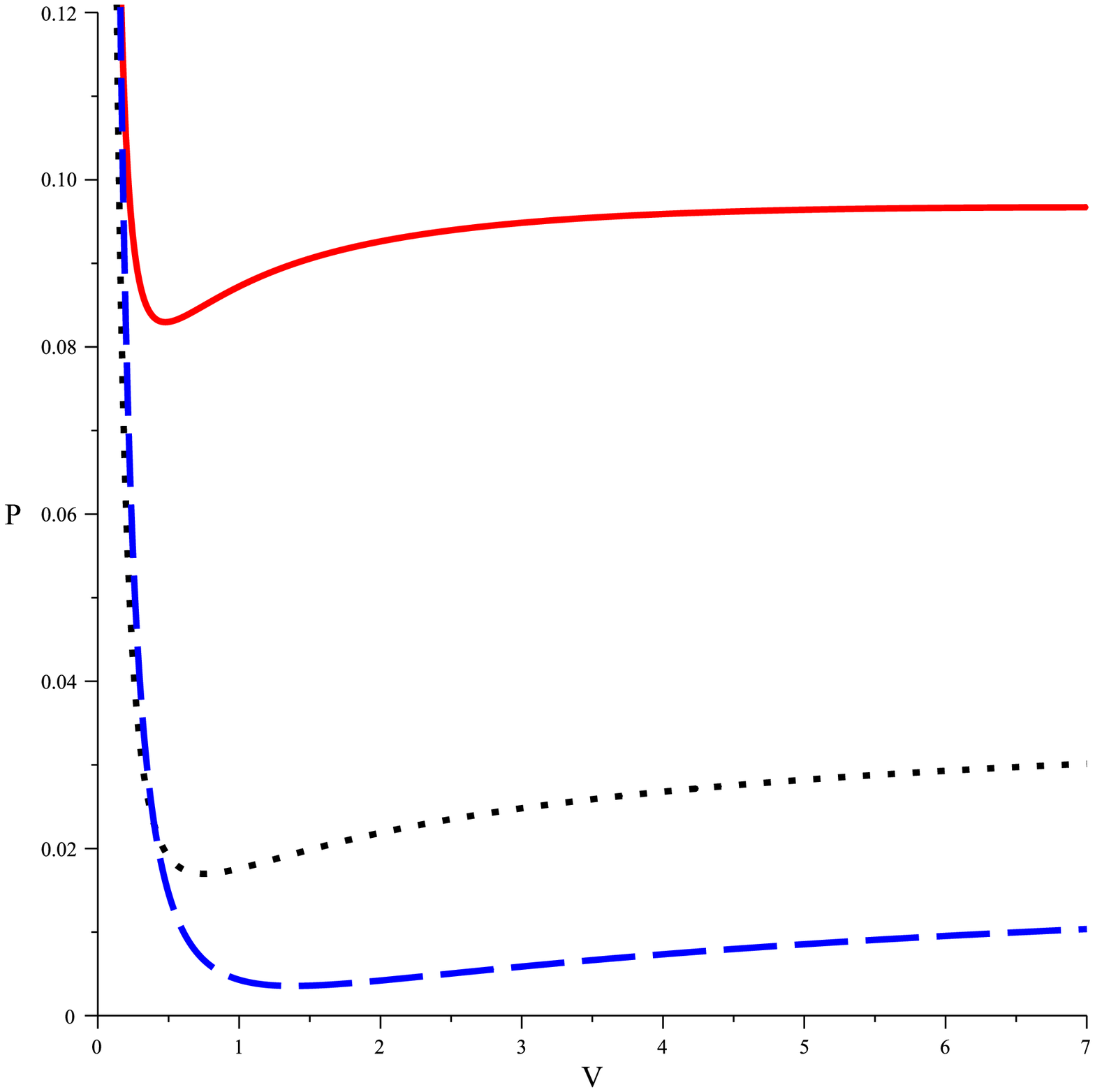} & \epsfxsize=6cm %
\epsffile{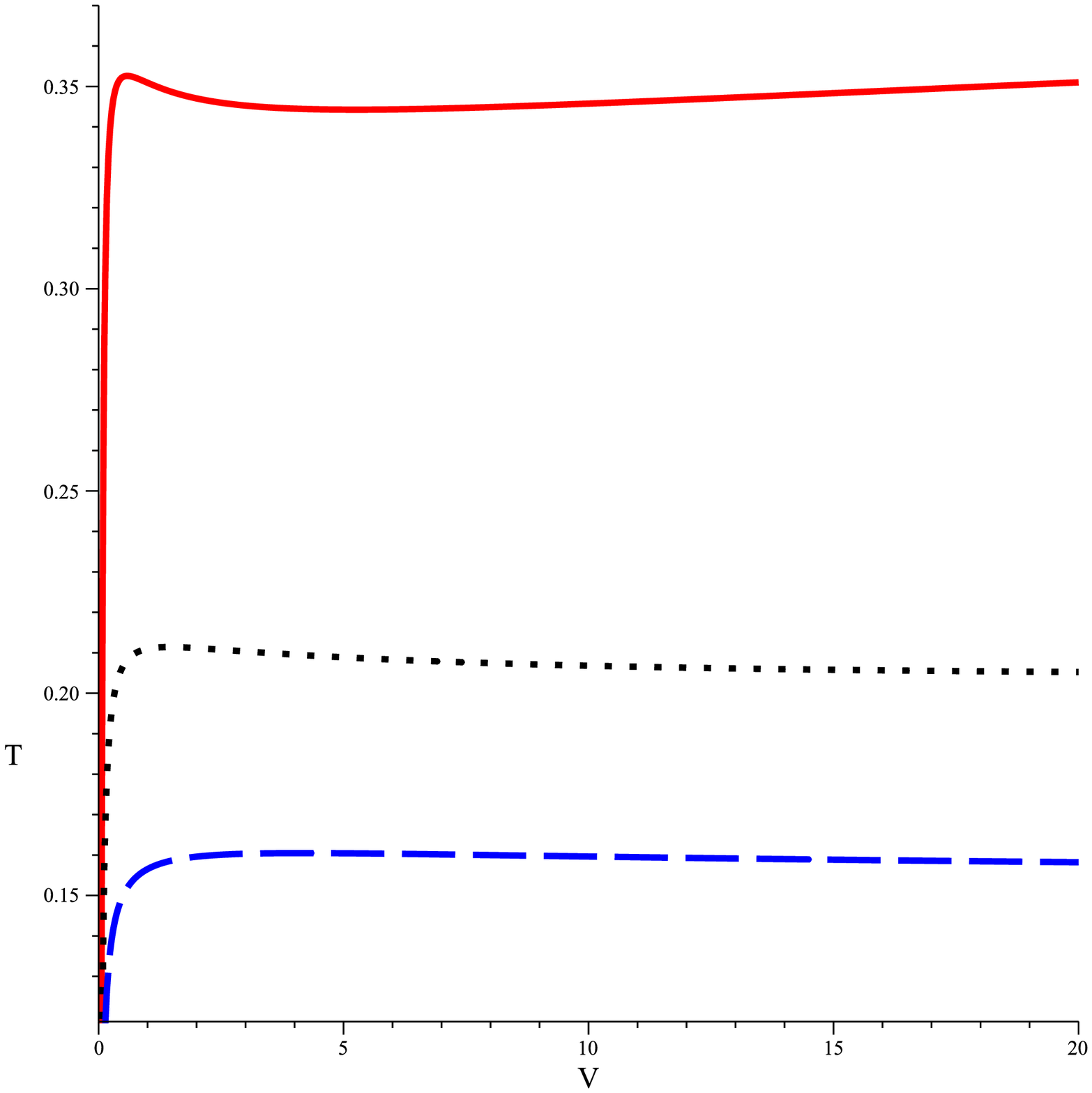} & \epsfxsize=6cm \epsffile{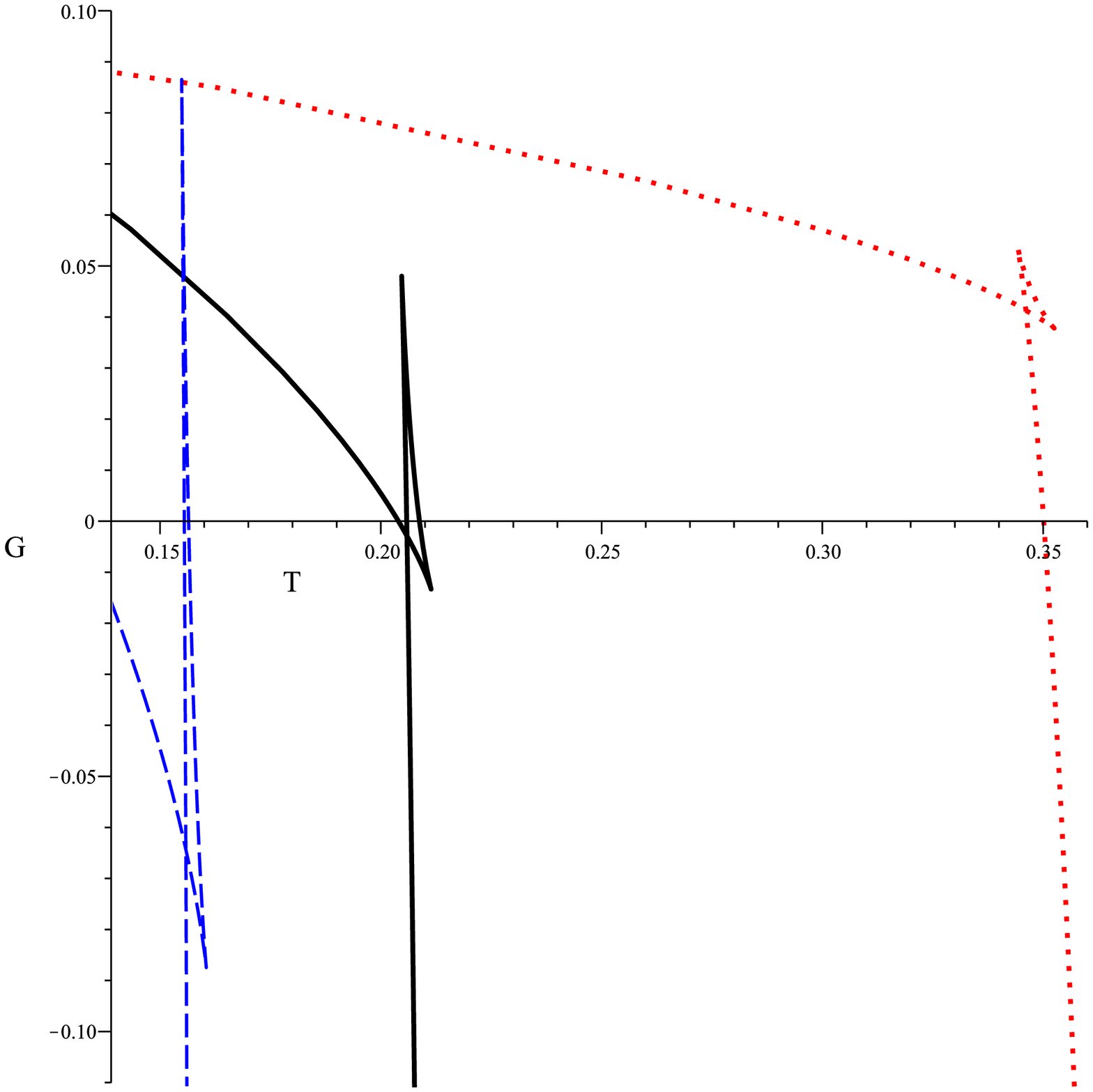}%
\end{array}
$%
\caption{\textbf{Lovelock-Maxwell gravity's rainbow:} $P-V$ for $T<T_{c}$
(left), $T-V$ for $P<P_{c}$ (middle) and $G-T$ for $P<P_{c}$ (right)
diagrams for $k=1$, $d=7$, $q=1$, $f(\protect\varepsilon)=g(\protect%
\varepsilon)=0.9$, and $\protect\alpha=0.1$ (continuous line), $\protect%
\alpha=0.5$ (dotted line) and $\protect\alpha=0.9 $ (dashed line).}
\label{PV-Max-Comp}
\end{figure}

\begin{figure}[tbp]
$%
\begin{array}{ccc}
\epsfxsize=6cm \epsffile{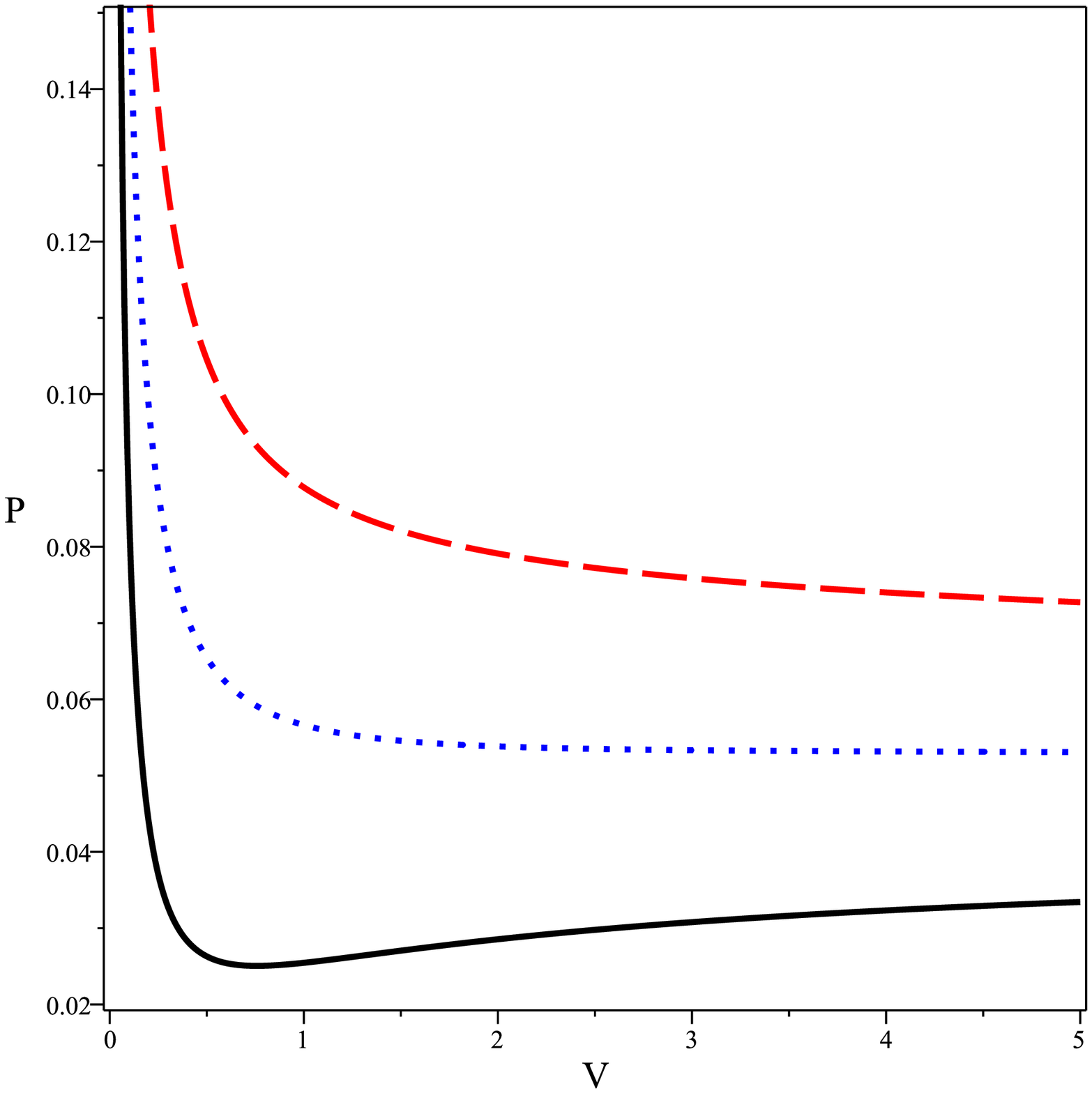} & \epsfxsize=6cm \epsffile{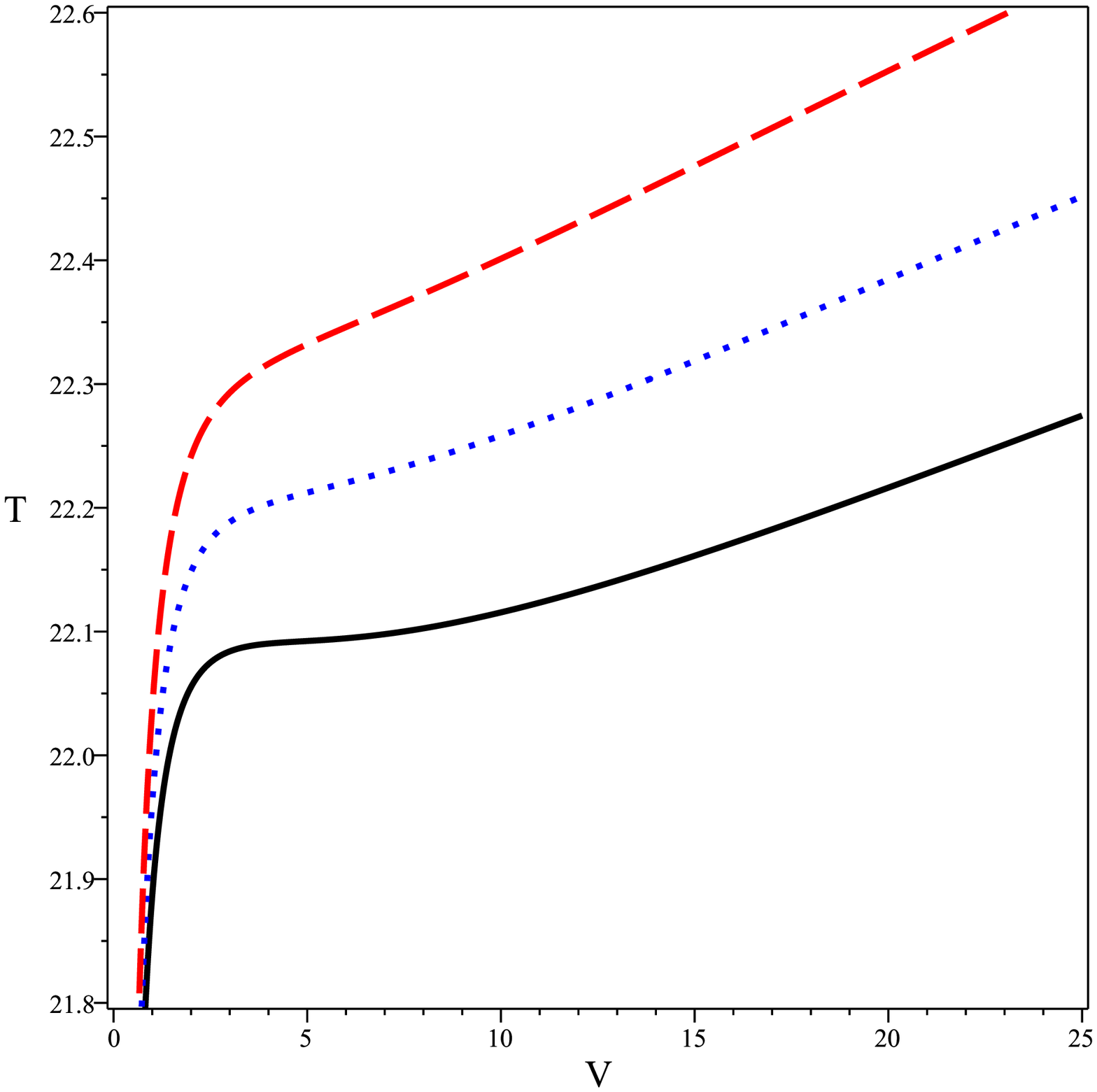} & %
\epsfxsize=6cm \epsffile{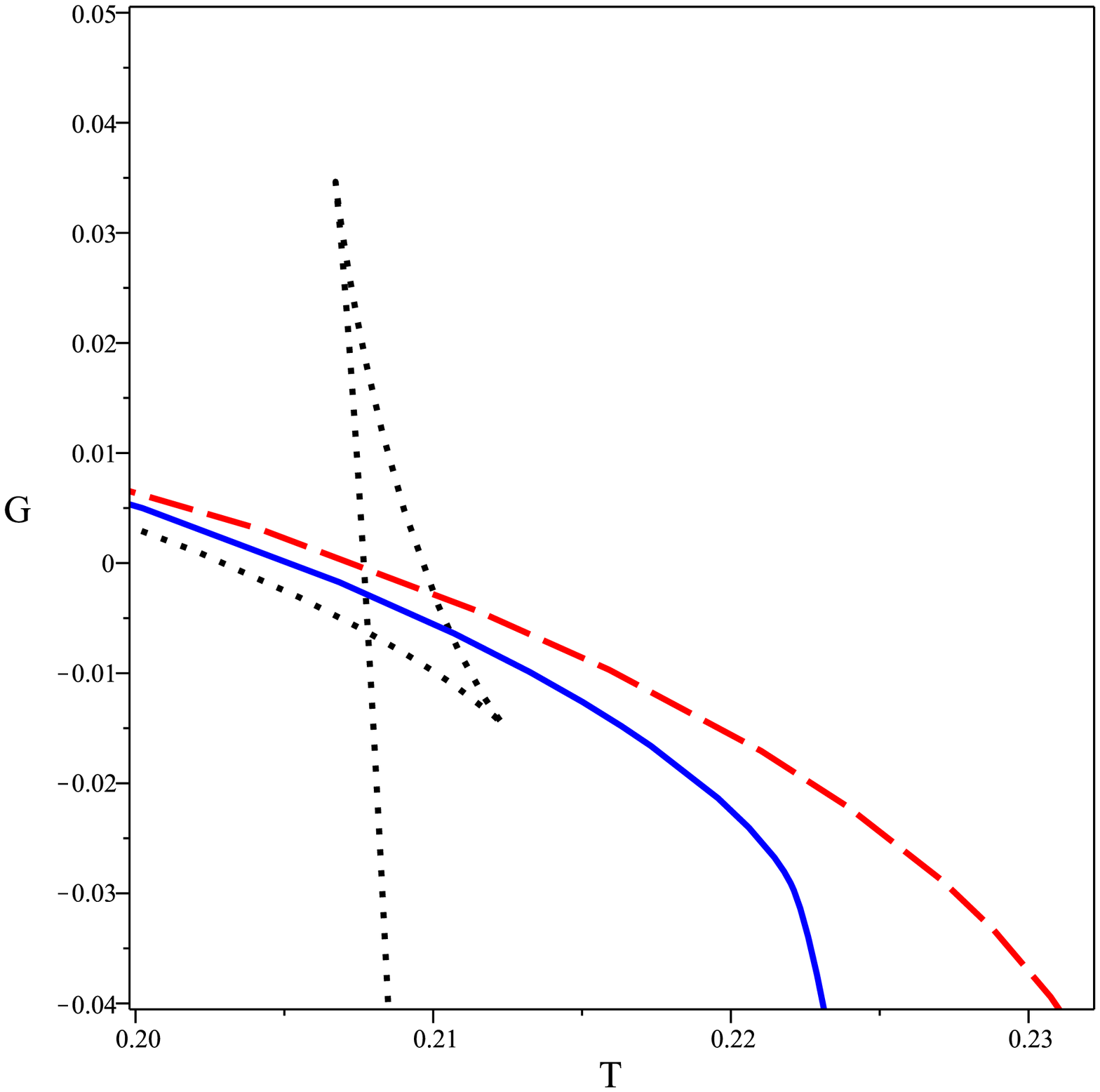}%
\end{array}
$%
\caption{\textbf{Lovelock-BI gravity's rainbow:} $P-V$ (left), $T-V$
(middle) and $G-T$ (right) diagrams for $k=1$, $d=7$, $q=1$, $f(\protect%
\varepsilon)=g(\protect\varepsilon)=0.9$ and $\protect\alpha=0.5$, $\protect%
\beta=0.5$. \newline
\textbf{Left panel:} $T<T_{c}$ (continuous line), $T=T_{c}$ (dotted line)
and $T>T_{c}$ (dashed line).\newline
\textbf{Middle and right panels:} $P<P_{c}$ (continuous line), $P=P_{c}$
(dotted line) and $P>P_{c}$ (dashed line).}
\label{PV-BI}
\end{figure}
\begin{figure}[tbp]
$%
\begin{array}{ccc}
\epsfxsize=6cm \epsffile{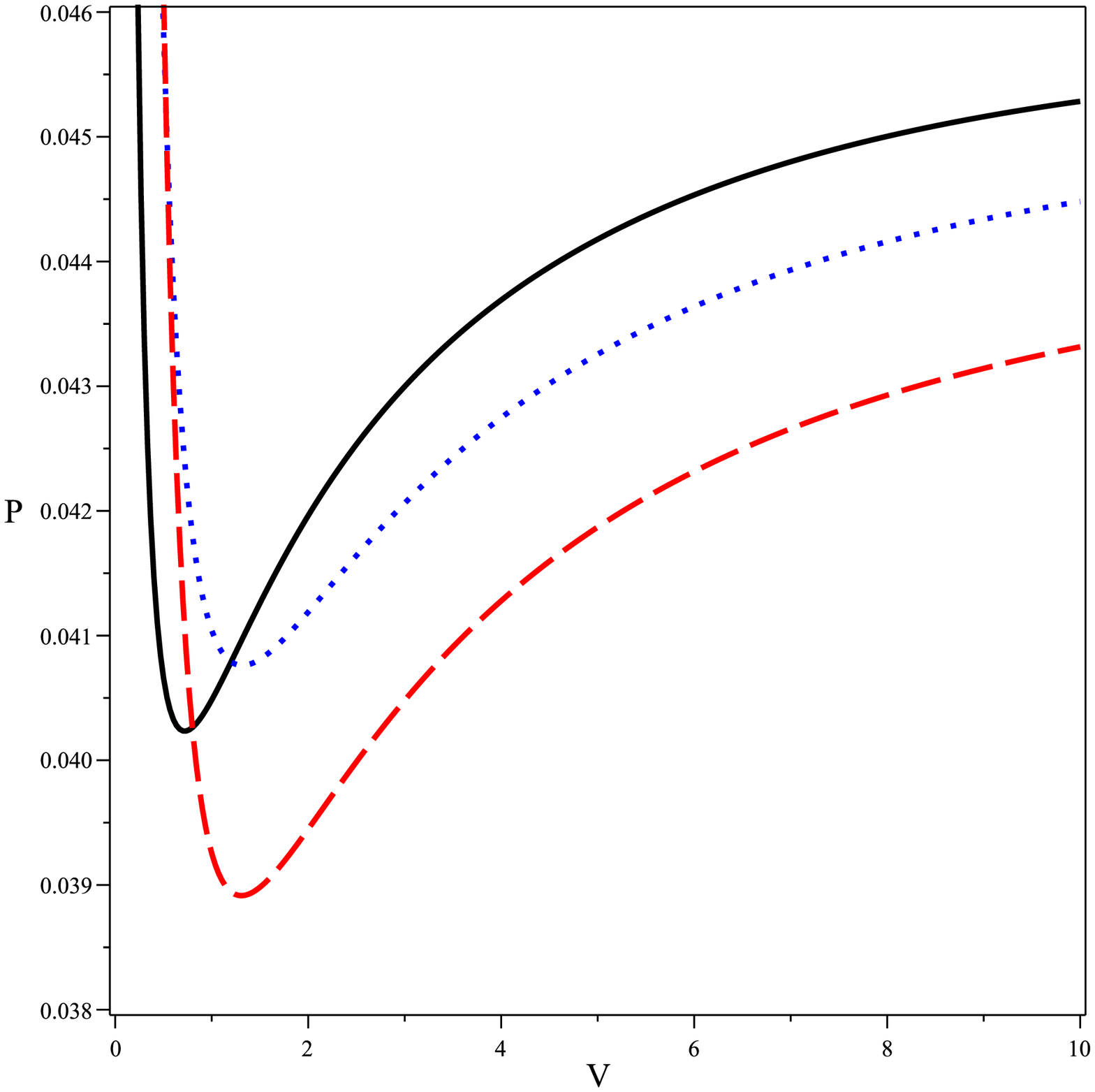} & \epsfxsize=6cm %
\epsffile{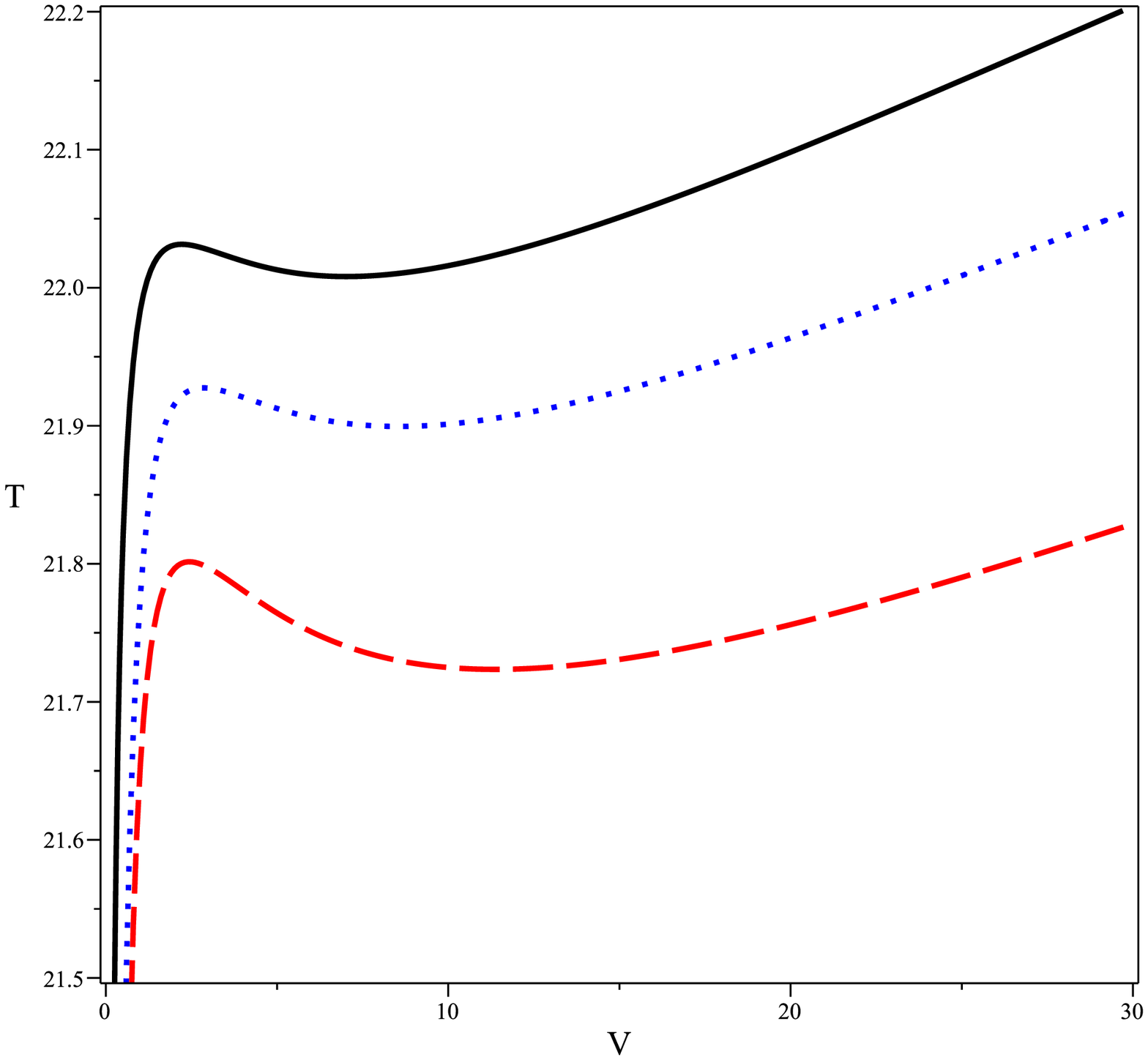} & \epsfxsize=6cm \epsffile{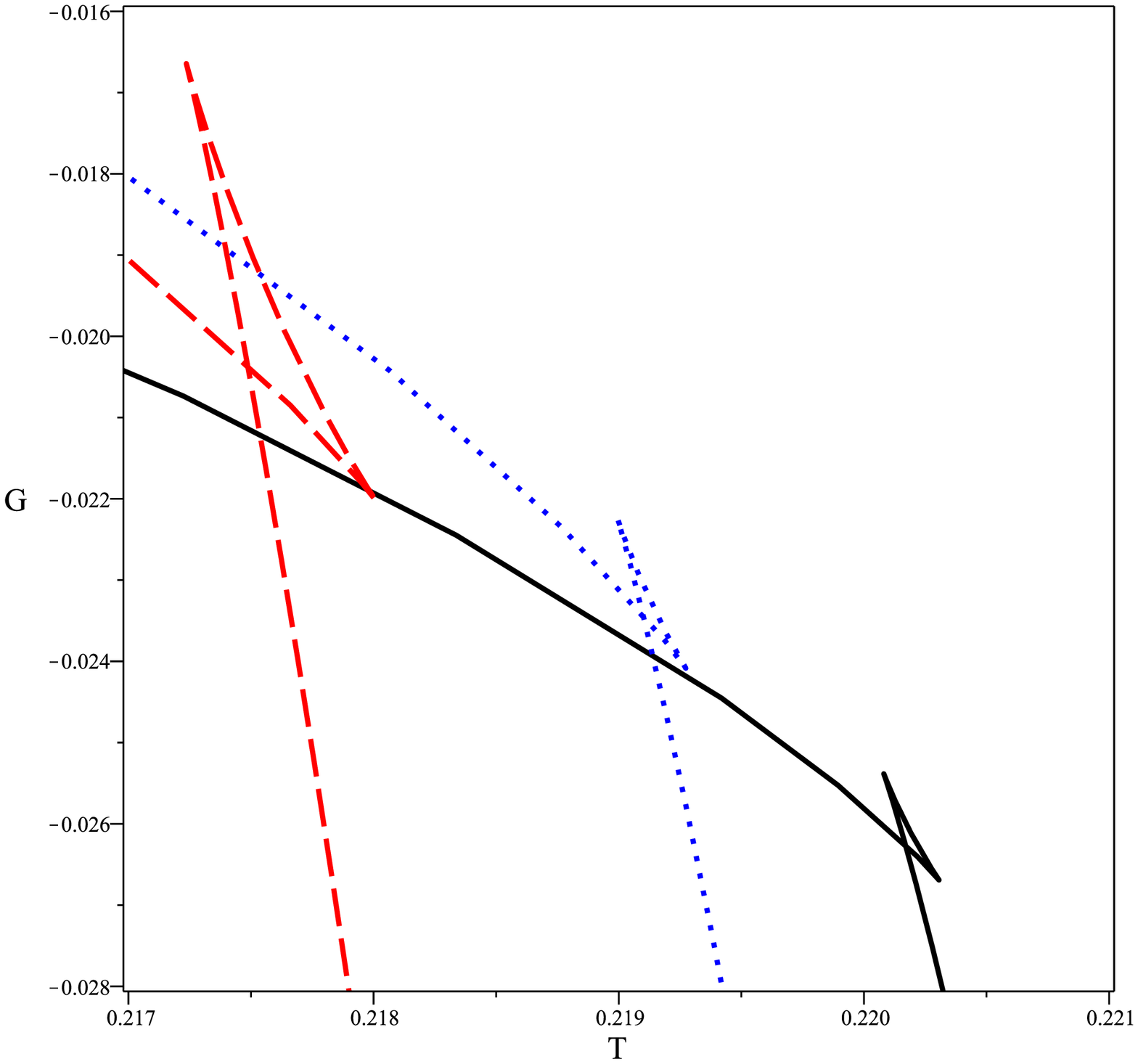}%
\end{array}
$%
\caption{\textbf{Lovelock-BI gravity's rainbow:} $P-V$ for $T<T_{c}$ (left),
$T-V$ for $P<P_{c}$ (middle) and $G-T$ for $P<P_{c}$ (right) diagrams for $%
k=1$, $d=7$, $q=1$, $f(\protect\varepsilon)=g(\protect\varepsilon)=0.9$, $%
\protect\alpha=0.1$, and $\protect\beta=0.1$ (continuous line), $\protect%
\beta=0.5$ (dotted line) and $\protect\beta=5$ (dashed line).}
\label{PV-BI-Comp}
\end{figure}
\begin{figure}[tbp]
$%
\begin{array}{ccc}
\epsfxsize=5cm \epsffile{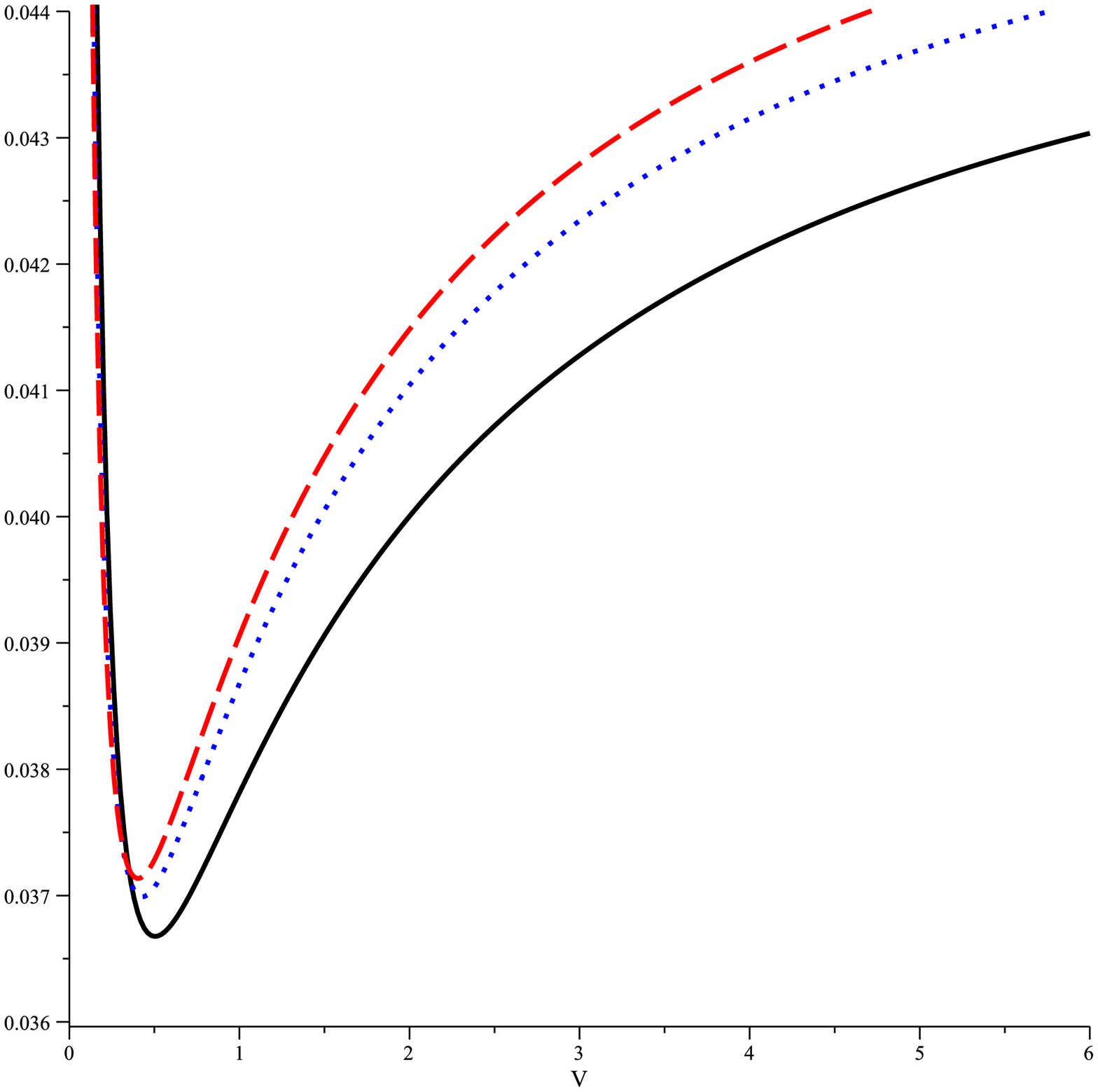} & \epsfxsize=5.5cm \epsffile{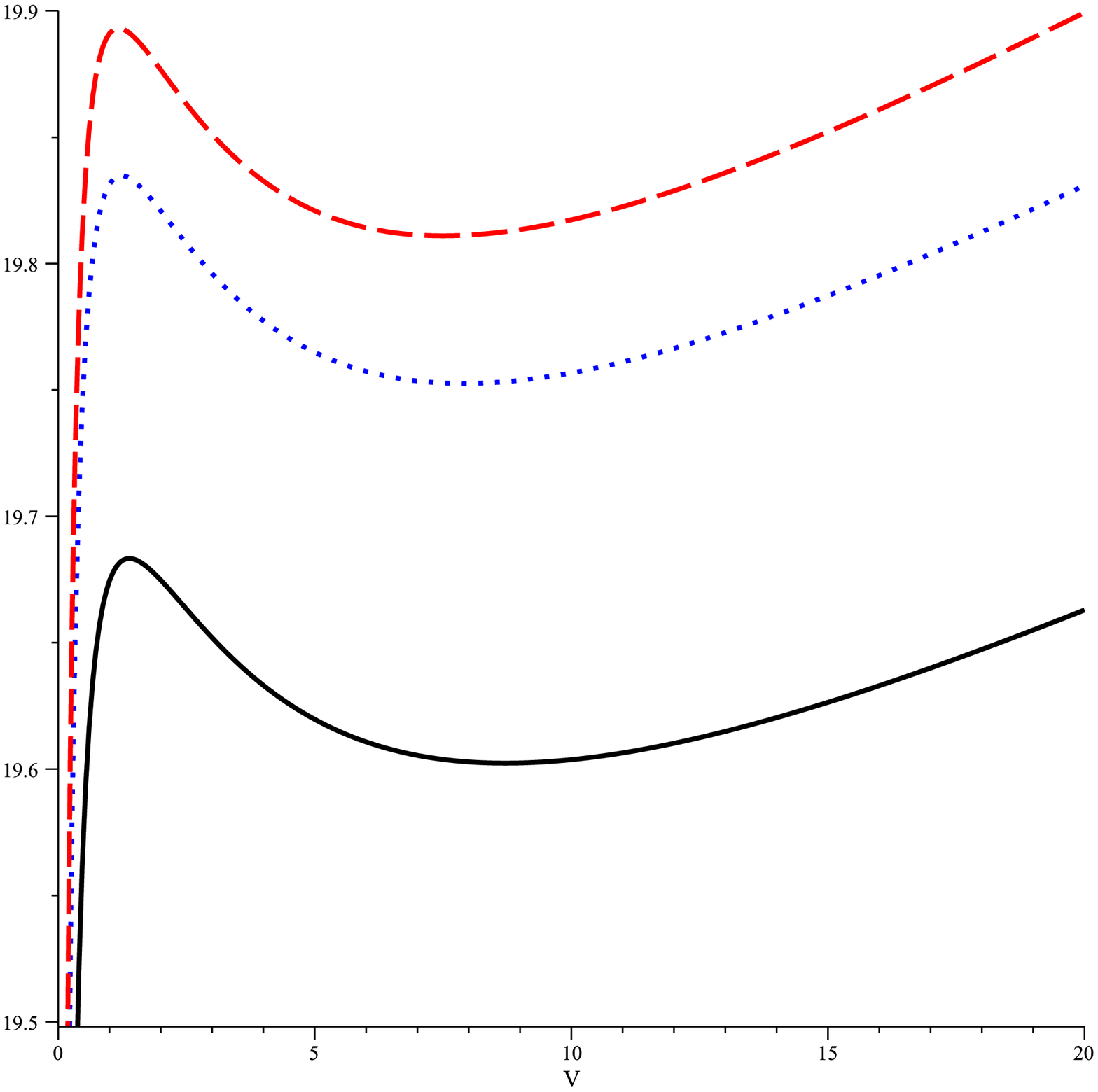} & %
\epsfxsize=5.5cm \epsffile{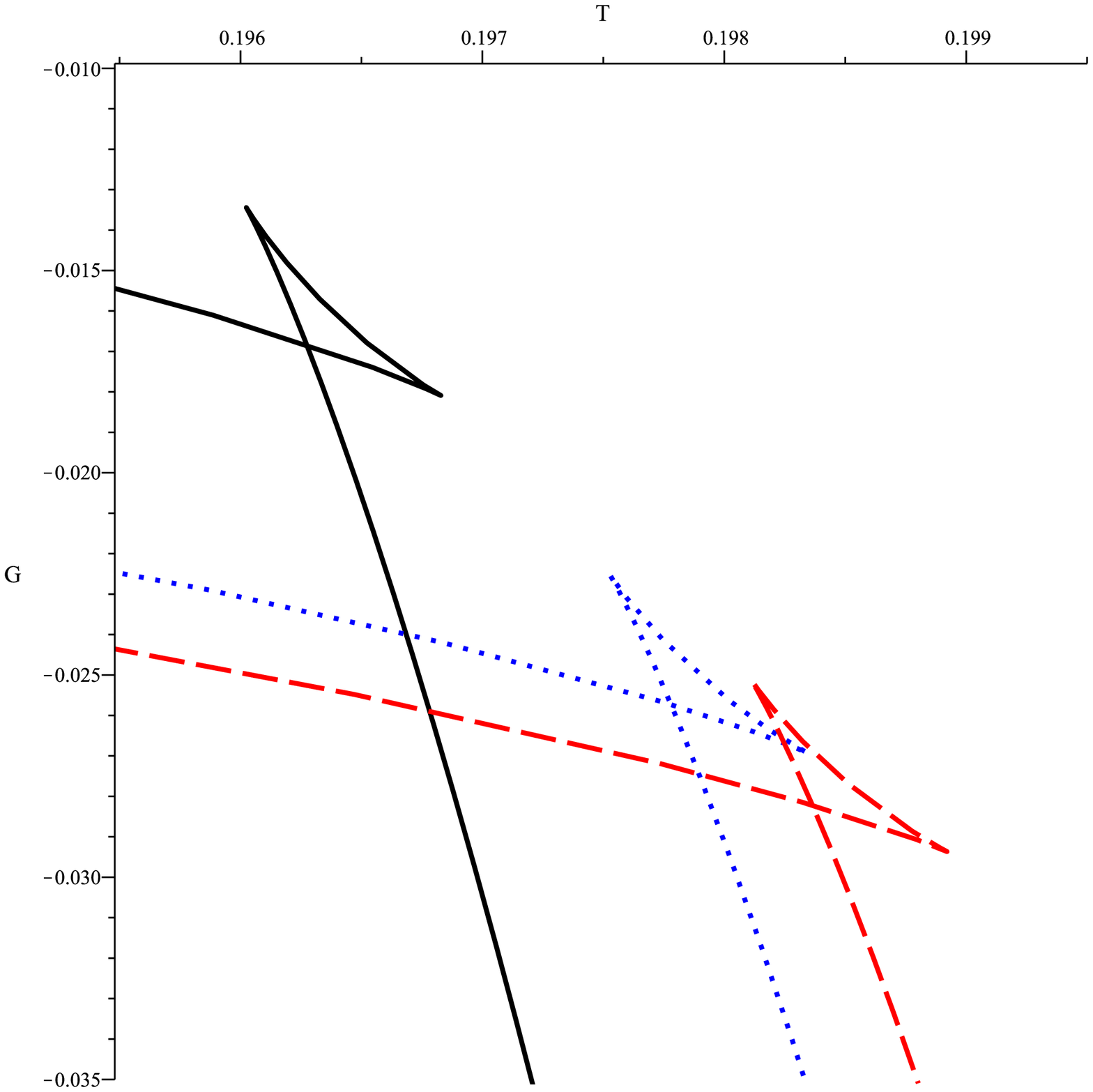}%
\end{array}
$%
\caption{\textbf{Lovelock-BI gravity's rainbow:} $P-V$ for $T<T_{c}$ (left),
$T-V$ for $P<P_{c}$ (middle) and $G-T$ for $P<P_{c}$ (right) diagrams for $%
k=1$, $d=7$, $q=1$, $\protect\alpha=0.5$, $\protect\beta=0.05$, $f(\protect%
\varepsilon)=1$ and $g(\protect\varepsilon)=0.8$(continuous line), $g(%
\protect\varepsilon)=1$ (dotted line) and $g(\protect\varepsilon)=1.2$
(dashed line).}
\label{PVTVGT-gE}
\end{figure}
\begin{figure}[tbp]
$%
\begin{array}{ccc}
\epsfxsize=5cm \epsffile{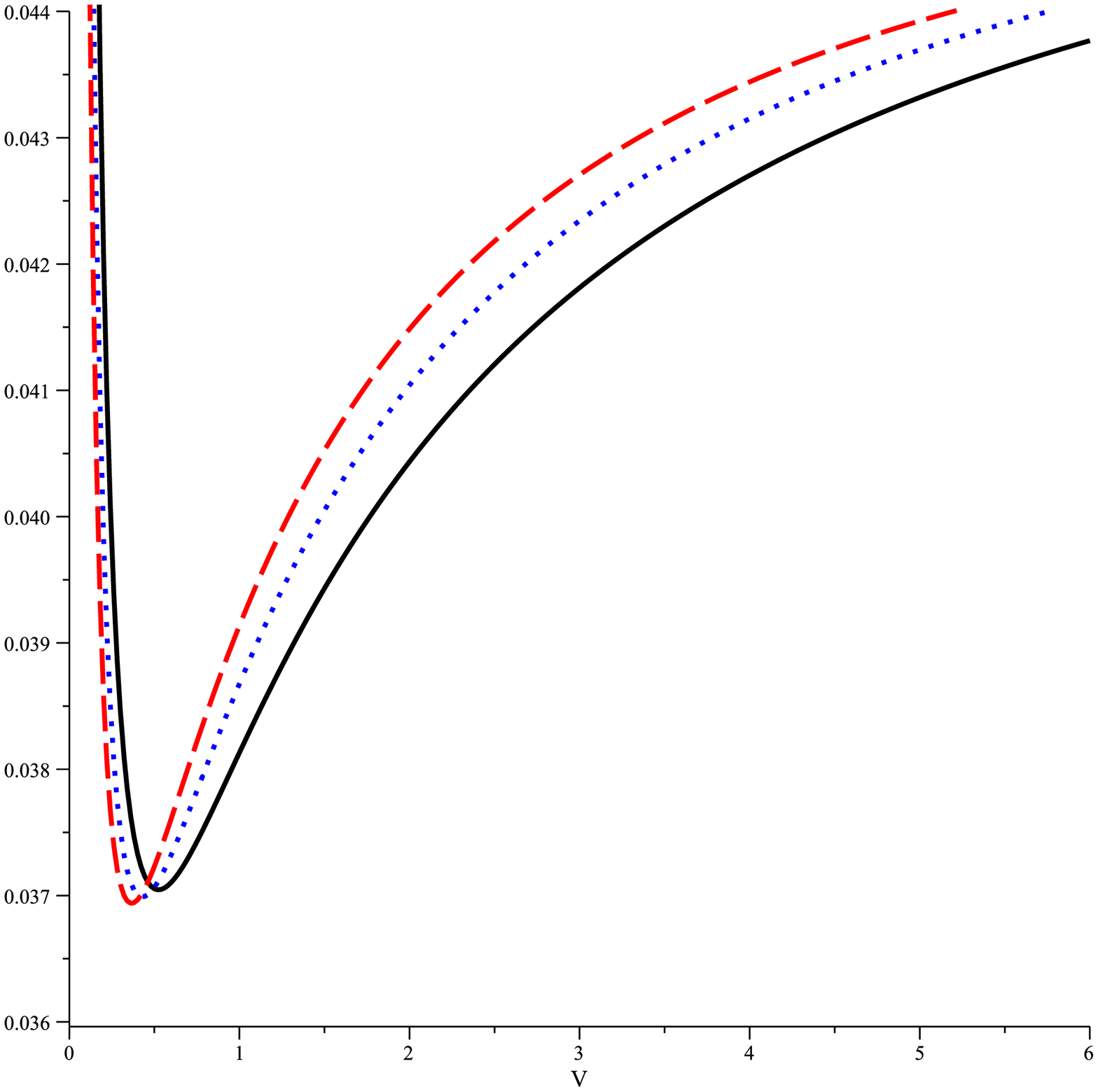} & \epsfxsize=5.5cm \epsffile{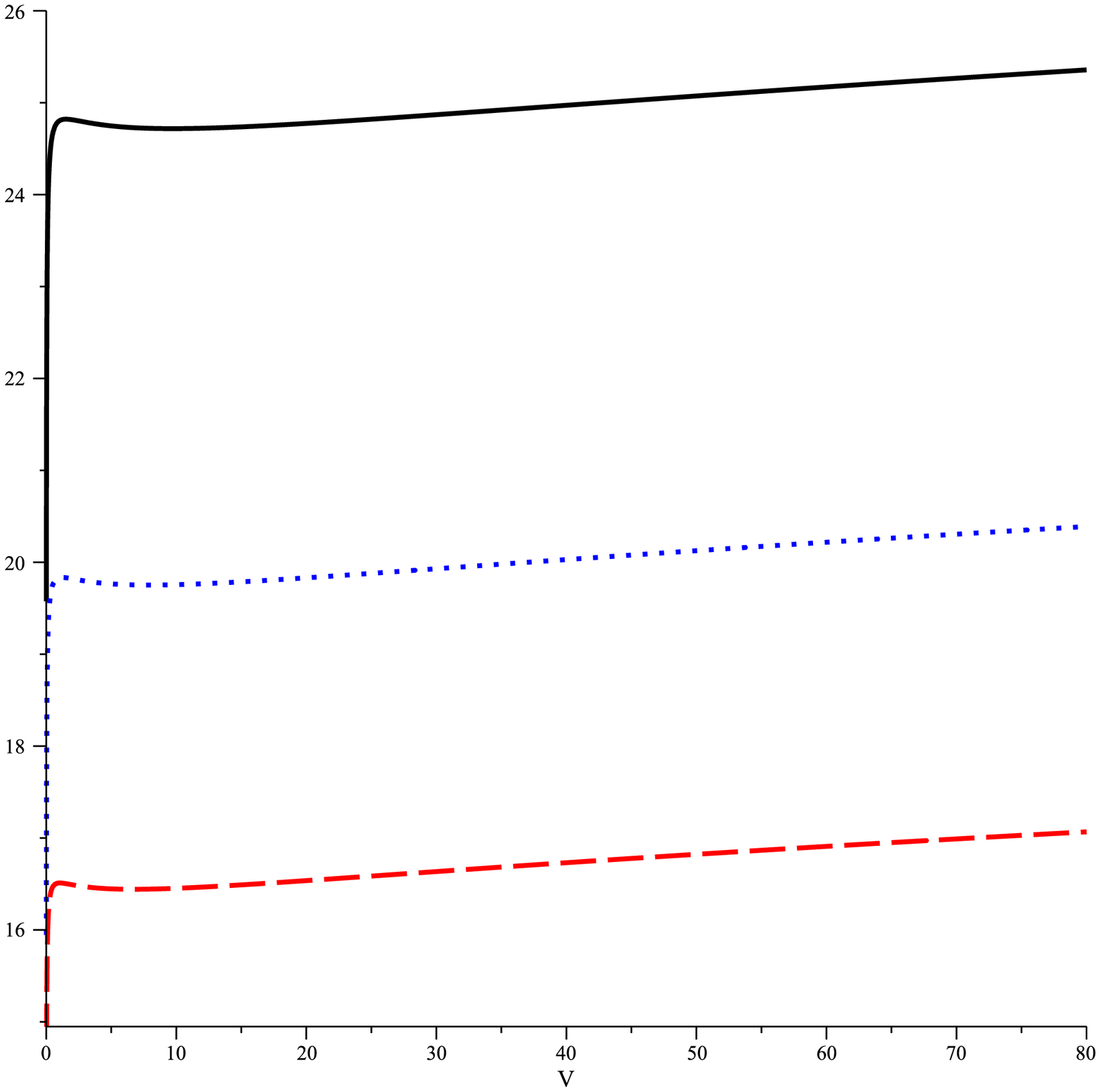} & %
\epsfxsize=5.5cm \epsffile{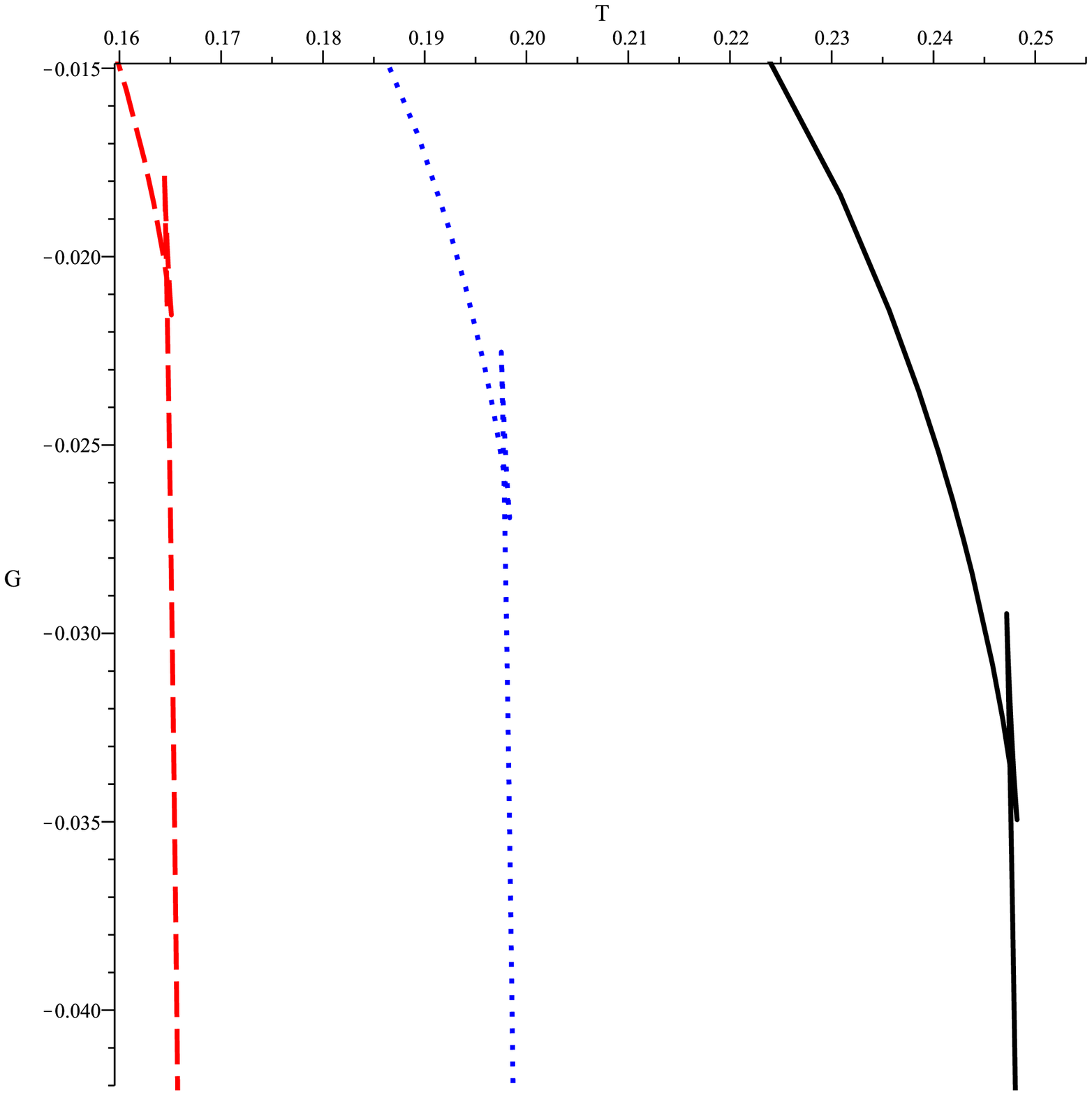}%
\end{array}
$%
\caption{\textbf{Lovelock-BI gravity's rainbow:} $P-V$ for $T<T_{c}$ (left),
$T-V$ for $P<P_{c}$ (middle) and $G-T$ for $P<P_{c}$ (right) diagrams for $%
k=1$, $d=7$, $q=1$, $\protect\alpha=0.5$, $\protect\beta=0.05$, $g(\protect%
\varepsilon)=1$ and $f(\protect\varepsilon)=0.8$(continuous line), $f(%
\protect\varepsilon)=1$ (dotted line) and $f(\protect\varepsilon)=1.2$
(dashed line).}
\label{PVTVGT-fE}
\end{figure}


\begin{center}
\begin{tabular}{ccccc}
\hline\hline
$\alpha $ & $V_{c}$ & $T_{c}$ & $P_{c}$ & $\frac{P_{c}V_{c}}{T_{c}}$ \\
\hline\hline
$0.0010$ \  & $0.9223$ \  & $0.4800$ \  & $0.1858$ \  & $0.3571$ \\ \hline
$0.0100$ & $0.9505$ & $0.4681$ & $0.1792$ & $0.3639$ \\ \hline
$0.1000$ & $1.2680$ & $0.3784$ & $0.1303$ & $0.4366$ \\ \hline
$0.5000$ & $4.7578$ & $0.2205$ & $0.0516$ & $1.1134$ \\ \hline
$0.9000$ & $18.2741$ & $0.1664$ & $0.0299$ & $3.2801$ \\ \hline
\end{tabular}
\\[0pt]
\vspace{0.1cm} Table II: Lovelock-Maxwell solutions: $k=1$, $q=1$, $%
f(\varepsilon)=0.9$, $g(\varepsilon)=0.9$ and $d=7$. \vspace{0.5cm}

\begin{tabular}{ccccc}
\hline\hline
$d$ & $V_{c}$ & $T_{c}$ & $P_{c}$ & $\frac{P_{c}V_{c}}{T_{c}}$ \\
\hline\hline
$7$ \  & $4.7578$ \  & $0.2205$ \  & $0.0516$ \  & $1.1134$ \\ \hline
$8$ & $3.8650$ & $0.2886$ & $0.0872$ & $1.1674$ \\ \hline
$9$ & $3.2021$ & $0.3579$ & $0.1333$ & $1.1932$ \\ \hline
$10$ & $2.7139$ & $0.4282$ & $0.1907$ & $1.2087$ \\ \hline
\end{tabular}
\\[0pt]
\vspace{0.1cm} Table III: Lovelock-Maxwell solutions: $k=1$, $q=1$, $%
f(\varepsilon)=0.9$, $g(\varepsilon)=0.9$ and $\alpha =0.5$.

\begin{tabular}{ccccc}
\hline\hline
$f(\varepsilon)$ & $V_{c}$ & $T_{c}$ & $P_{c}$ & $\frac{P_{c}V_{c}}{T_{c}}$
\\ \hline\hline
$0.6$ & $5.7837$ & $0.3331$ & $0.05277$ & $0.9163$ \\ \hline
$0.8$ & $4.9952$ & $0.2487$ & $0.0520$ & $1.0445$ \\ \hline
$1.0$ & $4.5803$ & $0.1980$ & $0.0512$ & $1.1846$ \\ \hline
$1.2$ & $4.3399$ & $0.1642$ & $0.0504$ & $1.3327$ \\ \hline
\end{tabular}
\\[0pt]
\vspace{0.1cm} Table IV: Lovelock-Maxwell solutions: $k=1$, $q=1$, $%
g(\varepsilon)=0.9 $, $d=7$ and $\alpha =0.5$.

\begin{tabular}{ccccc}
\hline\hline
$g(\varepsilon)$ & $V_{c}$ & $T_{c}$ & $P_{c}$ & $\frac{P_{c}V_{c}}{T_{c}}$
\\ \hline\hline
$0.6$ & $21.1861$ & $0.2036$ & $0.0410$ & $4.2655$ \\ \hline
$0.8$ & $6.6801$ & $0.2176$ & $0.0495$ & $1.5193$ \\ \hline
$1.0$ & $3.8300$ & $0.2221$ & $0.0528$ & $0.9106$ \\ \hline
$1.2$ & $3.1450$ & $0.2232$ & $0.0537$ & $0.7572$ \\ \hline
\end{tabular}
\\[0pt]
\vspace{0.1cm} Table V: Lovelock-Maxwell solutions: $k=1$, $q=1$, $%
f(\varepsilon)=0.9$, $d=7$ and $\alpha =0.5$.

\begin{tabular}{ccccc}
\hline\hline
$\beta $ & $V_{c}$ & $T_{c}$ & $P_{c}$ & $\frac{P_{c}V_{c}}{T_{c}}$ \\
\hline\hline
$0.01000$ & $2.9612$ & $0.2234$ & $0.0539$ & $0.7144$ \\ \hline
$0.05000$ & $3.2919$ & $0.2222$ & $0.0530$ & $0.7851$ \\ \hline
$0.10000$ & $3.8407$ & $0.2213$ & $0.0523$ & $0.9069$ \\ \hline
$0.50000$ & $4.7025$ & $0.2206$ & $0.0516$ & $1.1009$ \\ \hline
$1.00000$ & $4.7439$ & $0.2206$ & $0.0516$ & $1.1102$ \\ \hline
\end{tabular}
\\[0pt]
\vspace{0.1cm} Table VI: Lovelock-BI solutions: $k=1$, $q=1$, $%
f(\varepsilon)=0.9$, $g(\varepsilon)=0.9$, $d=7$ and $\alpha =0.5$.
\end{center}

In Figs. \ref{PV-Max}-\ref{PVTVGT-fE} we show the critical behavior of the
system in both Lovelock-Maxwell and Lovelock-BI gravity's rainbow. Also, we
present five tables to investigate the effects of various parameters on the
critical point. According to these figures and related tables we find that
the Lovelock parameter ($\alpha$), nonlinearity parameter ($\beta$),
dimensionality ($d$) and rainbow functions affect the critical point,
significantly. In addition, variation of these parameters can change the
near universal ratio $\frac{P_{c}V_{c}}{T_{c}}$, considerably.

Regarding Figs. \ref{PVTVGT-gE} and \ref{PVTVGT-fE} with tables IV
and V, we can find the effects of changing rainbow functions with
more details. Based on Fig. \ref{PVTVGT-gE} and table V, we find
that the critical values
of pressure and temperature (volume) increase (decreases) by increasing $%
g(\varepsilon) $. While Fig. \ref{PVTVGT-fE} and table IV show
that increasing $f(\varepsilon)$ leads to decreasing all critical
values. Looking at the relation $\frac{P_{c}V_{c}}{T_{c}}$, we
find that it is an increasing function of $f(\varepsilon)$, as
opposed to increasing $g(\varepsilon)$ where such ratio is a
decreasing function. Therefore, one can adjust the values of these
parameters to cancel the effects of all free parameters and obtain
the universal ratio.

\section{Conclusion}

In this paper, we have analyzed the thermodynamics of black holes in
Lovelock gravity's rainbow. We have obtained black hole solutions for both
Lovelock-Maxwell and Lovelock-BI gravity's rainbow theories with different
horizon topologies. We have also found that the asymptotical behavior of the
solutions may be (a)dS or flat with an effective energy dependent
cosmological constant. We have calculated conserved and thermodynamic
quantities in the energy dependent background and found that although
rainbow functions may change these quantities, the first law of
thermodynamics is valid. In addition, we have examined thermal stability of
the solutions in the context of canonical ensemble. We have shown depending
on the chosen parameters, there is a critical horizon radius ($r_{+u}$), in
which for $r_{+}>r_{+u}$ the black holes are thermally stable.

At last, we have investigated the critical behavior and phase transition in
the extended phase space by considering the proportionality between $\Lambda$
and pressure. We have studied the effects of $\alpha$, $\beta$, $d$ and
rainbow functions on the critical values and found that they can change
critical point, significantly. In other words, we have shown that in
addition to Lovelock and nonlinearity parameters, rainbow functions affect
thermodynamic nature of the black hole. We have found that depending on the
values of rainbow functions, a phase transition can occur. While one could
adjust rainbow functions to obtain thermally stable black holes. In brief,
we should note that rainbow functions affect various aspects of a black hole
such as: strength of curvature and singularity, place and type of event
horizon, asymptotical behavior, thermodynamic quantities, thermal stability
and existence of phase transition.

It may be noted that various interesting systems have been analyzed using
Lovelock gravity. The quasinormal modes of black holes in Lovelock gravity
has been studied \cite{qs}. In this study, the WKB method was applied for
analyzing the quasinormal frequencies Lovelock gravity. It would be
interesting to analyze the quasinormal modes of black holes in Lovelock
gravity's rainbow. It is expected that the choice of the rainbow functions
will effect the behavior of these quasinormal modes. A scalar-tensor version
of Lovelock theory with a non trivial higher order galileon term involving
the coupling of the Lovelock two tensor with derivatives of the scalar
galileon field has been construed \cite{t1}. It would be interesting to
construct this theory using Lovelock gravity's rainbow, and then use it for
analyzing black hole solutions. It has also been demonstrated that a black
remnant forms in gravity's rainbow \cite{a1}, and this has important
phenomenological consequences for the detection of mini black holes at the
LHC \cite{f1}. It would be interesting to investigate the formation of such
black remnants in Lovelock gravity's rainbow. In fact, it has been
demonstrated that such remnants will form for all black objects in usual
gravity's rainbow \cite{f4}. It will be interesting to investigate if such a
result holds for all black holes in Lovelock gravity's rainbow.


\begin{acknowledgements}
We gratefully thank the anonymous referee for enlightening
comments and suggestions which substantially helped in improving
the quality of the paper. We also acknowledge M. Momennia for
reading the manuscript. SHH and AD wish to thank Shiraz University
Research Council. This work has been supported financially by the
Research Institute for Astronomy and Astrophysics of Maragha,
Iran.
\end{acknowledgements}

\end{document}